\documentclass[reqno, 11pt]{amsart}

\usepackage{fullpage} 
\usepackage{srcltx} \usepackage{pdfsync}
\usepackage{calc,amsfonts,amsthm,amscd,epsfig,psfrag,amsmath,amssymb,enumerate,graphicx}

\numberwithin{equation}{section}
\usepackage{enumerate}
\renewcommand\r {\rho}
\newcommand{\tc}{\, | \,}
\newcommand\wt{\widetilde}
\newcommand\z {\zeta}

\newcommand\cL{\mathcal L}
\newcommand\si{\sigma}
\newcommand\cC{\mathcal C}
\newcommand\cS{\mathcal S}
\newcommand\Var{{\rm {Var}}}

\newcommand\cG{\mathcal G}

\newcommand\tmix{{\rm T_{mix}}}
\newcommand\tmixinf{{\rm T^\infty_{mix}}}
\newcommand\G {\Gamma}
\newcommand\bbN {{\mathbb N}}
\newcommand\bbR {{\mathbb R}}
\newcommand\bbZ {{\mathbb Z}}
\renewcommand\O {\Omega}

\newcommand{\sumtwo}[2]{\sum_{\substack{#1 \\ #2}}} 
\newcommand{\inftwo}[2]{\inf_{\substack{#1 \\ \smallskip\\#2}}} 
\newcommand\bbP{{\mathbb P}}
\newcommand\bbE{{\mathbb E}}
\newcommand\bE{{\bf  E}}
\newcommand\rb{{\rm b}}
\newcommand\Mu{\mu_{\bfp}}
\newcommand\bfp{{\bf p}}
\newcommand\rw{{\rm w}}

\newcommand\gap {{\rm gap}}

\newcommand\bP{{\bf P}}

\newcommand\NN {{\mathbb N}}
\newcommand\RR {{\mathbb R}}

\newcommand\ZZ {{\mathbb Z}}

\usepackage{amssymb, amsmath}
\newtheorem{rem}{Remark}

\newtheorem{prop}{Proposition}
\newtheorem{theo}{Theorem}
\newtheorem{lemma}{Lemma}
\newtheorem{definition}{Definition}
\title[]{}

\begin{document}

\title[]{``Zero'' temperature stochastic 3D Ising model and dimer covering fluctuations: a
first step towards interface mean curvature motion}
\author[P.~Caputo]{Pietro Caputo}
\address{P. Caputo, Dipartimento di Matematica,
  Universit\`a Roma Tre, Largo S.\ Murialdo 1, 00146 Roma, Italia.
  e--mail: {\tt caputo@mat.uniroma3.it}}
\author[F.~Martinelli]{Fabio Martinelli} 
\address{F. Martinelli, Dipartimento di Matematica,
  Universit\`a Roma Tre, Largo S.\ Murialdo 1, 00146 Roma, Italia.
  e--mail: {\tt martin@mat.uniroma3.it}} 
\author[F.~Simenhaus]{Fran\c{c}ois Simenhaus}
\address{F. Simenhaus, Dipartimento di Matematica,
  Universit\`a Roma Tre, Largo S.\ Murialdo 1, 00146 Roma, Italia.
  e--mail: {\tt simenhaus@mat.uniroma3.it}}
\author[F.L.~Toninelli]{Fabio Lucio Toninelli} 
\address{F. L. Toninelli, CNRS and ENS Lyon, Laboratoire de Physique\\
  46 All\'ee d'Italie, 69364 Lyon, France.  e--mail: {\tt
    fabio-lucio.toninelli@ens-lyon.fr}}

\thanks{This work was
  supported by the European Research Council through the ``Advanced
  Grant'' PTRELSS 228032}

\begin{abstract}
 We consider the Glauber dynamics for the Ising model with ``$+$'' boundary
 conditions, at zero temperature or at temperature which goes to zero
 with the system size   (hence the quotation marks in the title). In dimension $d=3$ we prove that an initial
domain of linear size $L$ of ``$-$'' spins disappears within a 
time $\tau_+$ which is at most $L^2(\log L)^c$ and at least
$L^2/(c\log L)$,
for some $c>0$. The proof of the upper bound proceeds via comparison with an auxiliary
dynamics which mimics the motion by mean curvature that is expected
to describe, on large time-scales, the evolution of the interface
between ``$+$'' and ``$-$'' domains. The analysis of the auxiliary
dynamics requires recent results on the fluctuations of the height
function associated to dimer
coverings of the infinite honeycomb lattice. Our result, apart from the
spurious logarithmic factor, is the first rigorous confirmation of the
expected behavior $\tau_+\simeq const\times L^2$, 
conjectured on
heuristic grounds \cite{Li,FH}. In dimension $d=2$, $\tau_+$
can be shown to be of order $L^2$ without
logarithmic corrections: the upper bound was proven in
\cite{FSS} and here we provide the lower bound. For $d=2$, 
we also prove that the spectral gap of the generator behaves like $c/L$ for $L$ large,
 as conjectured in \cite{BM}.
\end{abstract}

\maketitle

\thispagestyle{empty}

\section{Introduction}
A long standing open problem in the mathematical analysis
of the stochastic Ising model \cite{cf:notefabio} can be described as follows. 

Consider the
standard $\pm 1$ spin Ising model at inverse
temperature $\beta$ and zero external magnetic field in a cubic box 
$\Lambda\subset \ZZ^d$, $d\ge 2$, of side $L$, with homogeneous, e.g.\ ``plus'', boundary
conditions outside $\Lambda$ (see Section
\ref{sec:model} for the precise definition). Denote by $\pi_\Lambda^+$ the corresponding Gibbs
measure and assume that $\beta$ is larger
(or much larger) than the critical value $\beta_c$ where the ``plus'' and
the ``minus'' phases start to coexist. 
On the spin
configuration space $\Omega_\Lambda$ consider also the continuous time Markov
chain, reversible w.r.t.\ the Gibbs measure $\pi_\Lambda^+$, in which each
spin $\sigma_x$, $x\in \Lambda$, with rate one chooses a new value
from $\pm 1$ with probabilities given by the conditional Gibbs measure
given the current values of the spins outside the site $x$. Such a chain is known in the literature as \emph{Gibbs sampler}
or \emph{Glauber chain} and, because of reversibility, its unique
stationary distribution coincides with $\pi_\Lambda^+$. Let $\mu_t^\sigma$ denote the distribution of
the chain at time $t$ when the initial configuration at time $t=0$ is
$\sigma\in \{-1,1\}^\Lambda$ and let 
$T_{\rm mix}$  be the minimum time such that the variation distance
between $\mu_t^\sigma$ and  $\pi_\Lambda^+$ is smaller than, e.g.,
$1/(2e)$ for any $\sigma$. Because $\Lambda$ is a finite set and the
chain is ergodic, the
Perron-Frobenius theorem guarantees that $T_{\rm mix}$ is finite and it is not too
difficult to prove that $T_{\rm mix}\le e^{c \beta L^{d-1}}$ for some
constant $c$ (see e.g.\ \cite{cf:notefabio}). Despite the fact that
the above upper bound is known to be the correct
order of growth of $T_{\rm mix}$ when the boundary conditions are
absent, i.e.\ free (see e.g.\ \cite{cf:Thomas, cf:notefabio}), it is known
that the
presence of the 
homogeneous ``plus'' boundary conditions drastically modifies
the whole process of relaxation to equilibrium. In this case the conjectured correct growth of $T_{\rm mix}$ as a
function of $L$ is given by the Lifshitz's law $T_{\rm
  mix}=O(L^2)$ (see
\cite{Li} and \cite{FH}). 

Unfortunately, \emph{any} polynomial upper bound on $T_{\rm
  mix}$ has escaped so far a rigorous analysis and the best known
result, following a
recent breakthrough \cite{MT}, is confined to the two dimensional
case $d=2$ and it is of the form
$T_{\rm mix}\le \exp(c\log(L)^2)$ for any $\beta>\beta_c$ (see \cite{LMST}).        

The conjectured $L^2$ growth of $T_{\rm mix}$ is related to the
shrinking of a ``spherical'' bubble of the ``minus'' phase under the
influence of the ``plus'' boundary conditions. On a macroscopic scale one
expects \cite{FH,Spo} that the dynamics of the bubble 
follows a \emph{motion by mean curvature} which, if true, implies
immediately that in a time $O(L^2)$ the bubble disappears and
equilibrium is achieved.  

Some of the above questions and in particular the validity of the
Lifshitz's law can be formulated
(and its mathematical justification remains highly non-trivial) even at zero temperature ($\beta=+\infty$), a situation that
was considered in great detail in \cite{FSS} (see also
\cite{CSS}, \cite{Spo}). Similarly one can consider, as we do
in the sequel, an almost zero temperature, i.e.\ $\beta=\beta(L)$ and increasing so fast with
$L$ that thermal
fluctuations become irrelevant on the relaxation process. In the
extreme case $\beta=+\infty$,
the stationary distribution is concentrated on the ``plus'' configuration
(i.e.\ all spins equal to $+1$) and the Glauber
chain evolves towards it without ever increasing the spin energy. The
mixing time $T_{\rm mix}$ becomes closely related to the
hitting time of the ``plus'' configuration starting from all minuses  and it is not too hard to prove,
by induction on the dimension $d$ (starting from the case $d=2$, which
requires non-trivial work \cite{FSS}), an upper bound of the form $T_{\rm
  mix}\le c L^d$. However the inductive argument, which in $d=3$
simply boils down to comparing the true evolution of the original
cubic bubble of minuses with an auxiliary chain in which the $L$ two-dimensional square layers of the bubble disappear one after the other starting e.g.\ from the top
one, completely neglects the interesting cooperative effect in which
the whole bubble, starting from the corners,
is eroded by a ``mean curvature effect'' which enhances considerably its shrinking. In other words, the Lifshitz's law in $d=3$
cannot be proved or even approached
closely without considering in detail how a two-dimensional curved interface
separating the pluses from the minuses evolves in time.              

The first main contribution of the present work is a strategy to
attack and solve the above
problem at zero temperature for $d=3$ (with logarithmic corrections)
and for $d=2$ (with different constants in the upper and lower
bounds). 
We refer to Section \ref{sec:results} for the precise statements.
In the challenging three dimensional case our method involves the use and
adaptation to our specific situation of
two different and beautiful sets of results concerning:
\begin{enumerate}[(i)]
\item the Gaussian Free Field-like equilibrium fluctuations of \emph{random monotone surfaces} and
  their connection with \emph{random dimer coverings} of the two-dimensional
  hexagonal lattice (see
  \cite{CK, KenyonCMP, Kenyon, KOS, Sheffield}); 
\item the mixing time of a Glauber chain for monotone surfaces \cite{Wilson}.    
\end{enumerate}
The key point of our approach is to prove that the evolution of the Glauber chain is
dominated by that of another effective chain which follows a motion by mean
curvature, only slowed down by a logarithmic factor (in the radius of
the bubble). In turn the evolution of the bounding chain is constructed by
``peeling off'' at time $t$ a layer of logarithmic width from the
bubble of radius $R_t$. The peeling process is realized by letting relax to equilibrium a
mesoscopic spherical cap of radius $O(\sqrt{R_t}\times polylog(R_t))$
and height $O(polylog(R_t))$ (where $polylog(x)$ stands for a suitable
polynomial of $\log x$) centered at each
point of the surface of the bubble. Results (i) above are essential in order to prove the
peeling effect while the results (ii) prove that the overall effect
occurs on the correct time scale $O(R_t \times polylog(R_t))$. 
We strongly believe that our approach can be helpful in
solving other related problems. 

The second contribution, this time for the model in dimension
 $d=2$, are upper and lower bounds, \emph{linear}
in $L^{-1}$ and \emph{uniform} in $\beta>c\log L$, $c$ large enough, on the \emph{spectral gap} (see \eqref{eq:72})
of
the Glauber chain. In \cite{BM} for $d=2$ the upper bound was proved (apart from
logarithmic corrections and with constants depending on $\beta$) for any $\beta>\beta_c$ and it was conjectured to
be the correct behavior of the spectral gap (to be compared with the
$L^2$ scaling of the mixing time). In our case the proof of the lower
bound (the most interesting one) has an analytic flavor. We first unitarily transform the original Markov generator of the
Glauber chain acting on $\ell^2(\Omega_\Lambda,\pi_\Lambda^+)$ into a new matrix
acting on $\ell^2(\Omega_\Lambda)$ with the flat (i.e.\ counting) measure, whose off-diagonal elements corresponding to
spin transitions which \emph{do not conserve the energy} vanish very fast as
$\beta\to \infty$. Such a property, for $\beta \ge c\log L$,  allows us to write the new matrix
into a block-form plus a remainder whose norm is very small with
$L$. Since each block describes a suitably killed Glauber chain (killing occurs as soon as the spin energy decreases) the desired bound follows by a
probabilistic analysis of the killing time for each block.  
\tableofcontents
\section{Model and preliminaries}
\label{sec:model}
Given $x,y\in\ZZ^d$, we write $x\sim y$ if $x$ is a nearest neighbor of $y$.
To each point $x\in \ZZ^d$ is assigned a spin $\sigma_x$ which takes values in $\{-1,+1\}$. Given $\Lambda\subset \ZZ^d$, we let $$\partial \Lambda:=\{x\in \ZZ^d\setminus \Lambda:\exists y\in\Lambda\mbox{\;such that\;}x\sim y\}$$ and $\Omega_\Lambda:=\{-1,+1\}^\Lambda$
be the set of all possible configurations in $\Lambda$. 
The Gibbs measure  in a finite domain $\Lambda$ for a boundary
condition  $\eta\in\Omega_{\partial \Lambda}$ at inverse temperature $\beta>0$
is given by
\begin{eqnarray}
  \label{eq:32}
  \pi^\eta_\Lambda(\sigma)=\frac{e^{-\beta H^\eta_\Lambda(\sigma)}}{Z^\eta_\Lambda}
\end{eqnarray}
with
\begin{eqnarray}
  \label{eq:33}
  H^\eta_\Lambda(\sigma):=-\sumtwo{x,y\in\Lambda}{x\sim y}\sigma_x\sigma_y-\sumtwo{x\in\Lambda,y\in\partial \Lambda}{x\sim y}\sigma_x\eta_y
\end{eqnarray}
and 
\begin{eqnarray}
  \label{eq:34}
  Z^\eta_\Lambda:=\sum_{\sigma\in \Omega_\Lambda}e^{-\beta H^\eta_\Lambda(\sigma)}.
\end{eqnarray}
When no danger of confusion arises, we will write just $\pi$ for $\pi^\eta_\Lambda$.
We will consider also the case $\beta=+\infty$, in which case   $\pi^\eta_\Lambda$ is taken to be the uniform measure over all configurations
$\sigma\in\Omega_\Lambda$ which minimize $H^\eta_\Lambda$.

The dynamics we consider is the Glauber dynamics
$\{\sigma^\xi(t)\}_{t\ge0}$, 
where $\xi=\sigma^\xi(0)$ is the initial condition.
To each $x\in\Lambda$ is associated an independent Poisson clock of
rate $1$. When the clock labeled $x$ rings, say at time $s$, one
replaces $\sigma_x$ with a value sampled from the probability
distribution
$\pi_{x,\sigma^\xi(s)}(\cdot)$, where 
\begin{eqnarray}
  \label{eq:113}
\pi_{x,\sigma}(\cdot):=\pi(\cdot|\sigma_y,y\ne x).   
\end{eqnarray}
We denote 
the law of
$\sigma^\xi(t)$
for a given  time $t$ as $\mu^\xi_t$. It is well known that
$\{\sigma^\xi(t)\}_{t\ge0}$
is a Markov process, reversible with respect to the equilibrium
measure $\pi$. 

We  are interested in the order of magnitude of the 
``mixing time'', defined for some $\epsilon\in(0,1)$ as
\begin{eqnarray}
  \label{eq:76}
  \tmix(\epsilon)=\inf\bigl\{t>0:\sup_{\sigma\in\Omega_\Lambda}\|\mu^\sigma_t-
\pi\|\le \epsilon\bigr\}
\end{eqnarray}
where 
$$\|\mu-\nu\|=\frac12\sum_{\sigma\in\Omega_\Lambda}|\mu(\sigma)-\nu(\sigma)|$$
denotes the total variation distance between two probability measures.
When $\epsilon=1/(2e)$, we will just write $\tmix$ for
$\tmix(\epsilon)$.
 Another quantity we will focus on is the
spectral gap,
\begin{eqnarray}
  \label{eq:72}
  \gap=\gap^\eta_\Lambda=\inf\frac{\pi^\eta_\Lambda(f(-\mathcal L)f)}{\Var_{\pi^\eta_\Lambda}(f)}
\end{eqnarray}
where $\Var_{\pi^\eta_\Lambda}$ denotes the variance
w.r.t.\ $\pi^\eta_\Lambda$, $\mathcal L$ is the generator of the dynamics,
\begin{eqnarray}
  \label{eq:106}
  \mathcal L f(\sigma)=\sum_{x\in\Lambda}\left(\pi_{x,\sigma}(f)-f(\sigma)
\right)
\end{eqnarray}
(a more explicit expression for the generator is given in Section 
\ref{sec:LBgap3D})
 and the infimum is
taken over non-constant functions
$f:\Omega_\Lambda\mapsto\RR$. The inverse of the gap (respectively the mixing
time) measures the speed of
convergence 
to equilibrium in $L^2(d\pi)$ (resp. in total variation norm).
Also, it is well known that
\begin{eqnarray}
  \label{eq:90}
  \sup_{\sigma\in\Omega_\Lambda}\|\mu_t^\sigma-\pi\|\le e^{-\lfloor t/\tmix\rfloor},
\end{eqnarray}
which actually holds for any reversible Markov process.

\subsection{Preliminary results}

\subsubsection{Monotonicity and global coupling}
\label{sec:monotone}
For notational clarity, in this subsection we indicate explicitly the dependence on  the boundary condition $\eta$ in
$\mu^\xi_{t,\eta}$, the law of the dynamics $\sigma^\xi_{\eta}(t)$ at time $t$, $\xi$ being the initial condition.
The Glauber dynamics enjoys the following well known monotonicity property. One can introduce a partial order in $\Omega_\Lambda$ by
saying that $\sigma\le \sigma'$ if $\sigma_x\le \sigma'_x$ for every $x\in\Lambda$. 
Then,  one has
\begin{eqnarray}
  \label{eq:93}
  \mu^\xi_{t,\eta}\preceq \mu^{\xi'}_{t,\eta'} \mbox{\;if \;}\xi\le\xi',\eta\le \eta'
\end{eqnarray}
 where $\preceq$ denotes stochastic domination (one writes $\mu\preceq
 \nu$ if $\mu(f)\le \nu(f)$ for every increasing function $f$,
 i.e.\ $f(\sigma)\le f(\si')$ whenever $\si\le\si'$; an
 event will be called increasing if its characteristic function is increasing).
In particular, letting $t\to\infty$ one obtains 
\begin{eqnarray}
  \label{eq:94}
\pi_\Lambda^\eta\preceq \pi_\Lambda^{\eta'}.  
\end{eqnarray}
Also, it is possible to realize on the same probability space 
the trajectories of the Markov chain corresponding to distinct initial conditions $\xi$ and/or distinct boundary conditions $\eta$, in such a way that, with probability one,
\begin{eqnarray}
  \label{eq:92}
  \sigma^\xi_{\eta}(t)\le \sigma^{\xi'}_{\eta'}(t) \mbox{\;for every\;} t\ge 0,\mbox{\;if \;}\xi\le\xi',\eta\le \eta'.
\end{eqnarray}
This will be referred to as the ``monotone global coupling'': its law will be denoted by $\bbP$ and the corresponding expectation by $\bbE$.

Throughout the paper we will apply several times the above monotonicity properties: for brevity, we will simply say ``by monotonicity...'' instead of referring explicitly to 
 \eqref{eq:93}-\eqref{eq:92}.

\subsubsection{Comparing $\beta=+\infty$ and $\beta\ge C \log L$}
In this section, it is understood that $\Lambda$ is the cubic domain 
\begin{eqnarray}
  \label{eq:89}
\Lambda_L:=\{-
L,\ldots,
L\}^d
  \end{eqnarray}
 of side $2L+1$ and that $\eta_x=+$ for every $x\in\partial \Lambda_L$, so that we omit
$\Lambda$ and $\eta$ from the notations.

In this work, we consider the situation where $\beta$ grows with $L$, and in
particular
\begin{eqnarray}
  \label{eq:7}
  \beta\in(C\log L,\infty]
\end{eqnarray}
for a sufficiently large $C$.

Note that, for $\beta=+\infty$, the equilibrium measure concentrates on the all
``$+$'' configuration: $\pi^{\infty}(\sigma)={\bf 1}_{\sigma\equiv +}$. 
 Given $c>0$, one can choose $C$ large enough so
that
\begin{eqnarray}
  \label{eq:8}
  \|\pi-\pi^{\infty}\|\le \frac1{L^c}
\end{eqnarray}
for every $\beta$ in the range \eqref{eq:7}
(this follows from easy Peierls estimates). Moreover, for every
initial condition $\sigma
\in\Omega$,
one can find a coupling between the dynamics at
$\beta=+\infty$ and $\beta>C\log L$ such that they coincide until time
$L^c$ with probability at least $1-1/L^c$ (just compare the transition rates).
In particular, this implies that 
\begin{eqnarray}
  \label{eq:9}
   \|\mu_t^{\sigma}-\mu_t^{\sigma,\infty}\|\le
   \frac1{L^c}
\;\;\mbox{for every}\;\; t\le L^c.
\end{eqnarray}
We will see later (cf. Remark \ref{rem:dinuguali}) that, as an immediate consequence of Theorem
\ref{th:d3}, one actually has the estimate 
\begin{eqnarray}
  \label{eq:18}
  \lim_{L\to\infty}\sup_{\sigma\in\Omega_{\Lambda_L}}\sup_{t\ge0}  \|\mu_t^{\sigma}-\mu_t^{\sigma,\infty}\|=0.
\end{eqnarray}
\smallskip

We define the random time $\tau_+$ as the first time 
when all spins are
``$+$'', starting from the all ``$-$'' configuration:
\begin{eqnarray}
  \label{eq:4}
  \tau_+:=\inf\{t>0: \sigma^-_x(t)=+1\; \mbox{for every\;} x\in\Lambda_L\}.
\end{eqnarray}
Observe that, under the global coupling, if $\beta=+\infty$, then $\sigma^\xi_x(t)=+1$ for every 
$t\ge \tau_+$ and for every initial condition $\xi$.

\medskip

{\bf Notational conventions}
\begin{itemize}
\item For notational clarity, quantities referring to the $\beta=+\infty$
dynamics will have a superscript $\infty$ (we will write for instance
$\mu^{\xi,\infty}_t$ and $\tmixinf$), while (for lightness of
notation) we will not put a superscript $\beta$ when $\beta<\infty$;

\item our main focus will be on the case of the all ``$+$''  boundary condition 
($\eta_x=+1$ for every $x\in\partial \Lambda$) and
we will just write $\eta\equiv +$ in this case;

\item we let $\bP$ denote the law of the process $\{\sigma^-(t)\}_{t\ge0}$
at $\beta=+\infty$, started from the ``$-$'' configuration, with boundary condition $\eta\equiv +$;

\item given $\sigma\in\Omega_\Lambda$ and $x\in\Lambda$, we will denote by
$\sigma^{(x)}\in\Omega_\Lambda$ 
the configuration obtained by flipping the spin at $x$ to $-\si_x$.

\end{itemize}

\medskip

Based on the observations \eqref{eq:8} and \eqref{eq:9}, 
one can use the time $\tau_+$ to estimate the mixing time for
$\beta$ large:
\begin{lemma}
\label{th:t+tmix}
 For every $\epsilon\in(0,1)$ one has
\begin{eqnarray}
  \label{eq:6}
  \tmixinf(\epsilon)=\inf\{t>0:\bP(\tau_+>t)\le\epsilon\},
\end{eqnarray}
where we recall that $\bP$ is the law of the process $\{\sigma^-(t)\}_{t\ge0}$ for $\beta=+\infty$, with ``$+$'' boundary conditions.

Moreover, 
fix $\epsilon\in(0,1)$ and $\delta>0$ and assume that, for some $c_0>0$, $\tmixinf(\epsilon)\le L^{c_0}$ for $L$
sufficiently 
large.
  If $\beta\in(C\log L,+\infty)$ with $C$ sufficiently large, then for
  every $L$ large one has
  \begin{eqnarray}
    \label{eq:10}
    \tmix(\epsilon+\delta)\le \tmixinf(\epsilon)
  \end{eqnarray}
and
 \begin{eqnarray}
    \label{eq:11}
    \tmix(\epsilon-\delta)\ge \tmixinf(\epsilon)
  \end{eqnarray}

\end{lemma}
\smallskip

{\sl Proof of Lemma \ref{th:t+tmix} }.
Recall that $\pi^\infty(\sigma)={\bf 1}_{\sigma\equiv +}$  so that
$\|\mu^{\xi,\infty}_t-\pi^{\infty}\|=1-\mu_t^{\xi,\infty}(\sigma\equiv
+)\le 1-\bP(\sigma^-(t)\equiv
+)$ (where the inequality follows from monotonicity) and
\eqref{eq:6}
is immediate.

To prove \eqref{eq:10}-\eqref{eq:11}, it is sufficient to choose 
$C$ sufficiently large so that 
\eqref{eq:8} and \eqref{eq:9} hold for $c=c_0$ and to take $L$
sufficiently large so that $2/L^{c_0}<a$.

\hfill$\stackrel{\tiny\mbox{Lemma \ref{th:t+tmix}}}{\qed}$

\section{Results}
\label{sec:results}
\subsection{Three-dimensional model}
Consider the 3D Ising model 
in the cubic box $\Lambda_L=\{-
L,\ldots,
L\}^3$  with
boundary condition $\eta\equiv+$ and $\beta=+\infty$.
The main result of this work is:
\begin{theo}
\label{th:d3} 
  There exists a  positive constant $c $ such that for $L\ge 2$
  one has
    \begin{eqnarray}
      \label{eq:5}
      \bP\left[\frac{L^2}{c \log L}\le \tau_+\le c L^2(\log L)^c\right]\ge 1-\frac cL.
    \end{eqnarray}

\end{theo}

\begin{rem}
\rm \label{rem:dinuguali}
\begin{enumerate}\

\item As we mentioned in the introduction, the previously known
  upper bound for $\tau_+$ was of order $L^3$ \cite[Th. 1.3]{FSS}, while \eqref{eq:5}
matches the heuristically expected behavior, except for the
logarithmic factor.
On the other hand, we are not aware of previously known lower bounds,
except for the trivial estimate $\tau_+\ge c L$;

\item via Lemma \ref{th:t+tmix}, once we prove \eqref{eq:5} we also have
that 
\begin{eqnarray}
  \label{eq:12}
 \frac{L^2}{c \log L}\le \tmix\le c\,L^2(\log L)^c
\end{eqnarray}
for $\beta\ge C\log L$ and $C$ large;

\item 
To prove \eqref{eq:18}, observe first of all that
\begin{eqnarray}
  \label{eq:91}
  \|\mu^\sigma_t-\mu^{\sigma,\infty}_t\|\le \|\mu^\sigma_t-\pi\|+\|\mu^{\sigma,\infty}_t-\pi^\infty\|+\|\pi-\pi^\infty\|.
\end{eqnarray}
The first two terms are estimated through \eqref{eq:90}, taking say $t\ge L^3$ and using the upper bound
\eqref{eq:12} for the mixing time. The proof is concluded thanks to \eqref{eq:8}-\eqref{eq:9} (if $C$ is chosen such that these bounds hold for $c=3$).

\end{enumerate}
\end{rem}

\subsection{Two-dimensional model}
For the 2D Ising model 
in the cubic box $\Lambda_L=\{-
L,\ldots,
L\}^2$  with
boundary condition $\eta\equiv+$ and $\beta=+\infty$ we have
\begin{theo}
  \label{th:d2} There exist positive constants $c,\gamma$
  such that
  \begin{eqnarray}
    \label{eq:2}
    \bP\left( c L^2\le \tau_+\le(1/c) L^2\right)\ge 1-e^{-\gamma L}.
  \end{eqnarray}
for every $L\ge1$.
\end{theo}
\begin{rem}
\label{rem:d2tmix}  \rm Thanks to Lemma \ref{th:t+tmix}, this implies that
$c_1L^2 \le \tmix\le c_2 L^2$ for $\beta\ge C\log L$.
We mention that the bound $\bP(\tau_+\ge (1/c)
L^2)\le \exp(-\gamma L)$ was proven in  \cite[Th. 1.3]{FSS}; as for the lower
bound for $\tau_+$, the authors of \cite{FSS} proved the following weaker result: for
every $\delta>0$,
$\bP(\tau_+\le L^2/(\log L)^{1+\delta})$ tends to zero as
$L\to\infty$. For a simplified version of the dynamics, the authors of \cite{CSS}
proved the sharp behavior $\tau_+\sim const\times L^2$.

\end{rem}

In two dimensions, always with ``$+$'' boundary conditions, we also have sharp bounds on the spectral gap:
\begin{theo}
\label{th:d2gap}
There exist positive constants $C,c$ such that 
for every 
$\beta\in[C\log L,+\infty)$, one has
\begin{eqnarray}
  \label{eq:1}
  \frac{c}L\le \gap\le \frac{1}{c L}.
\end{eqnarray}

\end{theo}
The reason why we require $\beta<\infty$ 
 is that for $\beta=+\infty$ the equilibrium measure is concentrated
on a single configuration (the all ``$+$'' configuration), and the spectral gap \eqref{eq:72} 
is not well defined
(all functions have zero variance with respect to $\pi$).

\begin{rem}
\label{rem:FSS}
  \rm
On the basis of the analysis of a one-dimensional toy model of evolution of a bubble
of ``$-$'' phase inside the ``$+$'' phase,
it was conjectured in \cite[Sec. 7]{BM} that in dimension $d=2$ the spectral gap behaves like $const/L$ for
every $\beta>\beta_c$. Our result \eqref{eq:1} is the first
confirmation of this conjecture, in the $\beta\to\infty$ limit. Let us
remark also that the authors of \cite{BM} showed,
via the construction of a suitable test function, that $\gap\le
c_1(\beta)(\log L)^{c_2(\beta)}/L$ for every $\beta>\beta_c$. However, when $\beta$ grows with
$L$ as in \eqref{eq:7}, the constants $c_{1,2}(\beta)$ possibly
diverge. Therefore, even the upper bound in \eqref{eq:1} is a
qualitative improvement over known results.

\end{rem}

The proof that $\gap\ge c/L$ for $\beta\ge C\log L$ extends easily to
the three-dimensional model (cf. Section  \ref{sec:LBgap3D}) but, as
we discuss in a moment, there is no reason to believe that this is the
correct behavior of the spectral gap in $d=3$.

\medskip
Two interesting questions which are left open by the above results are
the following:
\begin{enumerate}

\item 
What is the order of magnitude of the spectral gap in dimension $d=3$? One
  expects it to be much larger than in dimension $d=2$, possibly of order
  $(\log L)^{-c}$ for some $c>0$, or even of order $1$
  \cite[Sec. 7]{BM}, \cite{FH};

\item How to remove the logarithmic corrections in the upper and lower bounds for $\tmix$ in
  dimension 
$d=3$? 
A suggestive indication that indeed $\tmix \ge c L^2$ comes from the
following argument. If one defines the ``entropy constant'' (or
modified Log-Sobolev constant) as
\begin{eqnarray}
  \label{eq:137}
c_{ent}:=  \sup \frac{\pi(f\log f)}{\pi(\log f (-\cL)f)}
\end{eqnarray}
where the supremum is taken over positive functions $f$ which verify
$\pi(f)=1$, 
it is possible to exhibit a test function which gives $c_{ent}\ge c
L^2$ for some constant $c$ which is positive, uniformly for $\beta \ge C\log L$ with $C$
large. 
On the other hand, it has been conjectured \cite{TM} (and verified
in various explicitly solvable examples) that for every
reversible Markov process the mixing time is lower bounded by
$c_{ent}$ times some universal constant. 
\end{enumerate}

The organization of the paper is as follows.  Theorem \ref{th:d3} is proven in Section
\ref{sec:UBMT} (upper bound on $\tau_+$) and in Section \ref{sec:LBMT} (lower bound). The upper bound requires precise estimates on the
fluctuations of dimer coverings of the hexagonal (or honeycomb)
lattice, see Section
\ref{app:kenyon}, and on the mixing time of a dynamics on plane
partitions or monotone sets, proven in Section \ref{app:wilson}. 
As for the 2D model, Theorem \ref{th:d2} is proven in
Section \ref{sec:TM2D}
and Theorem \ref{th:d2gap}  in Section \ref{sec:GAP2D}.

\medskip

{\bf More notational conventions}
\begin{itemize}
\item 
In the proof of the results, $c,c', c''$ etc. denote
positive
and finite constants (independent of $L$ and $\beta$) which are not
necessarily the same at each occurrence.

\item If $u\in \RR^d$, it will be understood that $u^{(a)}, a=1,\ldots,d$ are its Cartesian coordinates.

\item $d(\cdot,\cdot)$ will denote the Euclidean distance in $\RR^d$.

\item  $\partial B$ will denote the geometric boundary of a set
  $B\subset \RR^d$, except when $B\subset \ZZ^d$, in which case
$\partial B:=\{x\in \ZZ^d\setminus B:\exists y\in B \mbox{\;such that
  \;} d(x,y)=1\}$.

\item $\mathbb N$ denotes the set of non-zero integers $\{1,2,\ldots\}$.
\end{itemize}

\section{The mixing time in dimension $d=3$ (upper bound)}
\label{sec:UBMT}
\subsection{A basic tool: dynamics of discrete monotone sets}
\label{sec:wilson}
\begin{definition}
\label{def:monset}
We say that $V\subseteq\NN^3$ is a {\sl 
positive  monotone set} if $z\in V$ implies $y\in V$ whenever $y\in \NN^3$ and
$y^{(a)}\le z^{(a)}, a=1,2,3$.  The collection of all positive monotone sets
will be denoted by $\Sigma^+$, which is partially ordered with respect
to inclusion.
When $V\in\Sigma^+$ is a finite set, it is usually called a {\sl plane partition},
the two-dimensional generalization of an ordinary partition (or
{\sl Young diagram}). 
\end{definition}

Given  $V^+,V_0,V^-$ in $\Sigma^+$ such that $V^-\subseteq V_0\subseteq V^+$,
we define a dynamics $\{V_t^{V_0}\}_{t\ge 0}$ on $\Sigma^+$ such that 
$V^-\subseteq V_t^{V_0}\subseteq V^+$ for all times, with initial condition
$V_{t=0}^{V_0}=V_0$. Let $\Lambda:=V^+\setminus V^-$ and associate
to each $z\in \Lambda$ an
independent Poisson clock of rate one. When the clock labeled $z$
rings at some time $s$: 
\begin{itemize}
\item if both sets $V_{s,z,-}:=V_s\setminus\{z\}$ and $V_{s,z,+}:=V_s\cup\{z\}$
belong to $\Sigma^+$, then we
replace $V_s$ with $V_{s,z,a}$, with $a$ chosen between ``$+$'' and ``$-$''
with equal probabilities $1/2$;

\item otherwise, we keep $V_s$ unchanged.

\end{itemize}

The link with the $\beta=+\infty$ dynamics of the Ising model is straightforward: Consider the Ising dynamics in the domain $\Lambda:=V^+\setminus
V^-$, with initial condition 
\begin{eqnarray}
  \label{eq:95}
  \xi_z=\left\{
    \begin{array}{lll}
      - & \mbox{if}&z\in V_0\cap \Lambda\\
+ & \mbox{if} & z\in \Lambda\setminus V_0
    \end{array}
\right.
\end{eqnarray}
and with boundary conditions
\begin{eqnarray}
  \eta_z=\left\{
    \begin{array}{lll}
      + & \mbox{if}&z\in \NN^3\setminus  V^+\\
- & \mbox{if} & z\in (\ZZ^3\setminus\NN^3)\cup V^-
    \end{array}
\right.
\end{eqnarray}
and let $M_t:=\{z\in \NN^3:\sigma^\xi_z(t)=-\}$. Then, $M_t$ has the
same law as $V_t^{V_0}$. 

To avoid any risk of confusion we choose different notations for the 
Ising dynamics and the monotone set dynamics: we call $\nu_t^{V_0}$ 
the law of $V_t^{V_0}$ and $\rho:=\rho_{V^\pm}$ its (reversible) invariant measure.
Of course, $\rho_{V^\pm}$ is just the uniform measure over the
positive monotone sets
$V\in\Sigma^+$ such that $V^-\subseteq V\subseteq V^+$.  The monotonicity properties of the Glauber dynamics discussed in Section \ref{sec:monotone} immediately give:
\begin{prop}
\label{th:mondyn}
If $V^\pm,W^\pm\in \Sigma^+$ and $V^\pm\subseteq W^\pm$, then
$\rho_{V^\pm}\preceq \rho_{W^\pm}$.
 \end{prop}

The next key result
quantifies the mixing time of the monotone set dynamics as a function
of the shape of $\Lambda$:
\begin{theo}
\label{th:wilson}
  Let $$D:=\max_{y,z\in \Lambda}\left\{d\left((z^{(1)},z^{(2)}),(y^{(1)},y^{(2)})\right)\right\}$$ and
  $$H:=\max\{|z^{(3)}-y^{(3)}|:z,y\in
  \Lambda\mbox{\;\;and\;\;}y^{(a)}=z^{(a)},a=1,2\}.$$ If $H\le D$, the mixing
  time of the monotone set dynamics is $O(H^2 D^2(\log D)^2).$
\end{theo}

We believe that the correct behavior is $O(D^2\log D)$. The spurious
factor 
$H^2$ is potentially dangerous, so we will take care to apply
Theorem \ref{th:wilson} only in situations where $H$ is small, say of order
of a power of $\log D$: this is a crucial step for the proof of the upper bound \eqref{eq:12}, which
differs
from the expected behavior $O(L^2)$ only by logarithmic corrections.
We give the proof of Theorem \ref{th:wilson} in Section \ref{app:wilson}.

\subsection{Upper bound on the mixing time}

\label{sec:UBMTbis}

Let 
$$S_{r}:=\ZZ^3\cap B_{r},$$ where $$B_{r}:=\{x\in \RR^3:d(x,0)\le
r\}$$ is the ball of radius $r$ centered at the origin. 
By monotonicity, the claim
\eqref{eq:5} follows if we prove the following result 
for the dynamics in
$S_{5L}$ with ``$+$'' boundary conditions, started from ``$-$'':
\begin{prop}
  \label{th:sfere}
Consider the dynamics in $S_{5L}$ with ``$+$'' boundary conditions on $\partial S_{5L}$.
There exists $c>0$ such that for every $L\ge2$ and $t=c\,L^2(\log L)^c$ one has
\begin{eqnarray}
  \label{eq:13}
 \bP\left(\exists x\in S_{5L}\setminus S_{L} \mbox{\;such that\;}\sigma^-_x(t)=-\right)\le c/L.
\end{eqnarray}

\end{prop}
The reason why this result implies \eqref{eq:5}  is that the set
$S_{5L}\setminus
S_L$ contains a cube, call it $Q$, which is a translate of $\Lambda_L$
and, by monotonicity, the marginal in $Q$ of the evolution in $S_{5L}$
is stochastically dominated by the evolution in $Q$ started
from ``$-$'', with ``$+$'' boundary
conditions on $\partial Q$.

In order to get Proposition \ref{th:sfere}, we prove
\begin{prop}
\label{th:sfere2}
Consider the dynamics in $S_{L}$ with ``$+$'' boundary conditions
  and let $x\in S_L\setminus S_{L-1}$. There exists $c>0$ such that for every $L\ge2$ 
  \begin{eqnarray}
    \label{eq:16}
    \bP\left(\sigma^-_x(t)=+\;\mbox{for all times}\;t\in[c\,L(\log
    L)^c,L^3]\right)\ge
1-c/L^{4}.
  \end{eqnarray}
\end{prop}

\smallskip
Proposition \ref{th:sfere} then easily follows:

{\sl Proof of Proposition \ref{th:sfere} (assuming Proposition
  \ref{th:sfere2})}
By the union bound, one then deduces that 
\begin{eqnarray}
  \label{eq:14}
  \bP\left(\exists x\in S_L\setminus S_{L-1}\mbox{\;and\;} t\in[c\,L(\log L)^c,L^3]\mbox{\; such that \;}
  \sigma^-_x(t)=-\right)
\le c/L^{2}.
\end{eqnarray}

For the dynamics in $S_{5L}$ we prove, by induction on
$i=1,\ldots,4L$, that
\begin{eqnarray}
  \label{eq:96}
  \bP(\mathcal A_i):=\bP\left(
\exists x\in S_{5L}\setminus S_{5L-i}\mbox{\;and\;} t\in[5c\,i L(\log (5L))^c,L^3]\mbox{\; such that\;} \sigma^-_x(t)=-
\right)\le \frac {c\,i}{L^2},
\end{eqnarray}
which for $i=4L$ implies \eqref{eq:13}.
For $i=1$, this follows from Eq. \eqref{eq:14}.
If the claim holds for some $1\le i<4L$, then
\begin{eqnarray}
  \label{eq:97}
  \bP(\mathcal A_{i+1})\le \bP(\mathcal A_i)+
\bP(\mathcal A_{i+1} |\mathcal A_i^{\tt c})\le \frac {c\,i}{L^2}
+
\bP(\mathcal A_{i+1} |\mathcal A_i^{\tt c})
\end{eqnarray}
where for any event $A$ we let $A^{\tt c}$ denote its complement.
Next, by monotonicity, 
\begin{eqnarray}
  \label{eq:98}
\bP(\mathcal A_{i+1} |\mathcal A_i^{\tt c})\le
\bP\left(  \exists x\in S_{5L-i}\setminus S_{5L-i-1},
t\in [5c\,(i+1) L(\log (5L))^c,L^3]
\;:\; \hat \sigma_x(t)=-\right)
\end{eqnarray}
where $\hat \sigma^-(t)$ is the evolution  in $ S_{5L-i}$
with ``$+$'' boundary conditions, which starts from ``$-$'' at time
$t=5c\,i L(\log (5L))^c$.
Thanks to  monotonicity we can restart from ``$-$'' the evolution at time
$t=5c\,i L(\log (5L))^c$ and we used the hypothesis
$\mathcal A_i^{\tt c}$ to freeze all spins outside $S_{5L-i}$ to the value
``$+$'' in the time interval $[5ci L (\log (5L))^c,L^3]$.
Via a trivial time translation, the upper bound  \eqref{eq:14} (applied with $L$ replaced by $5L-i$) shows that the right-hand side of 
\eqref{eq:98} is smaller than $c/L^2$ ($c$ being the same constant
which appears in \eqref{eq:97}) which, together with 
\eqref{eq:97}, completes the inductive proof.

\hfill$\stackrel{\tiny\mbox{Proposition \ref{th:sfere}}}{\qed}$

{\sl Proof of Proposition \ref{th:sfere2}.} Clearly, it is sufficient
to prove the claim for $L$ large enough. In the following, $\delta$
will denote a small positive universal constant (conditions on its
smallness
will be specified later).
Given a point $x\in \RR^3$, let $(r:=\|x\|,\theta,\phi)$ be its spherical coordinates, where $\theta\in[0,\pi]$ is the polar angle and $\phi\in[0,2\pi)$ is
the azimuthal angle, with the convention that the half-plane $\{z\in \RR^3:z^{(1)}\ge 0,z^{(2)}=0\}$ 
corresponds to $\phi=0$ and that $\theta=0$ identifies the positive
$z^{(3)}$ axis. Note that the portion of 
$\partial B_L$ contained in the octant $\mathcal C^+:=\{(\theta,\phi)\in(0,\pi/2)^2\}$ is a
monotone surface, where:
\begin{definition}
 \label{def:monotone}
A smooth surface $\Gamma\subseteq \RR^3$ 
is said to be monotone if for every $x\in \Gamma$ all three components of the
normal vector ${\bf n}_x$ at
$x$ are non-zero and have the same sign. 
\end{definition}

It is convenient to introduce a few geometric definitions:
\begin{definition}
\label{def:geom}
For a given $x\in S_L$, let 
\begin{enumerate}[(a)]
\item 
$\mathbb L$ be the infinite half-line which starts from  the origin of
$\RR^3$ and
goes through $x$
\item $x'$ be the intersection of $\mathbb L$ and $\partial
B_L$
\item ${\Pi}$ be the plane perpendicular to $\mathbb L$ which meets
  $\mathbb L$ at distance
$L-(\log L)^{3/2}$ from the origin (the exponent $3/2$ is somewhat
arbitrary, and it could be replaced by $1+\delta$ for any $\delta>0$)
\item 
$\Upsilon$ be the half-space not containing the origin and delimited by ${\Pi}$
\item 
$\mathcal M$ be the spherical cap $B_L\cap \Upsilon$.
\end{enumerate}
Also, decompose the boundary of $\mathcal M$ as $\partial \mathcal M=\Gamma_1\cup
\Gamma_2$, where $\Gamma_1=\partial \mathcal M\cap \partial 
B_L$ is the ``curved portion of the boundary'' and $\Gamma_2=\partial
\mathcal M\cap
{\Pi}$ is a disk.
  
\end{definition}

Thanks to the discrete symmetries of
the sphere $S_L$, it is clearly enough to prove \eqref{eq:16} for  points $x\in S_L\setminus S_{L-1}$ such that
$\theta\in[0,\pi/2]$ and $\phi\in[0,\pi/4]$. Actually, we claim that
it is enough to restrict to $\theta\in[0,\pi/2-\delta]$ and $\phi\in[0,\pi/4]$, if $\delta>0$ is
small enough. Indeed, if we interchange the role of the first and
third coordinates of $\RR^3$, the points with $\phi\in[0,\pi/4]$ and 
$\theta$ close to $\pi/2$ are mapped into points with $\theta\le
\pi/4$ and $\phi$ small.

It is convenient to distinguish three cases (see Figure \ref{fig:sfera}):

\begin{figure}[htp]
\begin{center}
\includegraphics[width=0.6\textwidth]{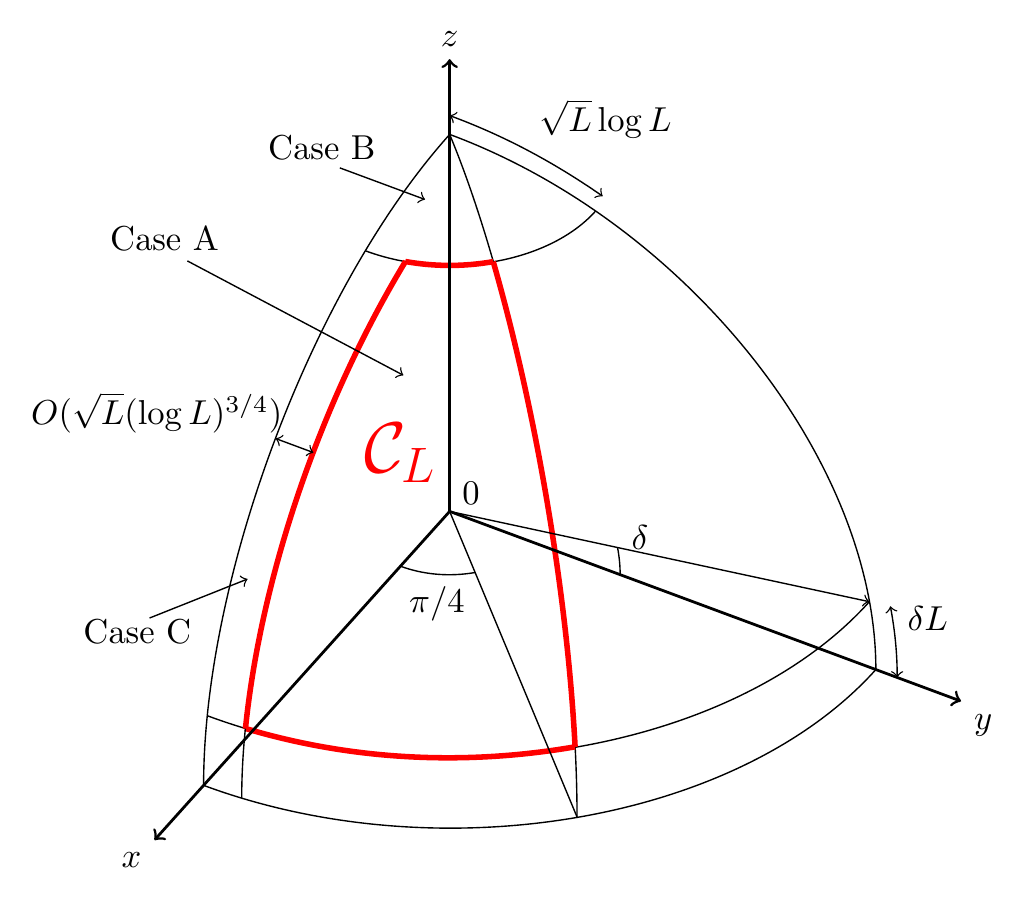}
\end{center}
\caption{The monotone octant $\mathcal C^+$ of the surface of the sphere of radius $L$. By symmetry, we need to prove \eqref{eq:16} only in the case where 
$\phi\le \pi/4$ and $x'$ (cf. Definition \ref{def:geom}(b))  does not fall in the strip of width $\delta L$, adjacent to the $(x,y)$ plane.
Case {\bf A} corresponds to $x'$ in $\mathcal C_L$ (the subset of the surface of the sphere delimited by the thick line). 
Case {\bf B} corresponds to $x'$ at geodesic distance at most $\sqrt L \log L$ from the north pole. Finally, 
case {\bf C} corresponds to $x'$ in the strip of width $(1+\delta)\sqrt{2L}(\log L)^{3/4}$ to the left of $\mathcal C_L$.
For reasons of graphical clarity, proportions are not respected in the drawing. 
}
\label{fig:sfera} 
\end{figure}
\begin{enumerate}[]
\item  {\bf Case A:} $x\in S_L\setminus S_{L-1}$ is such that $(\theta,\phi)\in
 \mathcal C_L$ where 
 \begin{eqnarray}
   \label{eq:15}
\mathcal C_L:=
\left([0,\pi/2-\delta]\times[0,\pi/4]\right)\cap\left\{(\theta,\phi):\theta\ge
\frac{\log L}{\sqrt{L}},\;\phi\sin(\theta)\ge \frac{(1+\delta)\sqrt2}{\sqrt L}(\log
L)^{3/4}\right\}  .
 \end{eqnarray}
 Remark that the lower bound on $\phi\sin(\theta)$ just means that,
 moving on $\partial B_L$ along a line of constant $\theta$, the
 distance between  $x'$ 
(cf. Definition \ref{def:geom}(b)) and the set of points in $\partial B_L$ where
$\phi=0$ is at least $(1+\delta)\sqrt {2L}(\log L)^{3/4}$.

\begin{lemma} 
Under the condition $(\theta,\phi)\in\mathcal C_L$, one has:
\label{th:geomovvio}\ 
\begin{enumerate}
\item The disk
$\Gamma_2$ has 
radius $\sqrt{2L}(\log L)^{3/4}(1+o(1))$
\item $\Gamma_1$ is contained in the monotone octant $\mathcal C^+
=(0,\pi/2)^2$.
\end{enumerate}
As a consequence, both
$\Gamma_1$ and $\Gamma_2$ are  monotone surfaces, cf. Definition \ref{def:monotone}.
\end{lemma}
{\sl Proof of Lemma \ref{th:geomovvio}.}
Statement (a) and monotonicity of $\Gamma_2$
require just elementary geometric considerations.
As for statement (b), it is easy to see that the condition $(\theta,\phi)\in \mathcal C_L$
implies
that the geodesic distance of $x'$ along $\partial  B_L$ from the boundary of $\mathcal
C^+$ is larger than $(1+\delta)\sqrt{2L}(\log L)^{3/4}(1+o(1))$.
 The fact that $\Gamma_1$ is contained (for $L$
large enough) in $\mathcal C^+$ then just follows (thanks to the fact
that $\delta$ is strictly positive) from statement (a).

\hfill$\stackrel{\tiny\mbox{Lemma \ref{th:geomovvio}}}{\qed}$

\smallskip

We can now continue the proof of Proposition \ref{th:sfere2} under the
condition $(\theta,\phi)\in\mathcal C_L$. Let 
$$\hat {\mathcal M}:=\mathcal M\cap \ZZ^3$$ be the collection of lattice sites contained
in $\mathcal M$ (cf. Definition \ref{def:geom}(e)). 
 It is clear that
the site $x\in S_L\setminus S_{L-1}$ under consideration belongs to
$\hat {\mathcal M}$ and that, 
thanks to Lemma \ref{th:geomovvio}(a), 
\begin{eqnarray}
  \label{eq:21}
  D:=\max_{y,z\in
  \hat {\mathcal M}}\left\{d\left((z^{(1)},z^{(2)}),(y^{(1)},y^{(2)})\right)\right\}\le 2\sqrt{ 2L}(\log L)^{3/4}(1+o(1)).
\end{eqnarray}

By monotonicity, it is enough to prove \eqref{eq:16} for a modified
dynamics where only the spins $y\in\hat {\mathcal M}$ evolve, starting from the
``$-$'' configuration, with boundary conditions given by $\eta_y=+$ for
$y\notin S_L$ and $\eta_y=-$ for $y\in S_L\setminus \hat {\mathcal M}$. For
lightness of notation, we still call such dynamics $\sigma^-(t)$.  Let
\begin{eqnarray}
  \label{eq:99}
  V^+:=S_L\cap \NN^3,\;\;\;V^-:=V^+\setminus \hat {\mathcal M},
\end{eqnarray}
and observe that
$V^\pm$ are positive monotone sets in the sense of Definition \ref{def:monset}, as
guaranteed by the monotonicity of the surfaces $\Gamma^1,\Gamma^2$ (cf. Lemma
\ref{th:geomovvio}). Also, call $M_t:=\{z\in \NN^3:\sigma^-_z(t)=-\}$:
thanks to the discussion in Section \ref{sec:wilson}, we have that
$M_t$ is a positive monotone set at all times $t\ge0$, and trivially
$V^-\subseteq M_t\subseteq V^+$. To be coherent with the  notations of Section \ref{sec:wilson},
we have in the present case $V_0=V^+$, $\Lambda=V^+\setminus V^-=\hat
{\mathcal M}$; also, the law
of $M_t$ is just $\nu_t^{V_0}$, with invariant measure $\rho_{V^\pm
  }$. Finally note that, from the definition of the spherical cap
$\mathcal M$, one has
\begin{eqnarray}
  \label{eq:20}
H:=\max\{|z^{(3)}-y^{(3)}|:z,y\in \hat {\mathcal M}\mbox{\;\;and\;\;}y^{(a)}=z^{(a)},a=1,2\} 
\le\frac{ (\log L)^{3/2}}{\cos \theta }\le c(\delta) (\log L)^{3/2}
\end{eqnarray}
(recall that we are working under the assumption that $\theta \le \pi/2-\delta$).
Putting together Theorem \ref{th:wilson} with the estimates
\eqref{eq:21}, \eqref{eq:20}, we get that the mixing time of 
the dynamics in $\hat {\mathcal M}$  is $O(L(\log L)^{13/2})$. 

We have the following key equilibrium estimate:
\begin{prop}
\label{th:kenyon} Recall that $ \rho_{V^\pm}$ is the uniform
distribution over all  positive monotone sets $V\in\Sigma^+$ such that $V^-\subseteq V\subseteq V^+$.
There exists $c>0$ such that for every $L\ge 2$  and whenever 
$(\theta,\phi)\in \mathcal C_L
$,
  \begin{eqnarray}
    \label{eq:17}
    \rho_{V^\pm}\left[\exists z\in \hat {\mathcal M}: d(z,V^-)\ge \frac14(\log
      L)^{3/2}
\mbox{\;\;and\;\;} z\in V
\right]\le \frac1c\exp(-c(\log L)^{3/2}).
  \end{eqnarray}
\end{prop}
In words, this result is saying that spins which are at distance of
order
$(\log L)^{3/2}$ away from the bottom of the spherical cap $\hat {\mathcal M}$,
where
the ``$-$'' boundary conditions act, are ``$+$'' with overwhelming probability.
It is crucial here that the estimate \eqref{eq:17} (i.e.\ the value of
 $c$) does not depend on
the spherical coordinates $(\theta,\phi)$ of the point $x$ under
consideration, i.e.\ on the slope of the plane $\Pi$ of Definition \ref{def:geom}(c).

Roughly speaking, \eqref{eq:17} follows from recent works on the
height fluctuations of random monotone interfaces associated to  dimer coverings
of the 
hexagonal lattice \cite{Kenyon,KenyonCMP,KenyonAIHP}, but it
requires some work to really prove the precise statement we need. The proof of Proposition \ref{th:kenyon} is given in detail in
Section
\ref{app:kenyon}.

\smallskip
Once we have Proposition \ref{th:kenyon}, the proof of \eqref{eq:16}
proceeds via a standard argument which we simply sketch: 
Eq. \eqref{eq:90}, together with the fact that the mixing time of
the dynamics in $\hat {\mathcal M}$ is $O(L(\log L)^{13/2})$, implies that
the variation distance between $\rho_{V^\pm}$ and the law of
$\sigma^-(t)$ 
is smaller than  $\exp(-c'(\log
L)^{3/2})$ for all $t>c\,L(\log L)^8$. Since the dynamics undergoes $O(L^{6})$ updates 
during the time interval $[c\,L (\log L)^8,L^3]$, 
via a union bound and the equilibrium
estimate \eqref{eq:17}
one gets that 
the probability in \eqref{eq:16} is lower bounded by
$$
1-L^{c''}\exp(-c(\log
L)^{3/2})\ge 1-c\,L^{-4}
$$
for $L$ large, which is the desired bound.
\begin{rem}
\rm 
\label{rem:cruciale}
 It is important to notice that exactly the same proof gives,
for some $c>0$ and for all $L\ge 2$,
  \begin{eqnarray}
    \label{eq:16rem}
\inftwo{x\in S_L\setminus S_{L-(3/4)(\log L)^{3/2}}:}{  (\theta,\phi)\in \mathcal C_L}  \bP\left(\sigma^-_x(t)=+\;\mbox{for all times}\;t\in[c\,L(\log
    L)^c,L^3]\right)\ge
1-\frac c{L^{4}}
  \end{eqnarray}
(just look at
the equilibrium estimate in Proposition
\ref{th:kenyon}).

\end{rem}

\bigskip

\item
{\bf Case B:}  $x\in S_L\setminus S_{L-1}$ is such that   $\theta\le L^{-1/2}\log L$. 

  Here, a rather rough argument suffices. Remark first of all that the
  vertical coordinate $x^{(3)}$ of $x$ belongs to the interval $[L-(1/2)(\log L)^2(1+o(1)),L]$. Call $\hat S:=\{y\in S_L:y^{(3)}\ge x^{(3)}\}$.  By
  monotonicity, to show \eqref{eq:16} it is sufficient to prove that 
\begin{eqnarray}
\label{eq:secc}
\bP(\tau_+> c L (\log L)^c)\le c/L^4
\end{eqnarray} 
for a modified
  dynamics where only spins in $\hat S$ evolve,
  starting from the ``$-$'' configuration, with boundary conditions given by $\eta_z=+$
  if $z\notin S_L$ and $\eta_z=-$ if $z\in S_L\setminus \hat
  S$. Note that $\hat S$ is a discrete spherical cap, and that the boundary conditions just described are ``$-$''
  below its base, and ``$+$'' elsewhere. Of course, $\tau_+$ in \eqref{eq:secc} is understood
  to be the first time when all spins in $\hat S$ are ``$+$''.

We note first of all that $\hat S$ is contained in a parallelepiped $Q$
whose base is a square of side  $\ell:=2\sqrt{L} \log L(1+o(1))$ and whose
height
is $h:=(1/2)(\log L)^2(1+o(1))$. Next, by monotonicity we see that $\tau_+$
is
stochastically increased if
we replace $\hat S$ with $Q$, with ``$-$'' boundary conditions below its base square, and
``$+$'' everywhere else.
One  can decompose $Q$ into $h$  horizontal squares
of side $\ell$, stacked one on top of the other. If $h$ were $1$, we would simply have the evolution 
for $\beta=+\infty$ of the two-dimensional Ising model with ``$+$''
boundary conditions in a square of side
$\ell$ (the ``$+$'' boundary conditions on the top face of $Q$ would
compensate exactly the ``$-$'' boundary conditions on the bottom face),
and we know \cite[Theorem 1.3 (a), case $d=2$]{FSS}  that in this case
one has $\bP(\tau_+\ge c \ell^2)\le \exp(-\gamma \ell)$ for some positive $\gamma$, if $c$ is
large enough.  Via a standard monotonicity argument (cf.  
\cite[Proof of Theorem 1.3 (a), case $d>2$]{FSS} for details) this implies that, for
the evolution in the parallelepiped $Q$, it is very unlikely that
$\tau_+$ exceeds $h\times c\ell^2$:
\begin{eqnarray}
  \label{eq:22}
\bP(\tau_+> c L (\log L)^c)=\bP\left[\tau_+> c\ell^2 h (\log L)^{c-4}(1/4+o(1))\right]\le
h\exp(-\gamma \ell)
\end{eqnarray}
if $c>4$, which is actually stronger than the estimate \eqref{eq:secc} we
wished to prove.
\bigskip

\item {\bf Case C:}
$x\in S_L\setminus S_{L-1}$ is such that
\begin{eqnarray}
  \label{eq:23}
\theta\in\left[\frac{\log L}{\sqrt L},\pi/2-\delta\right]
\mbox{\;\;and\;\;}  
\phi\sin\theta\in\left[0,
\frac{(1+\delta)\sqrt 2}{\sqrt L}(\log L)^{3/4}\right].
\end{eqnarray}
\begin{definition}
 Let $\tilde x$ be the point of $\partial B_L$ with angular coordinates
\begin{eqnarray}
  \label{eq:19}
(\theta,\phi+\phi'):=\left(\theta,\phi+\frac{\gamma}{\sqrt L}\frac{(\log L)^{3/4}}{\sin \theta}\right)
\end{eqnarray}
where $\gamma>0$ will be chosen later, and let $\bar x$ be a point
of $\ZZ^3$ of minimal distance from $\tilde x$.

For every $y\in \RR^3$, let $(\hat r(y),\hat \theta(y),\hat \phi(y))$ be the spherical coordinates of the vector $y+3\bar x$, i.e., the spherical coordinates
of $y$ with respect to the point $-3\bar x\in \ZZ^3$.
\end{definition}

\begin{lemma} 
\label{th:cartesiano}
We have, for the point $x\in S_L\setminus S_{L-1}$ under consideration,
  \begin{gather}
    \label{eq:24}
  \hat r(x)=4L-\frac{3\gamma^2}{8}(\log (4L))^{3/2}(1+o(1)),\\
\label{teta}\theta(1+o(1))\le \hat \theta(x)\le \theta\\
\label{fi}\hat \phi(x)\sin \hat \theta(x)=\phi\sin\theta+\frac{3\gamma}{2\sqrt{4L}}(\log (4L))^{3/4}(1+o(1)).  
\end{gather}
\end{lemma}
We omit the proof of Lemma \ref{th:cartesiano}, since it requires only tiresome but elementary computations: one first writes down the Cartesian coordinates of $\bar x$ and $x$, then one works out the  spherical coordinates of $x+3\bar x$; finally, the three statements  are obtained by expanding these spherical coordinates
for $L$ large, using the fact that $\phi'\ll 1$.

\medskip

We can now conclude the proof of \eqref{eq:16}, case {\bf C}. First of
all, remark that, thanks to \eqref{eq:24}, one has for $L$ large enough
and letting $S_{r,a}=\ZZ^3\cap B_{r,a}$, with $B_{r,a}$ the ball of
radius $r$ centered at $a$,
$$x\in S_{4L,-3\bar x}\setminus
S_{4L-(3/4) (\log (4L)^{3/2}),-3\bar x}$$
provided that
\begin{eqnarray}
  \label{eq:25}
  \frac{3\gamma^2}8<\frac34.
\end{eqnarray}
%
%
Secondly, \eqref{fi} implies that 
for $L$ large enough $$\hat \phi(x)\sin \hat \theta(x) \ge
\frac{(1+\delta)\sqrt2}{\sqrt{4L}}(\log (4L))^{3/4}$$ 
provided that
\begin{eqnarray}
  \label{eq:31}
  \frac{3\gamma}2>(1+\delta)\sqrt 2.
\end{eqnarray}
Note that conditions \eqref{eq:25} and \eqref{eq:31} are compatible if
$\delta$ is small enough, and choose a value for $\gamma$ which
satisfies both.
Finally, \eqref{teta}
shows that $\hat\theta\le \pi/2-\delta$.

All in all, thanks to Eq. \eqref{eq:16rem} in Remark \ref{rem:cruciale}
(applied with $L$ replaced by $4L$, and modulo a trivial translation
of the center of the sphere $S_{4L}$ from $0$ to $-3\bar x$), 
we have
\begin{eqnarray}
  \label{eq:26}
  \bP\left(\sigma_x(t)=+\mbox{\;\;for all times\;\;} t\in[c\,L(\log L)^c,L^3]\right)
\ge 1-c/(4L)^4
\end{eqnarray}
with $\bP$ the law of the dynamics in the sphere $S_{4L,-3\bar x}$ with ``$+$'' boundary conditions on $\partial S_{4L,-3\bar x}$, started from ``$-$''. Since $S_{4L,-3\bar x}\supset S_L$, by monotonicity
\eqref{eq:26} implies the same inequality when $\bP$ is the law of the dynamics in $S_L$, always started from ``$-$'' and with boundary ``$+$'' conditions on $\partial S_L$. 
The desired estimate \eqref{eq:16} is proven.
\end{enumerate}

\hfill$\stackrel{\tiny\mbox{Proposition \ref{th:sfere2}}}{\qed}$

\section{Estimates on height fluctuations of monotone interfaces}
\label{app:kenyon}
\begin{definition}
A subset $V\subseteq \ZZ^3$ is said to be {\sl monotone} if $x\in V$
implies
$y\in V$ whenever $y\in\ZZ^3$ and $y^{(a)}\le x^{(a)},a=1,2,3$. The collection of all monotone sets is denoted by $\Sigma$.  
\end{definition}
Remark
that this definition differs from that of {\sl positive monotone set}
(cf. Definition \ref{def:monset})  only  in that it is not required that
$V\subseteq\NN^3$. 
Given a {\sl positive} monotone set $V\in \Sigma^+$, in the following we
will always implicitly identify it with the monotone set
$V\cup(\ZZ^3\setminus \NN^3)\in\Sigma$. 

\begin{definition}
\label{def:v}
Given $V\in \Sigma$, we associate to it a {\sl vertical height
  function}, which we denote $v$.
The function $\{v_x\}_{x\in \ZZ^2}$ is defined as 
\begin{eqnarray}
  \label{eq:37}
  v_{x}:=\max\{x^{(3)}:(x^{(1)},x^{(2)},x^{(3)})\in V\} \mbox{\;\;if\;\;}x=(x^{(1)},x^{(2)}).
\end{eqnarray}  
\end{definition}
Observe that $v_x$ takes values in $ \ZZ\cup\{-\infty,+\infty\}$, and
that
\begin{eqnarray}
  \label{eq:50}
v_x\le v_y \mbox{\;if\;}y^{(a)}\le x^{(a)},a=1,2.  
\end{eqnarray}
We will denote $\mathcal
V$ the set of all possible functions $v:\ZZ^2\mapsto \ZZ\cup\{-\infty,+\infty\}$
which satisfy \eqref{eq:50}.
As discussed in Section \ref{sec:K}, one can identify $\mathcal V$
with
$\mathcal{D_H}\times \ZZ$, where $\mathcal{D_H}$ is the set of dimer
coverings of the infinite
honeycomb lattice $\mathcal H$.

\subsection{Proof of Proposition \ref{th:kenyon}}
\label{sec:eq17}
Recall the definition \eqref{eq:99} of the monotone sets $V^\pm$ in Section \ref{sec:UBMT}, and note
that the corresponding height functions $v^\pm\in \mathcal V$ coincide
outside some domain $U\subseteq \NN^2$ whose diameter is $O(\sqrt L
(\log L)^{3/4})$ (cf. Lemma \ref{th:geomovvio}(a)). Note that $U$ is
just the projection on the plane $(x,y)$ of the discrete spherical cap $\hat {\mathcal M}$.


The estimate \eqref{eq:17} is proven if we have, for some universal
constant
$c>0$,
\begin{eqnarray}
  \label{eq:38}
  \rho_{\left.v^-\right|_{ U^{\tt c}}}(A|\Gamma^1\cap \Gamma^2)\le \frac1c\exp(-c\,(\log L)^{3/2})
\end{eqnarray}
for $L\ge 2$
where
\begin{gather}
  \label{eq:39}
  A:=\left\{\exists x\in U:v_x\ge v_x^-+\frac{(\log L)^{3/2}}{4 \cos(\theta)}\right\}\\
\label{eq:41}
  \Gamma^1:=\{v^-_x\le v_x\mbox{\; for every\;} x\in U\}\\
 \label{eq:42}
   \Gamma^2:=\{v^+_x\ge v_x\mbox{\; for every\;} x\in U\}
\end{gather}
and $ \rho_{\left.\bar v\right|_{U^{\tt c}}}(\cdot)$ (for some $\bar v\in\mathcal V$) denotes the uniform measure over the elements $v$ of
$\mathcal V$ such that $v_x=\bar v_x$ for $x\notin U$. (Since monotone
sets and height functions are in one-to-one correspondence, we use the
same notation $\rho$ to denote the equilibrium uniform measure in both
cases). Note that the
event $A$ is increasing with respect to the natural partial order in
$\mathcal V$, where we say that $v\le w$ (with $v,w\in\mathcal V$)
if $v_x\le
w_x$ for every $x\in \ZZ^2$. To lighten notations, given $A\subseteq
\ZZ ^2$ we will write
$v_A\le
w_A$ if $v_x\le w_x$ for every $x\in A$.

We need the following monotonicity property, which is an immediate
consequence of
Proposition \ref{th:mondyn}:
\begin{lemma}
\label{th:monoton}
One has  
\begin{eqnarray}
  \label{eq:49}
\rho_{\left.w\right|_{U^{\tt c}}}(\cdot|a_U\le  w_U\le b_U)\preceq 
\rho_{\left.w'\right|_{U^{\tt c}} }(\cdot|a'_U\le w_U\le b'_U)  
\end{eqnarray}
 if $a,a',b,b',w,w'\in \mathcal V$ are such that $a_U\le a'_U$, $
b_U\le b'_U$ and $w_{U^{\tt c}}\le w'_{U^{\tt c}}$.
Moreover, $\rho_{\left.w\right|_{U^{\tt c}}}(\cdot)$ depends only on the value
of $w$ on $\partial U$.
\end{lemma}
Applying Lemma \ref{th:monoton}, we get
\begin{eqnarray}
  \label{eq:40}
  \rho_{\left.v^-\right|_{U^{\tt c}}}(A|\Gamma^1\cap \Gamma^2)\le   \rho_{\left.v^-\right|_{U^{\tt c}}}(A|\Gamma^1).
\end{eqnarray}
The key point is the following result, whose proof is given in Section
\ref{sec:K}:
\begin{theo}
\label{th:K}
There exists a probability measure $\nu$ on the elements $w\in\mathcal
V$ which satisfies the following properties:
\begin{gather}
  \label{eq:43}
  \nu(\exists x\in\partial U: w_x< v^-_x)\le
  \varepsilon_L    \\
   \label{eq:44}
    \nu(\rho_{\left.w\right|_{U^{\tt c}}}(A))\le \varepsilon_L\\
   \label{eq:45}
\nu(\rho_{\left.w\right|_{U^{\tt c}}}((\Gamma^1)^{\tt c}))\le \varepsilon_L,    
  \end{gather}
 where
 \begin{eqnarray}
    \label{eq:51}
\varepsilon_L:=(1/c) \exp(-c\,(\log L)^{3/2})
  \end{eqnarray}  
and $c$ is a universal positive constant.
\end{theo}

\medskip
We are now in a position to prove Proposition \ref{th:kenyon}.
We have
\begin{eqnarray}
  \label{eq:46}
 \nu(\rho_{\left.w\right|_{U^{\tt c}}}(A|\Gamma^1))&\ge &
 \nu\left[\rho_{\left.w\right|_{U^{\tt c}}} (A|\Gamma^1)|\;
v^-_{\partial U}\le w_{\partial U}
\right]\nu(v^-_{\partial U}\le w_{\partial U})\\\nonumber
&\ge& (1-\varepsilon_L) \rho_{\left.v^-\right|_{U^{\tt c}}} (A|\Gamma^1)
\end{eqnarray}
where we used \eqref{eq:43} and Lemma \ref{th:monoton}.
Therefore,
\begin{eqnarray}
  \label{eq:47}
 \rho_{\left.v^-\right|_{U^{\tt c}}}(A|\Gamma^1)&\le& 2
 \nu(\rho_{\left.w\right|_{U^{\tt c}}}(A|\Gamma^1))\\\nonumber
&=&
2 \nu(\rho_{\left.w\right|_{U^{\tt c}}}(A|\Gamma^1);\rho_{\left.w\right|_{U^{\tt c}}}(\Gamma^1)\le 1/2)+2
\nu(\rho_{\left.w\right|_{U^{\tt c}}}(A|\Gamma^1);\rho_{\left.w\right|_{U^{\tt c}}}(\Gamma^1)> 1/2)\\\nonumber
&\le& 2 \nu(\rho_{\left.w\right|_{U^{\tt c}}}(\Gamma^1)\le 1/2)+4\nu(\rho_{\left.w\right|_{U^{\tt c}}}(A))\\\nonumber
&\le& 2 \nu(\rho_{\left.w\right|_{U^{\tt c}}}((\Gamma^1)^{\tt c})\ge1/2)+4\varepsilon_L\le
4 \nu(\rho_{\left.w\right|_{U^{\tt c}}}((\Gamma^1)^{\tt c}))+4\varepsilon_L\le 8\varepsilon_L
\end{eqnarray}
where we used assumptions \eqref{eq:44}, \eqref{eq:45}, Markov's
inequality and the obvious bound $\rho_{\left.w\right|_{U^{\tt c}}}(A|\Gamma^1)\le \rho_{\left.w\right|_{U^{\tt c}}}(A)/\rho_{\left.w\right|_{U^{\tt c}}}(\Gamma^1)$. Recalling the definition of $\varepsilon_L$, 
this implies \eqref{eq:17}.
\hfill$\stackrel{\tiny\mbox{Eq. \eqref{eq:17}}}{\qed}$

\medskip

The idea behind Theorem \ref{th:K} is that, while the height
fluctuations of $v$ under $\rho$ for  the {\sl fixed
boundary conditions} imposed by $v^-|_{U^c}$ are hard to control, 
they are instead easily described if the boundary conditions are sampled from
the infinite measure $\nu$ described in Section \ref{sec:K}. Such
random boundary conditions are with high probability ``higher'' than
the deterministic ones (see \eqref{eq:45}) and an application of monotonicity 
(cf. the steps in  \eqref{eq:46}-\eqref{eq:47}) concludes the argument.

\subsection{Proof of Theorem \ref{th:K}: Dimers coverings and  height functions}
\label{sec:K}
We need first to recall some notions and results 
about the connection between monotone sets and dimer coverings of the
infinite two-dimensional hexagonal lattice (cf. for instance 
\cite{Kenyon,KenyonCMP,KenyonAIHP,KOS}).

\begin{figure}[htp]
\begin{center}
\includegraphics[width=0.9\textwidth]{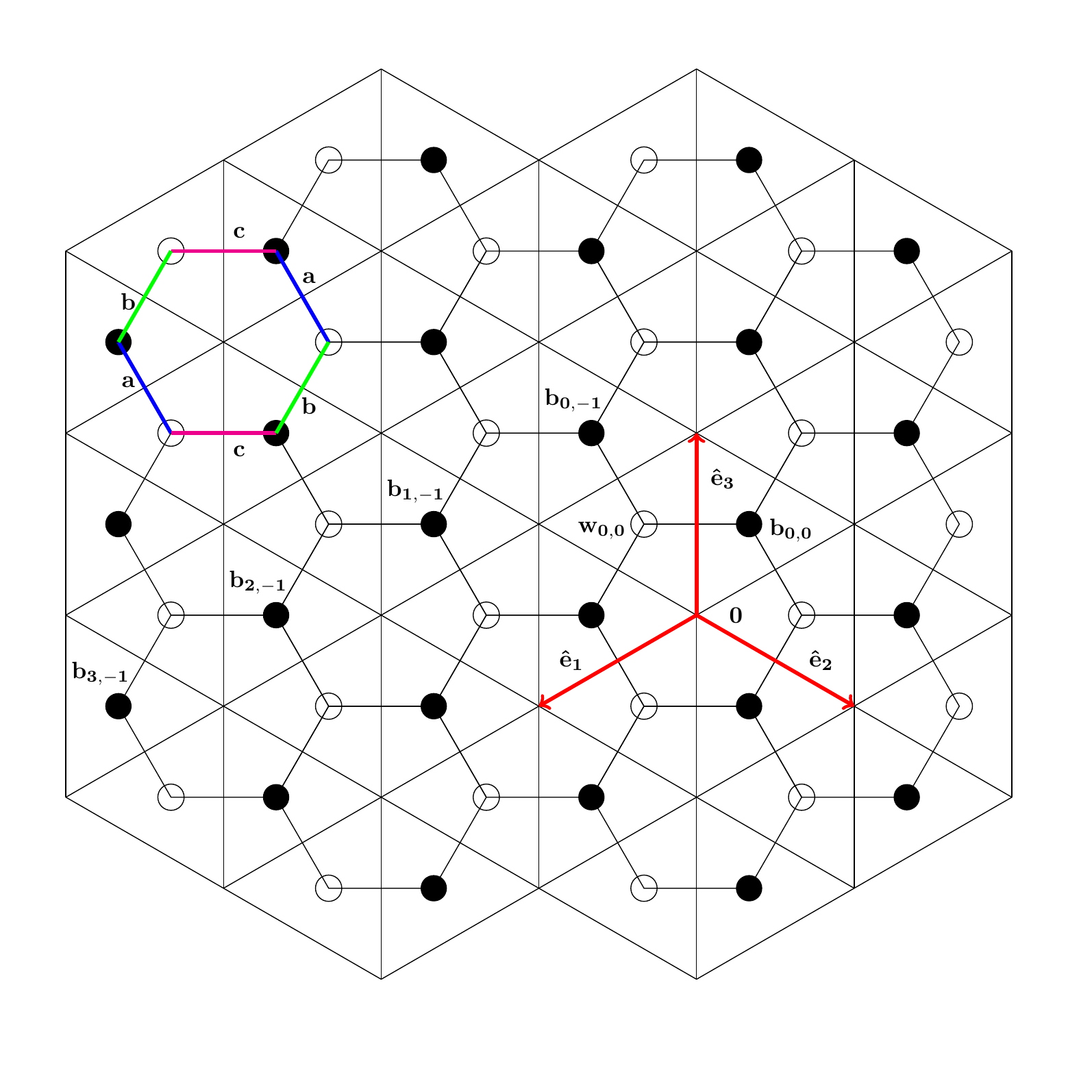}
\end{center}
\caption{A portion of the triangular lattice $\mathcal T$ and of the
  hexagonal lattice $\mathcal H$. }
\label{fig:hexa} 
\end{figure}

For $x\in \RR^3$,  let  $\pi_{111}(x)$ denote its orthogonal projection 
on the  $(111)$ plane $\{z\in \RR^3:z^{(1)}+z^{(2)}+z^{(3)}=0\}$.
Let also
\begin{eqnarray}
  \label{eq:35}
\mathcal T=\cup_{z\in \ZZ^3}\pi_{111}(z)  
\end{eqnarray}
and note that $\mathcal T$ is a two-dimensional triangular lattice,
which we consider as a graph by putting an edge between any two nearest neighbors
(whose mutual distance is $\sqrt{2/3}$). We also let the vector $\hat e_1$
(resp. $\hat e_2$ and $\hat e_3$) be the $\pi_{111}$ projection of the
vector which joins $(0,0,0)$ to $(1,0,0)$ (resp. to $(0,1,0)$ and to
$(0,0,1)$): of course, the $\hat e_i$ have norm $\sqrt{2/3}$.
See Figure \ref{fig:hexa}.

Let $\mathcal H$ be the
dual lattice of $\mathcal T$, again with edges between nearest
neighbors. $\mathcal H$
is a hexagonal lattice which we decompose as $\mathcal H=\mathcal H_B\cup
\mathcal H_W$, where both $\mathcal H_W$ and $\mathcal H_B$ are
translates of the triangular lattice $\mathcal T$, such that all
nearest neighbors of every vertex in $\mathcal H_W$ (resp. in
$\mathcal H_B$) belong to $\mathcal H_B$ (resp. to $\mathcal H_W$).
Vertices in $\mathcal H_W$ (resp. in $\mathcal H_B$) are said to be
{\sl white} (resp. {\sl black}).

{\bf Conventions}\footnote{
The conventions we adopt on the orientation of the axes and on the
labeling of the edges via the symbols $a,b,c$ do not coincide with those of \cite{Kenyon,KenyonCMP}, but
this is simply an irrelevant matter of convention.} We embed $\mathcal T$ and $\mathcal H$ in $\RR^2$, with the convention
that
$\pi_{111}(0)\in\mathcal T$ is mapped to $(0,0)$, that $\hat e_3$ is mapped to the
vertical vector $(0,\sqrt{2/3})$ and that $\hat e_1$ is obtained by
$\hat e_3$ by a counter-clockwise rotation of $(2/3)\pi$.
Also, given the two endpoints of the  edge of $\mathcal H$ which crosses $\hat e_3$, 
we decide that the one which has negative horizontal coordinate
(call it ${\rm w}_{0,0}$) belongs
to $\mathcal H_W$, while the other (call it ${\rm b}_{0,0}$) belongs to $\mathcal
H_B$. 
Observe that the vector ${\rm b}_{0,0}-{\rm w}_{0,0}$ is proportional
to $\hat e_2-\hat e_1$. 
We will label a site  $v\in\mathcal H_W$ (resp. in $\mathcal H_B$)
as $\rw_{x,y}$ if $v=\rw_{0,0}+x \hat
e_1+y\hat e_2$ (resp. as $\rb_{x,y}$ if $v=\rb_{0,0}+x \hat
e_1+y\hat e_2$).
Edges of $\mathcal H$ which are perpendicular to
$\hat e_1$ (resp. to $\hat e_2$ or $\hat e_3$) will be called edges
{\sl of type ``$a$''} (resp. of type ``$b$'' or ``$c$'').
See Figure \ref{fig:hexa}.

\medskip

Given a monotone set $V\in\Sigma$, we define
the height function $h:=h(V):=\{h_x\}_{x\in\mathcal T}$ as follows:
\begin{eqnarray}
  \label{eq:36}
  h_x:=\max_{z\in V}\{z^{(3)}:\pi_{111}(z)=x\}\in \ZZ.
\end{eqnarray}
We denote by $\mathcal W$ the set of all functions $h:\mathcal T\mapsto \ZZ$ such
that
there exists $V\in \Sigma$ with $h=h(V)$. 
It is easy to see that there is a one-to-one mapping between elements
of 
$\Sigma$ and elements of $\mathcal V$ (cf. Definition \ref{def:v}), and between elements of
$\mathcal V$ and elements of $\mathcal W$. 
In other words, the functions $\{v_x\}_{x\in\ZZ^2}$
(cf. Definition \ref{def:v}) and
$\{h_x\}_{x\in\mathcal T}$ are two equivalent ways to describe the height of the set
$V$ with respect to the horizontal plane.

A {\sl dimer covering} $M$ of the hexagonal lattice $\mathcal H$ is a
subset of the edges of $\mathcal H$ covering each vertex of $\mathcal
H$ exactly
once. Note that each edge in $M$ covers one black and one white
vertex. To each height function (or monotone set) $h\in \mathcal W$
is uniquely
associated a dimer covering, and conversely to a dimer covering one
can associate uniquely a height function, provided that one fixes
the height function at some arbitrary point $x\in \mathcal T$
(in other words, dimer coverings identify only gradients of the height
function).

\begin{figure}[htp]
\begin{center}
\includegraphics[width=0.9\textwidth]{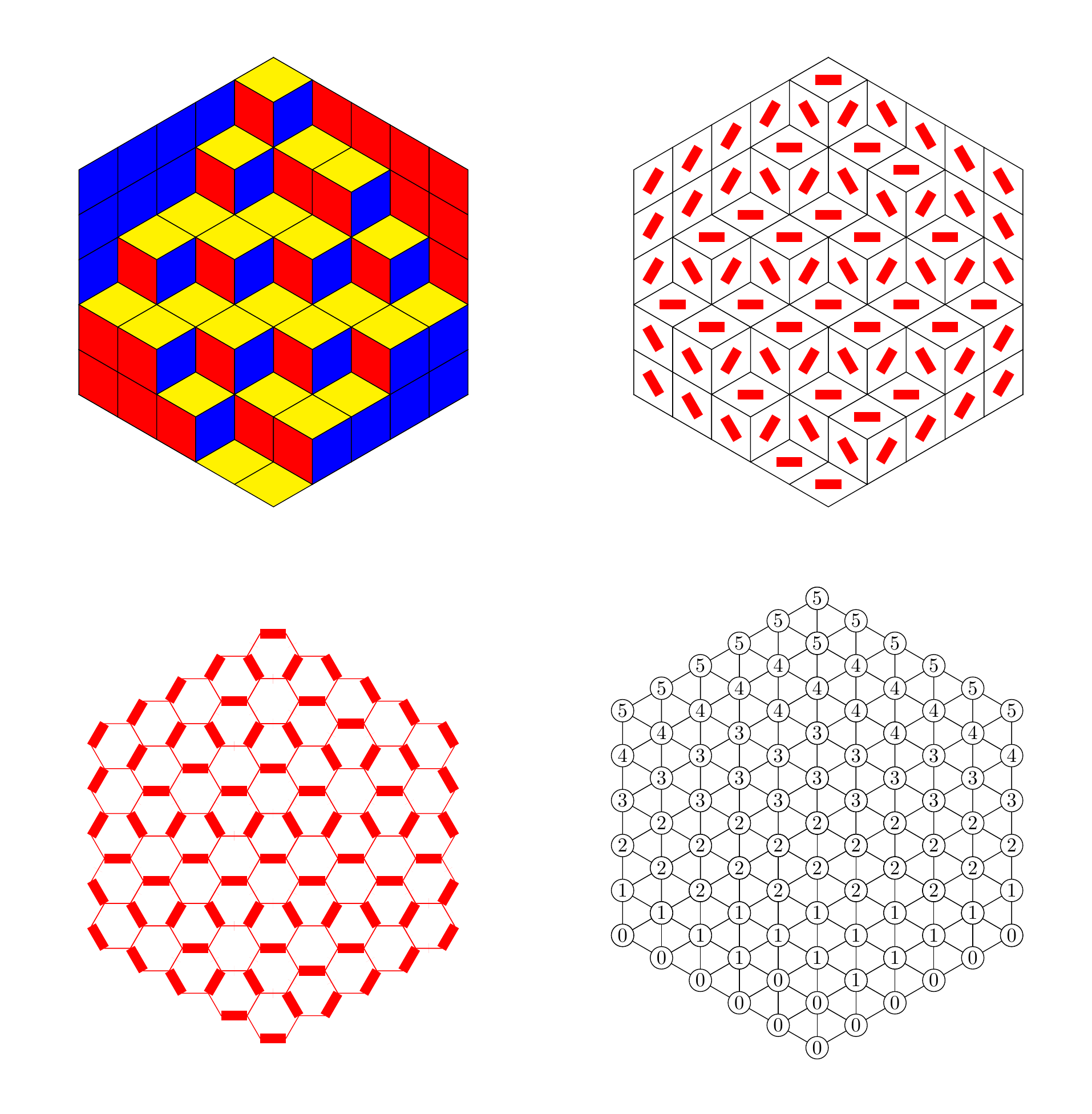}
\end{center}
\caption{A graphically convenient way to visualize a monotone set $V$ is
  to associate to every $x\in V$ a unit cube, centered at $x-(1/2,1/2,1/2)$.
In this way, a monotone set, when seen from the $111$ direction, appears
  as a tiling of a portion of the plane with three types of lozenges
  (top drawing on the left). If we mark a segment along the longer
  diagonal of each lozenge (top drawing on the right), we obtain a
  dimer covering of a portion of the hexagonal lattice (bottom,
  left). To each vertex of a lozenge, i.e.\ to every vertex in a
  triangular lattice, one associates its vertical height (bottom,
  right). One can check in the drawing that the dimer covering and the
height function are linked by the construction explained in the text.}
\label{fig:4} 
\end{figure}

The construction of the dimer covering $M$ given $h(V)$ goes as
follows (see also Figure \ref{fig:4}). Let $e$ be
an edge of $\mathcal H$, and let $(x,y)$ be the edge of $\mathcal
T$ which intersects $e$, with the convention that the vector
$x-y$ is $+\hat e_i$ for some $i\in\{1,2,3\}$. 
If $x-y=\hat e_1$ or $x-y=\hat e_2$, then we put a dimer on
the edge $e$ if $h_{x}=h_{y}-1$, and we do not put it if
$h_{x}=h_{y}$
(it is easy to see that these are the only two possibilities since $h$ is a monotone interface).
If $x-y=\hat e_3$, then we put a dimer on $e$ if
$h_{x}=h_{y}$ and we do not put it if $h_{x}=h_{y}+1$.
Note that the asymmetry between the indices $1,2,3$ is due to the fact
that we are computing heights with respect to the horizontal plane.
A more symmetric choice (but less convenient for our purposes) would
be to measure heights with respect to the $(111)$ plane.

Conversely, the construction of $h$ given a dimer covering $M$ goes as
follows. First we define a {\sl flux} $\omega$, i.e.\ a function on oriented
edges $e$ of $\mathcal H$, such that $\omega(e)=-\omega(-e)$. If
$e$ is oriented from the white to the black vertex, then:
\begin{itemize}
\item if $e$ is of type ``$a$'' or of type ``$b$'', then $\omega(e)=0$ if
  $e\notin M$
 and $\omega(e)=1$ otherwise

\item if $e$ is of type ``$c$'', then $\omega(e)=0$ if $e\in M$ and $\omega(e)=-1$ otherwise.
\end{itemize}
Next, we fix some arbitrary value $h_{\bar x}\in \ZZ$ at some point
$\bar x\in\mathcal T$. Finally, the difference $h_v-h_{\bar x}$ for $v\in\mathcal T$ is the total
flux of $\omega$ which crosses, from right to left, a path $\ell$ of edges of $\mathcal T$, which goes from
$\bar x$ to $v$. The fact that $h_v-h_{\bar x}$ does not depend on the
choice of the path $\ell$ is due to the fact that the flux $\omega$
has zero divergence, see \cite[Sec. 2.2]{Kenyon}.

Given ${\bf p}=(p_a,p_b,p_c)$ with  $p_a,p_b,p_c>0$ and
$p_a+p_b+p_c=1$, take a
triangle
of perimeter $1$ whose angles are $\theta_a:=\pi p_a,\theta_b:=\pi p_b,\theta_c:=\pi
p_c$ and let $k_a$ (resp. $k_b,k_c$) be the length of the side opposite to
$\theta_a$ (resp. $\theta_b,\theta_c$). To every choice of
${\bf p}$ as above, 
one can associate a
translation-invariant Gibbs measure $\Mu$ on dimer coverings of
$\mathcal H$.
Translation-invariance means
that,
if $A$ is a set of edges of $\mathcal H$, then $\Mu(A\subseteq
M)=\Mu(T(A)\subseteq M)$, where $T$ is a translation which maps
$\mathcal H$ into itself.

The precise statement is the following, whose different pieces were
proved in \cite{KOS,Sheffield,KenyonAIHP}:
\begin{theo}
\label{th:KOS}
There exists a unique translation-invariant law $\Mu$ on dimer
coverings, such that the
probability that a given edge of type ``$a$'' (resp. of type $``b",``c"$)
belongs to $M$ is $p_a$ (resp. $p_b,p_c$) and such that, conditionally
on the configuration $M_{X^{\tt c}}$ of the covering $M$ outside a given domain $X\subseteq \mathcal H$,
$\Mu$ is the uniform measure over all coverings $M_X$ of $X$ compatible
with $M_{X^{\tt c}}$, i.e., such that $M_X\cup M_{X^{\tt c}}$ is a covering of
$\mathcal H$ (we refer to this property as ``DLR property''). 

Explicitly, $\Mu$ is described as follows.
Define the matrix $K:=\{K(\rb,\rw)\}_{\{\rb\in \mathcal H_B,\rw\in \mathcal
  H_W\}}$ as follows:
\begin{itemize}
\item if
$\rm b$ is not a nearest neighbor of $\rm w$, then $K(\rb,\rw)=0$;

\item $K({\rb},{\rw})=k_a$
(resp. $k_b,k_c$) if the edge $(\rm b,\rm w$) is of type ``$a$'' (resp. of
type $``b",``c"$).
\end{itemize}
Define also the matrix  
$K^{-1}:=K^{-1}(\rw,\rb)_{\{\rw\in\mathcal H_W,\rb\in\mathcal H_B\}}$
as
\begin{eqnarray}
  \label{eq:55}
  K^{-1}(\rw_{x,y},\rb_{x',y'})=K^{-1}(\rw_{0,0},\rb_{x'-x,y'-y})
\end{eqnarray}
and
\begin{eqnarray}
  \label{eq:54}
  K^{-1}(\rw_{0,0},\rb_{x,y})=\frac1{(2\pi i)^2}\int_{\mathbb T}
\frac{z^{-y}w^{x}}{k_c+k_az+k_bw}\frac{dz}z\frac{dw}w
\end{eqnarray}
where the integral is taken over the two-dimensional torus $\mathbb
T:=
\{(z,w)\in \mathbb C^2:|z|=|w|=1\}$. Then,
  given a set of edges $X=\{(\rw_1,\rb_1),\ldots,(\rw_k,\rb_k)\}$ in
  $\mathcal H$, one
  has
  \begin{eqnarray}
    \label{eq:53}
    \Mu(X\subseteq M)=\left(\prod_{i=1}^k K(\rb_i,\rw_i)\right)
\det\left(K^{-1}(\rw_i,\rb_j)
\right)_{1\le i,j\le k}.
  \end{eqnarray}

\end{theo}
Note that $K$ is a weighted version of the adjacency matrix of
$\mathcal H$. It is immediate to check that  one has the relation 
$K K^{-1}=\mathbb I$, which justifies the notation
$K^{-1}$. The infinite matrix $K$ however does
not admit a unique inverse, as discussed for instance in \cite{KenyonCMP}.

\medskip

\begin{rem}\rm
\label{rem:pp}
If we fix deterministically the value $h_{\bar x}$ for some $\bar x\in\mathcal T$, then the function 
$\mathcal T\ni x\mapsto\Mu(h_x)-h_{\bar x}$ is linear in $x-\bar x$,
thanks to translation invariance of $\Mu$ . As a consequence, the set of points
\begin{eqnarray}
  \label{eq:52}
\{z\in \RR^3:\pi_{111}(z)\in \mathcal T, z^{(3)}=\Mu(h_{\pi_{111}(z)})
 \} 
\end{eqnarray}
is contained in some plane $\Pi_{\bf p}\subseteq \RR^3$ which of course contains the point $\bar z\in\RR^3$ such that
$\pi_{111}(\bar z)=\bar x$ and $\bar z^{(3)}=h_{\bar x}$. It is rather easy to check
that
the normal vector of the plane $\Pi_{\bf p}$ is parallel to $\bfp$. In
particular, the plane $\Pi_{\bf p}$ is monotone in the sense of
Definition \ref{def:monotone}.
  
\end{rem}

\subsubsection{Height fluctuations}
Set for lightness of notation $\Delta=(1/4)(\log L)^{3/2} $.

\begin{prop}
\label{th:flutt} Let $\bar x\in\mathcal T$ and fix $h_{\bar x}$ to some deterministic value.
  There exists $c>0$ such that for every $L\ge 2$ one has
  \begin{eqnarray}
    \label{eq:58}
    \Mu\left(\exists x\in \mathcal T\mbox{\;such that\;} d(x,\bar x)\le L
\mbox{\;and\;} |h_x-\Mu(h_x)|\ge \Delta/2\right)\le 
\frac1c\exp(-c (\log L)^{3/2})
  \end{eqnarray}
uniformly in $\bfp$ (where $d(\cdot,\cdot)$ denotes the Euclidean
distance in $\mathcal T$).
\end{prop}
{\sl Proof of Proposition \ref{th:flutt}.}
By translation invariance  of $\Mu$, we can assume without loss of generality
that
$\bar x=0$ and that we fixed $h_{\bar x}=0$.
If $d(x,0)\le L$, we can write $x=m\hat e_1+n\hat e_2$ for some
$m,n\in \{-L,-L+1,\ldots,L\}
$.
Therefore, via a union bound, to show \eqref{eq:58} it is enough to prove that, for every $|n|\le L$
\begin{eqnarray}
  \label{eq:60}
    \Mu(|h_{n \hat e_i}-\Mu (h_{n \hat e_i})|\ge \Delta/4)\le\frac1c\,\exp(-c(\log L)^{3/2})
\end{eqnarray}
for $i=1,2$. We consider for instance the case $n>0$ and $i=1$, the
other cases being  essentially identical. 
From the construction of the height function in Section \ref{sec:K}, we know that 
\begin{eqnarray}
  \label{eq:61}
  h_{0}-h_{n \hat e_1}=|M\cap \{(\rb_{1,0},\rw_{1,1}),\ldots (\rb_{n,0},\rw_{n,1})\}|=:\mathcal N_n,
\end{eqnarray}
i.e., it is just the number of dimers in the set $\{(\rw_{1,0},\rb_{1,1}),\ldots (\rw_{n,0},\rb_{n,1})\}$.
Thanks to the determinantal representation \eqref{eq:53}, one has \cite{Soshnikov}
that $\mathcal N_n$ has the same law as a sum of $n$ independent Bernoulli random variables $X_i,i\le n$, whose parameters $q_i:=P(X_i=1)$
are the eigenvalues of the matrix
$A:=\{k_a\,K^{-1}(\rw_{i,1},\rb_{j,0}  )\}_{1\le i,j\le n}$. 

In general, it is not easy to solve explicitly the double integral
\eqref{eq:54} which defines $K^{-1}$. However, there are special
values of $x,y$ for which $K^{-1}(\rw_{0,0},\rb_{x,y})$ takes an
easy form. In particular, one checks that\footnote{
The formula \eqref{eq:56} differs from the analogous one in
\cite[Sec. 6.3]{Kenyon} by the factor $(-1)^n$, probably due to a typo
there. In any case, the global sign is inessential for our
computation, since \eqref{eq:62} below depends only on the absolute value of
$K^{-1}$.
},
 for $n\in \ZZ\setminus \{0\}$, 
\begin{gather}
  \label{eq:56}
  K^{-1}(\rw_{0,0},\rb_{n,-1})=(-1)^n\frac{\sin(n p_a\pi)}{\pi n k_a}
\end{gather}
and that 
\begin{gather}
  \label{eq:48}
K^{-1}(\rw_{0,0},\rb_{0,-1})=\frac{p_a}{k_a}.
\end{gather}
As a consequence, the entries of the matrix $A$ are given by
\begin{eqnarray}
  \label{eq:63}
  A_{i,j}=a_{i-j}:=k_a\,K^{-1}(\rw_{0,0},\rb_{j-i,-1})=\left\{
  \begin{array}{ll}
(-1)^{j-i}\frac{\sin[(i-j)\pi p_a]}{\pi(i-j)}
&   i\ne j\\
p_a&i=j
  \end{array}
\right.
\end{eqnarray}
and then
\begin{eqnarray}
  \label{eq:62}
  \Var(\mathcal N_n)&=&\sum_{i=1}^n q_i(1-q_i)=Tr(A)-Tr(A^2)\\
&=&n a_0(1-a_0)-a_1^2(2n-2)-a_2^2(2n-4)-\ldots-2 a_{n-1}^2.
\end{eqnarray}
We will show in a moment that
\begin{eqnarray}
  \label{eq:64}
    \Var(\mathcal N_n)\le c\log n
\end{eqnarray}
for some $c<\infty$,
uniformly in $\bfp$, which allows to conclude the proof of
\eqref{eq:60}: via the exponential Tchebyshev inequality,
\begin{eqnarray}
  \label{eq:65}
  \Mu(\mathcal N_n-\Mu(\mathcal N_n)\ge \Delta/4)&=&P\left(\sum_{i\le n}(X_i-E(X_i))\ge \Delta/4\right)\\
&\le& e^{- \Delta/4}\prod_{i\le n}\left(q_i e^{(1-q_i)}+(1-q_i)e^{- q_i}
\right).
\end{eqnarray}
Using the inequality
$\exp(x)\le 1+x+ x^2$ which holds for $-1<x<1$ and the fact that $n\le
L$, one deduces
\begin{eqnarray}
  \label{eq:66}
  P(\mathcal N_n-\Mu(\mathcal N_n)\ge \Delta/4)&\le& 
e^{- \Delta/4}\prod_{i\le n}\left(1+ q_i(1-q_i)\right)\\
&\le& e^{-\Delta/4+ \Var(\mathcal N_n)}
\le e^{-(1/16)(\log L)^{3/2}+c\log L}.
\end{eqnarray}
Analogously, one obtains the same upper bound for $P(\mathcal N_n-\Mu(\mathcal N_n)\le -\Delta/4)$
and \eqref{eq:60} is proven.

It remains only to prove the estimate \eqref{eq:64}. Essentially the
proof can be found in \cite[Sec. 6.3]{Kenyon}, where however
uniformity with respect to $\bfp$ was not discussed, so we repeat
quickly the necessary steps here.
Observe first of all that
\begin{eqnarray}
  \label{eq:57}
  |a_i|\le \frac c{|i|+1}, \;\;i\in\ZZ
\end{eqnarray}
uniformly in $\bfp$.
Therefore,
\begin{eqnarray}
  \label{eq:59}
    \Var(\mathcal N_n)&=&n\left[a_0(1-a_0)-2\sum_{i=1}^\infty a_i^2 \right]+2\sum_{i=1}^{n-1}i\,a_i^2+2n\sum_{i=n}^\infty a_i^2\\
&\le& n\left[a_0(1-a_0)-2\sum_{i=1}^\infty a_i^2 \right]+c\log n
\end{eqnarray}
for some $c$ independent of $\bfp$. 
Finally, one observes that the quantity in square brackets is
identically equal to zero. To see this, let $f(x)=|x|(\pi-|x|)$ so that
$a_0(1-a_0)=f(\theta_a)/\pi^2$. One then observes that
\begin{eqnarray}
  \label{eq:67}
  c_k:=\frac1{2\pi}\int_{-\pi}^\pi f(x)e^{ikx}dx=
\left\{
  \begin{array}{lll}
    \frac{\pi^2}6&\mbox{if}& k=0\\
    -\frac{1+(-1)^k}{k^2}&\mbox{if}& k\in \ZZ\setminus\{0\}
  \end{array}
\right.
\end{eqnarray}
and then a straightforward computation shows that the Fourier identity
$f(\theta_a)=\sum_{k\in\ZZ}c_k\,e^{-ik\theta_a}$ is equivalent to
$a_0(1-a_0)=2\sum_{i\in\NN}a_i^2$.

\hfill$\stackrel{\tiny\mbox{Proposition \ref{th:flutt}}}{\qed}$

{\sl Proof of Theorem \ref{th:K}.} Recall the setting of Section \ref{sec:eq17}.
Let $u=(u^{(1)},u^{(2)})\in\partial U$, let $\bar
z:=(u^{(1)},u^{(2)},v^-_{u}+\lfloor \Delta/(2\cos(\theta))\rfloor)\in\ZZ^3$, and
$\bar x:=\pi_{111}(\bar z)\in\mathcal T$. Choose $\bfp$ to be parallel
to the normal to the plane ${\Pi}$ of Definition \ref{def:geom}(c) 
and consider the infinite volume distribution of height functions induced by $\Mu$, with the normalization $h_{\bar x}=v^-_{u}+\lfloor\Delta/(2\cos(\theta))\rfloor$.
Note that the plane $\Pi_{\bfp}$ defined in Remark
\ref{rem:pp} is obtained from  $\Pi$ by translating it for a distance
$\lfloor\Delta/(2\cos(\theta))\rfloor
+O(1)$ in the
positive vertical direction, so that $d(\Pi,\Pi_{\bf p})=\Delta/2+O(1)$.

For a given realization of the height function $h$, we let $Y$ be the
finite set of points
\begin{eqnarray}
  \label{eq:68}
  Y(h):=\{z\in \ZZ^3\mbox{\;such that \;}d(\pi_{111}(z),\bar x)\le L
\mbox{\;and\;} z^{(3)}=h_{\pi_{111}(z)}
 \} \subseteq \ZZ^3.
\end{eqnarray}
Thanks to Proposition \ref{th:flutt}, one has that
\begin{eqnarray}
  \label{eq:69}
  \mu_\bfp(\exists z\in Y(h)\mbox{\;such that \;}d(z,\Pi_\bfp)\ge\Delta/2)\le \frac1c\exp(-c(\log L)^{3/2})
\end{eqnarray}
with $c$ independent of ${\bf p}$.

Recall that to a height function $h\in \mathcal W$ there corresponds a
unique element $v=v(h)\in\mathcal V$ and
let $\nu$ be the law on $\mathcal V$ induced by $\mu_\bfp$. Note that
$v_{(u^{(1)},u^{(2)})}=v^-_u+\lfloor \Delta/(2\cos(\theta))\rfloor $
deterministically.
We show
now that $\nu$ satisfies the conditions
\eqref{eq:43}--\eqref{eq:45}.
The point is that, as observed at the beginning of Section \ref{sec:eq17}, the diameter of $U\cup \partial U$ is $o(L)$, so that the
$\pi_{111}$ projection of any point
$(x^{(1)},x^{(2)},v_{(x^{(1)},x^{(2)})})$, with $(x^{(1)},x^{(2)})\in
U\cup\partial U$, has distance from $\bar x$ smaller than $L$. As a consequence, given some $v=v(h)\in \mathcal V$,
\begin{eqnarray}
  \label{eq:70}
 B(v):= \{x\in \ZZ^3\mbox{\;such that \;} (x^{(1)},x^{(2)})\in
 U\cup\partial U
\mbox{\;and\;} x^{(3)}=v_{(x^{(1)},x^{(2)})}
 \}\subseteq Y(h)
\end{eqnarray}
and, from \eqref{eq:69}, 
\begin{eqnarray}
  \label{eq:71}
  \nu\left(\exists z\in B(v)\mbox{\;such that \;}d(z,\Pi_\bfp)\ge\Delta/2\right)\le \frac1c\exp(-c(\log L)^{3/2})
\end{eqnarray}
which immediately implies condition \eqref{eq:43}, since the distance
between $\Pi_{\bf p}$ and $\Pi$ is $\Delta$, and the graph of the
function $U\cup \partial U\ni x\mapsto v^-_x$ is within distance
$O(1)$ from the plane $\Pi$ (recall that $v^-$ is the vertical height
function of the monotone set $V^-$ defined in \eqref{eq:99} so that,
in $U$, $v^-$ is just the lattice approximation of the plane $\Pi$).
Conditions \eqref{eq:44}, \eqref{eq:45} are also immediate from \eqref{eq:71}, once one
realizes
that
$\nu(\rho_{v|{U^{\tt c}}}(O))=\nu(O)$ if $O$ is an event which depends only on
$\{v_x\}_{x\in U}$. This is because, thanks to the DLR property of
$\mu_\bfp$ (cf. Theorem \ref{th:KOS}), the measure $\nu(\cdot|\{v_x\}_{x\in U^{\tt c}})$ is the uniform
measure over the elements of $\mathcal V$ which coincide with $v$
outside $U$; in other words, 
$\nu(\cdot|\{v_x\}_{x\in U^{\tt c}})$ is nothing but $\rho_{v|_{U^{\tt c}}}(\cdot)$.

\hfill$\stackrel{\tiny\mbox{Theorem \ref{th:K}}}{\qed}$

\section{On the mixing time of a dynamics of monotone sets}
\label{app:wilson}
In this section we prove Theorem \ref{th:wilson}. Recall the
notation from Section \ref{sec:wilson}, in particular the definition
of $\Lambda=V^+\setminus V^-$, of $D$ (the diameter of the horizontal
projection of $\Lambda$) and of $H$ (the maximal vertical distance
between two points in $\Lambda$ with the same horizontal projection). Without loss of generality, we can assume that
$\Lambda$ is contained in  some 
parallelepiped $\G\subset\bbN^3$
defined by 
\begin{eqnarray}
  \label{eq:100}
  \G=\{z\in\bbN^3\,:\;1\le z^{(1)}\le a_1,1\le z^{(2)}\leq a_2\,,\; h_1\leq z^{(3)}\leq h_2\}\,,
\end{eqnarray}
with $a_i\le D$ and $h_2=\max\{z^{(3)},z\in\Lambda\}$,
 $h_1=\min\{z^{(3)},z\in\Lambda\}$.
Recalling Definition \ref{def:v},
one can identify uniquely any monotone set $V\in \Sigma^+$, such that $V^-\subset V\subset V^+$, via a 
vertical height function $v_{(x,y)}$, indexed by pairs $(x,y)$ of non-negative integers.
Since the configuration of $V$ outside $\Lambda$ is fixed and coincides with $V^+$, 
what matters is the collection  $v:=\{v_{(x,y)}\}_{1\le x\le a_1,1\le y\le a_2}$. This defines a {\sl plane partition} in the box $\G$ (cf. also Definition \ref{def:monset}), 
i.e.\ a collection of heights 
such that 
$v_{(x+1,y)}\leq v_{(x,y)}$, $v_{(x,y+1)}\leq v_{(x,y)}$.
As discussed in Section \ref{app:kenyon}, the bijective correspondence between $v$ and $V$ is given by 
$$v_{(x,y)} = \max\{z: (x,y,z)\in V\}\,.$$
Each pair $(x,y)$ can be identified with a unit square in the plane $z^{(3)}=0$, in such a way that
the center of this square is given by the point of $\bbR^3$ with coordinates $(x-\frac12,y-\frac12,0)$. Thus, we interpret $v_{(x,y)}$ as the height of the column at $(x,y)$.  The minimal and maximal sets $V^-, V^+$ correspond to minimal and maximal column heights, denoted  $v^-$ and $v^+$ respectively
(the same notation was used in Section \ref{sec:eq17}).  We define $\O
= \O(v^\pm)$ as the set of all plane partitions $v$ in the box $\G$
such that $v^-\leq v\leq v^+$. Thus, $\O$ is in one-to-one
correspondence with the set  $\{V\in \Sigma^+$ such that $V^-\subset
V\subset V^+\}$, and the measure $\r=\r_{V^\pm}$ of Section
\ref{sec:wilson} becomes now the uniform probability measure on $\O$ (we still call it $\r$).


\subsection{Column dynamics}\label{sec:columns}
The first step in the proof of Theorem \ref{th:wilson} consists in
establishing a mixing time upper bound (Lemma \ref{le:tmixbo} below) for a Markov chain that involves equilibration of full columns at each move.  The key idea here borrows from Wilson's analysis \cite{Wilson}
of the Luby-Randall-Sinclair Markov chain for lozenge tilings \cite{LRS}. 
The second step (see Section \ref{sec:compare}) is to show that  Lemma \ref{le:tmixbo} implies the upper bound on the
mixing time of the ``single spin-flip'' dynamics $\nu^{V_0}_t$ which
is under consideration  in
Theorem \ref{th:wilson}. A similar strategy was used in \cite{MS}
in the simpler context of the $(1+1)$-dimensional SOS model.

\begin{figure}[htp]
\begin{center}
\includegraphics[width=0.8\textwidth]{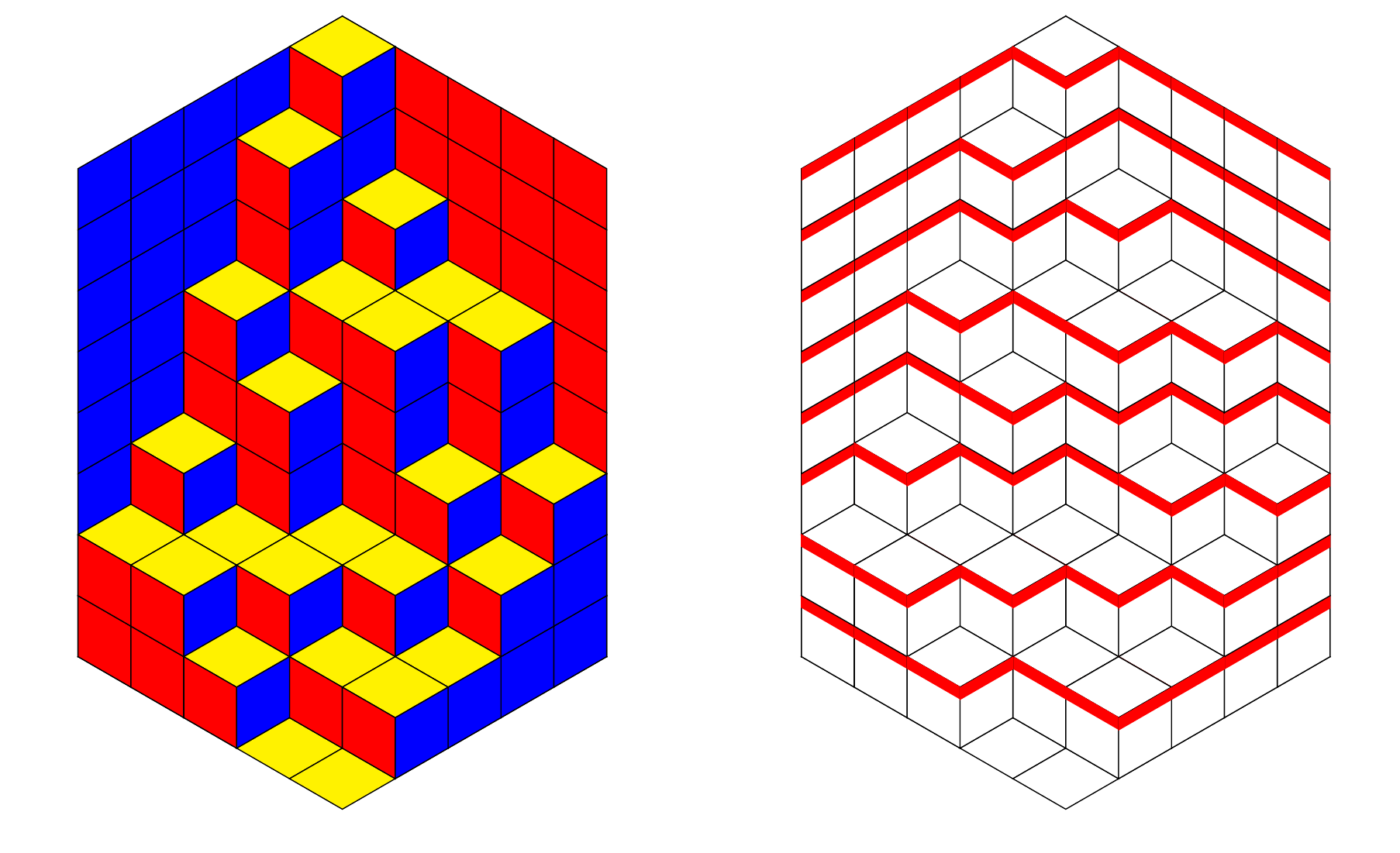}
\end{center}
\caption{In the left drawing, the ``column'' or ``plane partition''
  representation of a monotone set $V\in\Sigma^+$. In the right
  drawing,
the corresponding ``lattice path'' representation. The lattice path
$\phi^{(j)}$ is essentially the level set at height $j$ of the column
representation of $V$, 
seen from the $(111)$ direction. It is clear that, whenever we {\sl
  add} a cube to the plane partition on the left, one lattice path on
the right will be {\sl lowered}. This shows that  $v_1\le v_2$ is
reversed into
$\phi(v_1)\ge \phi(v_2)$.}
\label{fig:polymers} 
\end{figure}

A column $(x,y)$ is said to be even/odd if $(x-y)$ is even/odd.
Consider the continuous time Markov chain with state space $\O$, 
where at each arrival time of a Poisson process with parameter $1$ we flip a fair binary coin; 
if the coin is $0$ (resp.\ $1$) we update simultaneously all even (resp.\ odd) 
column heights $v_{(x,y)}$ with a sample from the distribution $\r$ conditioned on the 
current value of the height of the odd (resp.\ even) columns. Let $P_t(v,\cdot)$ denote the distribution at time $t$ of such Markov process
when the starting configuration is $v\in\O$. Note that the kernel $P_t=P_t(\cdot,\cdot)$ satisfies
\begin{equation}\label{expG}
P_t = e^{t\,\cG}\,,
\end{equation}
where the infinitesimal generator $\cG$ acts on functions $f:\O\mapsto\bbR$ by
\begin{equation}\label{cG}
[\cG f](v) = \frac12\, \r[\,f\tc {\rm odd}\,] (v)
+  \frac12\, \r[\,f\tc {\rm even}\,](v) - f(v)\,,
\end{equation}
where we use the notation $\r[\,f\tc  {\rm odd}\,] (v)$ (resp. $\r[\,f\tc  {\rm even}\,] (v)$) for the expectation with respect to  $\r$ conditioned on 
$\{v_{(x,y)}\,,\;(x,y)\;{\rm odd}\}$ (resp.
$\{v_{(x,y)}\,,\;(x,y)\;{\rm even}\}$). Note that sampling from $\r[\,\cdot\,\tc  {\rm odd}\,]$ 
amounts to pick uniformly at random a new configuration of even column heights
that is compatible with the current odd column heights (compatibility
here means that the configuration of all column heights is then a
plane partition $v$ that satisfies $v^-\le v\le v^+$).
 It is important to remark that the probability measure
 $\r[\,\cdot\,\tc  {\rm odd}\,]$ is a product of single column
 probability measures, i.e.\ conditionally on
 $\{v_{(x,y)}\,,\;(x,y)\;{\rm odd}\}$, all even columns become
 independent.  The same remarks apply to $\r[\,\cdot\,\tc  {\rm even}\,]$. 
  Clearly, $\r$ is the reversible invariant distribution of our Markov chain. 
\begin{lemma}
\label{le:tmixbo}
If $H\le D$, there exists a constant $c>0$ such that 
\begin{equation}
\sup_{v\in\O}\|P_t(v,\cdot)-\r\| \leq c\,D^4\,\exp{\Big(-\frac{t}{c\,D^2}\Big)}\,.
\label{tmixbo}
\end{equation}
\end{lemma}
\proof 
We use the well known ``lattice path 
representation'' of a plane partition; see e.g.\ \cite{LRS,Wilson}. 
Namely, any plane partition $v\in\O$ can be seen as a collection of 
$h_2-h_1+1$ non-intersecting lattice paths $\phi^{(i)}=\phi^{(i)}(v)$,
$i=h_1,\dots,h_2$, each of length $2D$ ($h_1,h_2$ are the same integers
that appear in
\eqref{eq:100}) which satisfy
\begin{eqnarray}
  \label{eq:102}
\phi^{(j)}_0=\phi^{(j)}_{2D}=j\,,\;\;
\phi^{(j)}_{x+1} -
\phi^{(j)}_{x}\in\{-1,+1\}\,,\;\;\phi^{(j)}_x< \phi^{(j+1)}_x\,  
\end{eqnarray}
for all $j=h_1,\dots,h_2\,,\;x=0,\dots,2D$.
The polymer $j$ describes the $j^{th}$ level set
$\{z\in V\cap \Lambda :z^{(3)}=j\}$.
For the precise construction of the paths, we refer to 
\cite[Section 5]{Wilson}, \cite[Section 2.1]{LRS} (see also Figure
\ref{fig:polymers} for a graphical construction).

In the plane partition-to-lattice path mapping, inequalities are reversed, i.e.\ if $v_1\le v_2$ then
$\phi(v_1)\ge \phi(v_2)$ (cf. Figure \ref{fig:polymers}).
We let $\phi^+:=(\phi^{+,(i)})_{i}$ and $\phi^-:=(\phi^{-,(i)})_{i}$
denote the lattice paths corresponding to the maximal and minimal
plane partitions $v^+,v^-$, so that $\phi^+\le \phi^-$. Then, the
condition $v^-\le v\le v^+$ gives 
\begin{eqnarray}
  \label{eq:101}
\phi^{+,(j)}_x\leq \phi^{(j)}_x\leq \phi^{-,(j)}_x \,,  
\end{eqnarray}
for all $j=h_1,\dots,h_2\,,\;x=0,\dots,2D$ or, more compactly, $\phi^+\le \phi\le \phi^-$.

Let $\wt\O$ denote the 
set of configurations $\phi=(\phi^{(j)}_x)_{x,j}$ of integer heights
$\phi^{(j)}_x\in\bbZ$, $j=h_1,\dots,h_2$ and $x=0,\dots,2D$ satisfying
the constraints \eqref{eq:102}, \eqref{eq:101}.
The construction above establishes a one-to-one correspondence between
the set $\O$ of plane partitions $v$ satisfying $v^-\leq v\leq v^+$ and the set $\wt\O$.
The image of the measure $\r$ is the uniform probability distribution on $\wt\O$ (which we call again $\r$ with some abuse of language). 
Moreover, it is not hard to see that the image of the Markov process
with ``full column moves'' under this map coincides with the continuous time Markov process with state space $\wt\O$ obtained as follows: at each arrival time of a Poisson process with parameter $1$ we flip a fair binary coin; 
if the coin is $0$ (resp.\ $1$) we update simultaneously all $\phi^{(i)}_x$, $i=h_1,\dots,h_2$, for $x$ even (resp.\ odd)
with a sample from the uniform distribution conditioned on the 
current values $\{ \phi^{(i)}_x, \;i=h_1,\dots,h_2, x \;{\rm odd }\}$ (resp.\ even).
With a slight abuse of notation we call again 
$P_t(\cdot,\cdot)$ the kernel of the Markov process on lattice path configurations, and its invariant measure is of course the uniform measure $\r$.
Moreover, we write again (cf. \eqref{cG}) $$\cG=\frac12\,\big( \r[\,\cdot\tc {\rm odd}\,] - 
\mathbb I
\big)
+  \frac12\,\big(\r[\,\cdot\tc {\rm even}\,] - 
\mathbb I
\big)\,,$$ for the generator of $P_t$, where $\mathbb I$ denotes the identity operator.  

Next, we turn to Wilson's coupling argument. Define $\Phi:\wt\O\mapsto\bbR$ by
$$
\Phi(\phi) = \sum_{x=1}^{2D-1} g(x)\sum_{j=h_1}^{h_2} \phi^{(j)}_x\,,\;\;\;g(x) = \sin\left(\frac{\pi x}{2D}\right)\,.
$$
To compute the action of $\cG$ on $\Phi$, fix some even $0<x<2D$, and observe that if the constraints 
$\phi^{+,(j)}_x\leq \phi^{(j)}_x\leq \phi^{-,(j)}_x$ were absent, one would have 
\begin{equation}\label{nobound}
\r\Big[\sum_{j=h_1}^{h_2} \phi^{(j)}_x\tc  {\rm odd}\,\Big] = \frac12\sum_{j=h_1}^{h_2} (\phi^{(j)}_{x-1}+ \phi^{(j)}_{x+1})\,.
\end{equation}
Note that the sum is important in \eqref{nobound} since the identity
has no reason to hold for a single $j$.
Now, the constraint \eqref{eq:101} is felt at $x$ in the path $j$ iff either 
$A_-(j,x):=\{\phi^{(j)}_{x}= \phi^{-,(j)}_{x}< \phi^{(j)}_{x-1}= \phi^{(j)}_{x+1}\}$ or $A_+(j,x):=\{
\phi^{(j)}_{x}= \phi^{+,(j)}_{x}> \phi^{(j)}_{x-1}= \phi^{(j)}_{x+1}\}$.
In the first case we have to compensate \eqref{nobound} with  $-1$ since 
$\phi^{(j)}_{x}$ cannot move and $\phi^{(j)}_{x}=\frac12(\phi^{(j)}_{x-1}+ \phi^{(j)}_{x+1})-1$. In the second case we have to compensate \eqref{nobound} with  $+1$ since 
$\phi^{(j)}_{x}$ cannot move and $\phi^{(j)}_{x}=\frac12(\phi^{(j)}_{x-1}+ \phi^{(j)}_{x+1})+1$.
In conclusion, 
\begin{equation}\label{withbound}
\r\Big[\sum_{j=h_1}^{h_2} \phi^{(j)}_x\tc  {\rm odd}\,\Big] = \frac12\sum_{j=h_1}^{h_2} (\phi^{(j)}_{x-1}+ \phi^{(j)}_{x+1})
+\sum_{j=h_1}^{h_2} (1_{A_+(j,x)}(\phi)-1_{A_-(j,x)}(\phi))\,.
\end{equation}
Clearly, for any $x$ even one has $\r\Big[\sum_{j=h_1}^{h_2} \phi^{(j)}_x\tc  {\rm even}\,\Big]= \sum_{j=h_1}^{h_2} \phi^{(j)}_x$.
Moreover, exactly the same equations hold when $x$ is odd, provided we change the conditioning from odd to even.
Therefore, using the notation $(\Delta\varphi)(x)=\frac12(\varphi(x-1)+\varphi(x+1))-\varphi(x)$ for the discrete Laplacian, 
\eqref{withbound} together with its analogue for odd $x$ imply that
\begin{equation*}
[\cG\,\Phi](\phi) = \frac12 \sum_{x=1}^{2D-1} g(x)\sum_{j=h_1}^{h_2} (\Delta\phi^{(j)})(x)
+ \sum_{x=1}^{2D-1} g(x)\sum_{j=h_1}^{h_2} (1_{A_+(j,x)}(\phi)-1_{A_-(j,x)}(\phi)).
\end{equation*}
Summing by parts and using $\Delta g = -\kappa_D\,g$, where $\kappa_D=1-\cos(\pi/(2D))$, one has
\begin{equation}\label{genero}
[\cG\,\Phi](\phi) = -\frac{\kappa_D}2\, \Phi(\phi)
+ \sum_{x=1}^{2D-1} g(x)\sum_{j=h_1}^{h_2} (1_{A_+(j,x)}(\phi)-1_{A_-(j,x)}(\phi)).
\end{equation}
Next, let $\phi^\xi(t)$ denote the state of the Markov process at time $t$ with initial condition $\xi\in\wt\O$ at time $0$.
When $\xi=\phi^+$ (minimal state in terms of lattice paths) or $\xi=\phi^-$
 (maximal state), we simply write $\phi^+(t)$ or $\phi^-(t)$.
Define $u(t)= \bbE[\Phi(\phi^-(t)) - \Phi(\phi^+(t))]\ge0$, where $\bbE$ denotes expectation with respect to the global monotone coupling of the 
lattice path Markov process with kernel $P_t$ (cf. Section \ref{sec:monotone} and \cite{Wilson}). 
From \eqref{expG} we infer
\begin{equation}\label{derivata}
\frac{d}{dt}\,u(t) = -\frac{\kappa_D}2\, u(t) + \psi(t)\,,
\end{equation}
with 
$$
\psi(t) = \sum_{x=1}^{2D-1} g(x)\sum_{j=1}^h \bbE\left[(1_{A_-(j,x)}(\phi^+(t))-1_{A_-(j,x)}(\phi^-(t)))
+ (1_{A_+(j,x)}(\phi^-(t))-1_{A_+(j,x)}(\phi^+(t)))\right]\,.
$$
By monotonicity of the coupling it is immediate to see that $\psi(t)\leq 0$. 
Therefore \eqref{derivata} implies $u(t)\leq u(0)\,e^{-\frac{\kappa_D}2\,t}$. 
Note that $u(0)$ can be upper bounded by the volume enclosed between the minimal and maximal plane partition, i.e.\
$u(0)\leq |V^+\setminus V^-| \leq D^2H\le D^3$ (recall that in
Theorem
\ref{th:wilson} we are assuming $H\le D$). It follows that $u(t)\leq D^3\,e^{-\frac{\kappa_D}2\,t}$. 

To finish the proof it suffices to observe that by monotonicity $\Phi(\phi^-(t)) - \Phi(\phi^+(t))\geq 0$ and 
$\phi^+(t)\neq\phi^-(t)$ iff $\Phi(\phi^-(t)) - \Phi(\phi^+(t))\geq 2\sin(\pi/2D)$, so that by Markov's inequality
$$ 
\bbP\left(\phi^+(t)\neq\phi^-(t)\right) \leq \frac{u(t)}{2\sin(\pi/2D)} \leq \frac{D^3}{2\sin(\pi/2D)} \,e^{-\frac{\kappa_D}2\,t}\,.
$$
We can now 
bound the total variation distance by the probability of no coupling
up to  time $t$, and using monotonicity and the bound above this gives, for any initial state $v$
$$
\|P_t(v,\cdot) - \r\|\leq \frac{D^3}{2\sin(\pi/2D)} \,e^{-\frac{\kappa_D}2\,t}\,,
$$
which is easily seen to imply the desired estimate.

\hfill$\stackrel{\tiny\mbox{Lemma \ref{le:tmixbo}}}{\qed}$

\subsection{Proof of Theorem \ref{th:wilson}}\label{sec:compare}
The proof of Theorem \ref{th:wilson} uses the bound of Lemma
\ref{le:tmixbo} together with a comparison argument that allows us to
translate the mixing time upper bound of the ``column dynamics'' to an upper bound
for the mixing time of the dynamics $\nu^{V_0}_t$ defined in Section
\ref{sec:wilson}.
In the plane partition language, the dynamics  $V^{V_0}_t$ will be called
$v_t^{v_0}$, with law $\nu_t^{v_0}$ and equilibrium distribution as usual denoted by $\r$: it has
initial condition $v_0\in\Omega$ ($v_0$ being the plane partition
corresponding to the monotone set $V_0$), to each column $(x,y)$
is associated an independent Poisson clock with parameter $1$ and the
evolution proceeds by local updates. By this we mean that when the
clock of the column $(x,y)$ rings, one replaces $v_{(x,y)}$ by
$\max\{v_{(x,y)}-1,v_{(x+1,y)},v_{(x,y+1)},v^-_{(x,y)}\}$  (with
probability $1/2$) or by 
 $\min\{v_{(x,y)}+1,v_{(x-1,y)},v_{(x,y-1)},v^+_{(x,y)}\}$
(with probability $1/2$). 
In other words, the chosen column performs a
simple symmetric random walk step, except that jumps which violate the plane partition constraints $v_{(x,y)}\geq \max\{v_{(x+1,y)},v_{(x,y+1)}\}$,  $v_{(x,y)}\leq \min\{v_{(x-1,y)},v_{(x,y-1)}\}$,  or the 
overall constraint $v^-_{(x,y)}\leq v_{(x,y)}\leq v^+_{(x,y)}$, are
rejected. 
We need to prove that the mixing time of this chain is $O(D^2H^2(\log D)^2)$. 

\smallskip

We start with a simple observation that allows us to reduce to the case of 
maximal ($v_0=v^+$) and minimal ($v_0=v^-$) initial conditions. 
\begin{lemma}
\label{le:monopm} For any $t>0$ and any $v_0,v_0'\in\O$:
$$\|\nu_t^{v_0} - \nu_t^{v_0'}\|\leq D^3\,\|\nu_t^{v^+} - \nu_t^{v^-}\|\,.$$
\end{lemma}
\proof 
Let $\bbP$ denote a monotone coupling of $\nu_t^{v_0},
\nu_t^{v_0'}$. Then, using the fact that each column $(x,y)$ has a
minimal height $v^-_{(x,y)}$ and a maximal height $v^+_{(x,y)}$ such
that $v^+_{(x,y)} -v^-_{(x,y)} \leq H\le D$ we have 
\begin{align*}
\|\nu_t^{v_0} - \nu_t^{v_0'}\|&\leq \bbP(v_t^{v_0}\neq v_t^{v_0'})
\leq \bbP(v_t^{v^+}\neq v_t^{v^-})\\ & \leq 
\sum_{(x,y)}\sum_{h= v^-_{(x,y)}}^{v^+_{(x,y)}}
[\bbP((v_t^{v^+})_{(x,y)}> h) - \bbP((v_t^{v^-})_{(x,y)}> h)]
\leq 
D^3\,\|\nu_t^{v^+} - \nu_t^{v^-}\|\,.
\end{align*}
\hfill$\stackrel{\tiny\mbox{Lemma \ref{le:monopm}}}{\qed}$

Thanks to Lemma \ref{le:monopm}, to prove Theorem \ref{th:wilson} it is sufficient to show that 
\begin{equation}
\label{mecoides}
\|\nu_t^{v^\pm} - \r\|\leq \frac1{4e D^3}\,,
\end{equation}
for some $ t=O(D^2H^2(\log D)^2)$.
Let us prove the statement for $\nu_t^{v^+}$, the argument for $\nu_t^{v^-}$ being identical. 
Consider i.i.d.\ Bernoulli($\frac12$) random variables $\z=(\z_1,\z_2,\dots)$
and let $T_\ell=\ell\,T$, with $T=c_1\,H^2\log D$ and $\ell\in\NN\cup\{0\}$, denote a partition of the time axis.
Furthermore, consider i.i.d.\ Poisson processes with parameter $1$ at each column $(x,y)$, and let 
$\{S_i(x,y)\}_{i,(x,y)}$ denote the corresponding collection of arrival times. 
Call $\{\wt S_i(x,y)\}_{i,(x,y)}$ the collection of arrival times obtained from $\{S_i(x,y)\}_{i,(x,y)}$
by deleting (or ``censoring'') all arrivals
$S_i(x,y)$ such that, for some $j\in\bbN$, $T_{j-1}\leq  S_i(x,y) < T_{j}$ and $(x,y)$ has the opposite parity as $\z_{j}$ (e.g.\ $(x,y)$ is odd and $\z_j=0$). 

By construction, if we start from the configuration $v^+$ at time $0$
and perform local updates using all the marks $\{S_i(x,y)\}_{i,(x,y)}$ up to time $t$ we obtain the distribution $\nu_t^{v^+} $. Let us call 
$\wt\nu_{\z,t}^{\,v^+}$ the distribution of the ``censored'' dynamics obtained in the same way but only using the marks 
$\{\wt S_i(x,y) \}_{i,(x,y)}$, for a fixed Bernoulli sequence $\z$. From the ``censoring inequality'' of Peres
and Winkler \cite{notePeres}, \cite[Th. 2.5]{MT} it follows that $\nu_t^{v^+}$  is stochastically dominated by 
$\wt\nu_{\z,t'}^{\,v^+}$, for any $t'\leq t$. Moreover, setting $
\wt\nu_t^{\,v^+} :=\bbE_\z\wt\nu_{\z,t}^{\,v^+}$, where $\bbE_\z$
denotes expectation over the random sequence $\z$, by linearity of the
expectation one sees that 
\begin{eqnarray}
  \label{eq:130}
  \nu_t^{v^+}\preceq \wt\nu_{t'}^{\,v^+}\mbox{\;\; for any \;}t'\leq t.
\end{eqnarray}

Let us fix $t_0=c_2\,s\,T$, where $s=c_3\,D^2\log D$ and $c_2,c_3$ are constants to be taken sufficiently large. Let $\{N(s), s\geq 0\}$ denote a Poisson process with parameter $1$, and write  $\bbP$ and $\bbE$ for the associated probability and expectation.
Monotonicity implies that the event 
$A=\{v\in\Omega\,:\;\nu_{t_0}^{v^+}(v)>\r(v)\}$ is increasing \cite[Th. 2.5]{MT}.
Therefore, the censoring inequality \eqref{eq:130} gives
\begin{eqnarray}
\label{pw2}
\|\nu_{t_0}^{v^+} - \r\|&=& \nu_{t_0}^{v^+}(A)-\r(A)
\leq \bbE(\wt \nu_{N(s)T}^{\,v^+} (A)|N(s)T<t_0)-\r(A)\\\nonumber
&
\leq &
\frac{\bbE\wt \nu_{N(s)T}^{\,v^+} (A)-\r(A)}{\bbP(N(s)T<t_0)}+\rho(A)
\frac{\bbP(N(s)T\ge t_0)}{\bbP(N(s)T<t_0)}\\\nonumber &\le&
\frac{\| \bbE\wt \nu_{N(s)T}^{\,v^+}-
\r\|}{1-\bbP(N(s)T\ge t_0)}+
\frac{\bbP(N(s)T\ge t_0)}{\bbP(N(s)T<t_0)}.
\end{eqnarray}
Taking $c_2$ large in the definition of $t_0$ above and using
standard estimates for the Poisson random variable we can make $\bbP(N(s)T\ge t_0)$
smaller than $D^{-4}$. Therefore, we see that thanks to  
Lemma \ref{le:tmixbo} and \eqref{pw2}, if $c_3$ in the definition of
$s$ is chosen large the claim \eqref{mecoides}
is a consequence of
\begin{equation}
\label{couple}
\|\bbE\,\wt \nu_{N(s)T}^{\,v^+} - P_s(v^+,\cdot)\|\leq \frac1{5e D^3}\,,
\end{equation}
where $P_s(\cdot,\cdot)$ is the kernel defined in Section
\ref{sec:columns}, which involves full column equilibrations (cf. \eqref{expG}).
To prove \eqref{couple} we need to compare column equilibration moves with local updates.
Let us 
use the notation $\r_0(v,\cdot)$ and  $\r_1(v,\cdot)$ for the probability kernels associated to 
$\r[\,\cdot\tc {\rm odd}\,](v)$ and $\r[\,\cdot\tc {\rm even}\,](v)$
respectively, 
see \eqref{cG}. That is, for any $f:\O\mapsto\bbR$ one has for
instance $\r[f\tc {\rm odd}](v)=\sum_{v'\in\O}\r_0(v,v')f(v')$. 
Define $P^v_{\z,n} = [\r_{\z_1}\cdots\r_{\z_n}](v,\cdot)$, and note that 
$Q^v_n:=2^{-n}\sum_{\z\in\{0,1\}^n}P^v_{\z,n}$ is nothing else but a discrete time version of the kernel $P_s(v,\cdot)$, i.e.\  $P_s(v,\cdot)=\bbE Q^v_{N(s)}$ where $\bbE$ and $N(\cdot)$ are as above. 
Since $s=c_3\,D^2\log D$, excluding an event of probability $O(D^{-p})$ for some large $p>0$ we can assume that 
$N(s)=n$ for some $n \leq 2s$. Recall that 
$\wt\nu_{\z,t}^{\,v^+}$
denotes the law $\wt \nu_{t}^{\,v^+}$
conditioned on the binary sequence $\z$. 
The previous remarks imply that it will be sufficient to prove the upper bound 
\begin{equation}
\label{rwbo2}
\sup_{\z\in\{0,1\}^n}\|\wt \nu_{\z,nT}^{\,v^+} - P^{v^+}_{\z,n}\|\leq
\frac1{6e D^3}\,,\quad \; n\leq 2s\,.
\end{equation}
Observe that, for the censored dynamics $\tilde \nu^{v^+}_{\zeta,t}$, in the time interval $T_{j-1}\leq t <T_{j}$, all columns with the same parity as $\z_j$ 
are independently updated by local moves, i.e.\ 
on columns of that parity we have independent continuous-time simple symmetric random
walks in segments (determined by the columns of opposite parity) of length bounded by $H$. 
It is standard
that for some constant $c>0$ the mixing time on each column is bounded by $c\,H^2$ and therefore after a time $T$ we have the bound, for any $\z_1\in\{0,1\}$, uniformly in the starting configuration  $v\in\O$: 
\begin{equation}
\label{rwbo}
\|\wt\nu_{\z,T}^{\,v}(\cdot)- \r_{\z_1}(v,\cdot) \|\leq c\, D^2\,\exp{\Big(-\frac{T}{c\,H^2}\Big)}\,.
\end{equation}
For general $n\in\bbN$, by recursive coupling of the distributions involved, using 
\eqref{rwbo} at each step, one has 
\begin{equation}
\label{rwbo1}
\|\wt\nu_{\z,nT}^{\,v^+}(\cdot)- [\r_{\z_1}\cdots\r_{\z_n}](v^+,\cdot) \|
\leq
1-\Big(1-c\,D^2\,\exp{\Big(-\frac{T}{c\,H^2}\Big)}\Big)^n\,.
\end{equation}
Since $T=c_1\,H^2\log D$ and $n\le 2s=O(D^2\log D)$, we see that the desired estimate
\eqref{rwbo2} follows for a suitable choice of $c_1$. 

\hfill$\stackrel{\tiny\mbox{Theorem \ref{th:wilson}}}{\qed}$

\section{The mixing time in dimension $d=3$ (lower bound)}
\label{sec:LBMT}
Here we prove the bound
\begin{eqnarray}
  \label{eq:118}
  \bP(\tau_+\le L^2/(c\log L))\le c/L 
\end{eqnarray}
for a
suitable constant $c$.

Let $C_L$ be the cube $\{1,\ldots,L/2\}^3$ (we assume for definiteness that $L$ is
even) with boundary conditions $\eta_x=+$ if $x\in\partial C_L$ with $\min(x^{(1)},x^{(2)},x^{(3)})=0$ and $\eta_x=-$ otherwise. In other words, the boundary conditions $\eta$ are ``$+$'' at the three faces of $\partial C_L$ which meet at the origin of $\ZZ^3$, and ``$-$'' at the other three.
Denote by $\{s^-(t)\}_t$ the zero-temperature  Glauber evolution started from the ``$-$'' configuration
(in order not to confuse it with the evolution $\sigma^-(t)$ in $\Lambda_L$) and by $\bP^\eta_{C_L}$ its law. One has
\begin{prop}
\label{prop:spigolo}
  There exists $c>0$ such that 
  \begin{eqnarray}
    \label{eq:3}
    \bP^\eta_{C_L}\left(\mbox{there exists\;} t< \frac{L^2}{c\log L} \mbox{\; such that\;}
s^-_{(1,L/2,1)}(t)=+\right)\le \frac cL.
  \end{eqnarray}
\end{prop}

{\sl Proof of Eq. \eqref{eq:118}, assuming Proposition \ref{prop:spigolo}}.
Since the set $\{x\in C_L: s^-_x(t)=+\}$ is a monotone set
(cf. Definition \ref{def:monset}) at all times, 
the event that $s^-_{(1,L/2,1)}(t)=-$
implies the event $s_y^-(t)=-$ for all $y\in C_L$ such that $y^{(2)}=L/2$.
Therefore, \eqref{eq:3} implies (using also symmetry among the three coordinate axes)
\begin{eqnarray}
  \label{eq:131}
      \bP^\eta_{C_L}\left(\exists\; t<\frac{ L^2}{c\log L}, x\in C_L \mbox{\;with \;}
\max(x^{(1)},x^{(2)},x^{(3)})=L/2 \mbox{\; such that\;}
s^-_x(t)=+\right)\le \frac {3c}L.
\end{eqnarray}
The cube $\Lambda_L=\{1,\ldots,L\}^3$ can be seen as the union of eight disjoint
sub-cubes  $C^{(i)}$ of side $L/2$, with $C^{(1)}=C_{L/2}$ while $C^{(i)},i=2,\ldots,8$ are suitable translations of $C_L$. Let as usual $\bP$ denote the law of the Glauber evolution
inside $\Lambda_L$, with ``$+$'' b.c., started from ``$-$'' and let $\bP'$ be 
law of the evolution, again started from ``$-$'', where the spin configuration inside each
$C^{(i)}$ evolves independently for different $i$, with b.c. given by $\eta_x=+$ for
$x\in \partial C^{(i)}\cap \partial \Lambda_L$ and $\eta_x=-$ for 
$x\in \partial C^{(i)}\cap \Lambda_L$. It is clear that, until the
random time
\begin{eqnarray}
  \label{eq:132}
  t_1:=\inf\{t>0: \exists i\in\{1,\ldots,8\}, x\in \partial C^{(i)}\cap \Lambda_L
\mbox{\;such that\;} \sigma^-_x(t)=+\}\,,
\end{eqnarray}
the two evolutions can be perfectly coupled, and that $t_1<\tau_+$.
On the other hand, thanks to \eqref{eq:131} and to the symmetry among the 
various cubes $C^{(i)}$ (up to suitable translations of the origin and reflections of the coordinate axes) one sees that 
\begin{eqnarray}
  \label{eq:133}
\bP\left(\tau_+< \frac{ L^2}{c\log L}\right) \le \bP\left(t_1<\frac{ L^2}{c\log L}\right)=
\bP'\left(t_1<\frac{ L^2}{c\log L}\right)\le 8\times\frac{3c}L.
\end{eqnarray}

\hfill$\stackrel{\tiny\mbox{Eq. \eqref{eq:118}, given Prop. \ref{prop:spigolo}}}{\qed}$

The main ingredient in the proof of Proposition \ref{prop:spigolo} is
a ``column dynamics'' $\tilde s^-(t)$ (analogous to the one used in Section
\ref{sec:columns}) defined as follows. Start from  the ``all
minus'' configuration in $C_L$, $\tilde s^-(0)\equiv -$. Assign  to each $x\in
\{1,\ldots,L/2\}^2$ an i.i.d. Poisson clock of rate $2$; when the clock labeled
$x=(x^{(1)},x^{(2)})$ rings, we assign  a new value to the collection of spins at sites $z\in C_L$ with
$(z^{(1)},z^{(2)})=(x^{(1)},x^{(2)})$, by sampling it from the
equilibrium distribution conditioned on the present value of all the other spins. In other words, we set
to equilibrium the column of horizontal coordinates $(x^{(1)},x^{(2)})$, conditionally on the value of the
neighboring columns.  For every $t$ we have the stochastic domination
\begin{eqnarray}
\label{eq:134}
s^-(t)\preceq \tilde s^-(t).
\end{eqnarray}
Indeed, for $k\in\NN$ let $s^{-,k}(t)$ be the following dynamics. Set
$s^{-,k}(0)\equiv-$ and, when the Poisson clock of the column labeled $x$
rings, repeat $k$ times the following procedure:
\begin{itemize}
\item flip a fair binary coin;

\item if the coin gives ``head'', then make a heat bath update at each
  of the
  sites of the column $x$ with vertical coordinate belonging to $2\NN$, one by one, starting from the bottom site;

\item if instead the coin  gives ``tail'', then make a heat bath
  update at each of the
  sites of the column $x$ with vertical coordinate belonging to $2\NN+1$, one by one, starting from the bottom site.

\end{itemize}
Here, ``making a heat-bath update'' at a site $z$ means updating
$\sigma_z$ according to the equilibrium conditioned on the value of
the spins outside $z$.
It is immediate to realize that the process $\{s^{-,1}(t)\}_t$ has the
same law as $\{s^-(t)\}_t$, and that the law of  $\{s^{-,k}(t)\}_t$
converges to that of $\{\tilde s^-(t)\}_t$ for $k\to\infty$. Also, the
stochastic domination $s^{-,k}(t)\preceq s^{-,k+1}(t)$ is an immediate
consequence of the Peres-Winkler censoring inequality \cite{notePeres}, \cite[Th. 2.5]{MT}.

Call $\tilde\bP$ the law of the column dynamics $\{\tilde s^-(t)\}_t$. One has
\begin{prop}
  \label{prop:coldyn} Fix $c>0$. For $L$ sufficiently large and 
$t<L^2/(c\log L)$ one has
\begin{eqnarray}
  \label{eq:135}
  \tilde\bP(\tilde s^-_{(1,L/2,1)}(t)=+)\le c' L^2 \,e^{-L^2/(64 t)}
\end{eqnarray}
for some finite constant $c'$.
\end{prop}

{\sl Proof of Proposition \ref{prop:spigolo}, given Proposition
  \ref{prop:coldyn}}.
Let
\begin{eqnarray}
  \label{eq:136}
  H:=\int_0^{L^2/(c\log L)} {\bf 1}_{\{s^-_{(1,L/2,1)}(t)=+\}} dt
\end{eqnarray}
so that, thanks to Proposition
  \ref{prop:coldyn} and to the stochastic domination \eqref{eq:134}
  one has
  \begin{eqnarray}
    \label{eq:138}
    \bE^\eta_{C_L}(H)\le c' \,  L^{4-(c/64)}.
  \end{eqnarray}
Note that the desired bound  \eqref{eq:3} can be rewritten as
$\bP^\eta_{C_L}(H>0)<c/L$.
One has
\begin{eqnarray}
  \label{eq:139}
  \bP^\eta_{C_L}(H>0)&=&\bP^\eta_{C_L}(H\ge L^{-c/128})+\bP^\eta_{C_L}(0<H<L^{-c/128})\\\nonumber
&\le&
c'\,L^{4-(c/128)}+\bP^\eta_{C_L}(0<H<L^{-c/128})
\end{eqnarray}
where in the first term we applied Markov's inequality.
Choosing $c$ sufficiently large, one can make both terms in the last
expression smaller than $c/L$. Indeed, in order that $0<H<L^{-c/128}$,
there must be two times $s,t$ with $0<t-s<L^{-c/128}$ such that 
the Poisson clock associated to the site $(1,L/2,1)$ rings both at
times $s$ and $t$. Via a simple union bound, and using the exponential
form of the law of the intervals between two successive rings, one
sees that the probability of such event is $O(L^{2-c/128})$.

\hfill$\stackrel{\tiny\mbox{Prop. \ref{prop:spigolo}, given Prop. \ref{prop:coldyn}}}{\qed}$

{\em Proof of Proposition \ref{prop:coldyn}}.
Since the set $\{x\in C_L:\tilde s^-_x(t)=+\}$
is a monotone subset of $C_L$ at all times, one can identify it
(recall Section \ref{sec:columns} and Fig.~\ref{fig:polymers}) with the set of $L/2$ paths
$\{\phi^{(j)}(t)\}_{j=1,\ldots,L/2}$ of length $L+1$, where 
the $j^{th}$ path is the collection
$\{\phi^{(j)}_x(t)\}_{x=0,\ldots,L}$ and the following relations are
satisfied:
\begin{eqnarray}
  \label{eq:102bis}
\phi^{(j)}_0(t)=\phi^{(j)}_{L}(t)=j\,,\;\;
\phi^{(j)}_{x+1}(t) -
\phi^{(j)}_{x}(t)\in\{-1,+1\}\,,\;\;\phi^{(j)}_x(t)< \phi^{(j+1)}_x(t)\, .
\end{eqnarray}
At time $t=0$ one has $\{x\in C_L:\tilde s^-_x(0)=+\}=\emptyset$
and therefore 
$\phi^{(j)}_x(0)=j+ x$
if $x\le L/2$ and $\phi^{(j)}_x(0)=j+L-x $ if $x\ge L/2$.
 Defining 
 \begin{eqnarray}
   \label{eq:140}
h_x(t):=\frac2L\sum_{j=1}^{L/2}\left[\phi^{(j)}_x(t)-j\right],
 \end{eqnarray}
$u(t,x):=\tilde \bE(h_x(t/2))$
and reasoning like in Section \ref{sec:columns} (see also \cite[Sec. 5]{Wilson}), one
sees that $u(t,x)$ satisfies the
discrete heat equation
\begin{eqnarray}
  \label{eq:141}
\left\{
  \begin{array}{ll}
    \frac{d}{dt}u(t,x)=(\Delta
    u)(t,x):=\frac{u(t,x+1)+u(t,x-1)-2u(t,x)}2,\;\; x=1,\ldots,L-1
\\
u(0,x)=x\,{\bf 1}_{x\le L/2}+(L-x)\,{\bf 1}_{x>L/2} 
  \end{array}
\right.
\end{eqnarray}
with Dirichlet boundary conditions $u(t,0)=u(t,L)=0$.
On the other hand, from the representation of monotone sets as collections of paths, one sees that 
\begin{eqnarray}
  \label{eq:142}
  \tilde \bP(\tilde s^-_{(1,L/2,1)}(t)=+)
&=&
\tilde \bE\left({\bf 1}_{\{\phi^{(1)}_{L-1}(t)-1=-1\}}\right)
=\tilde
\bE\left(\frac{1-(\phi^{(1)}_{L-1}(t)-1)}2\right)\\\nonumber
&\le&
\sum_{j=1}^{L/2} \tilde
\bE\left(\frac{1-(\phi^{(j)}_{L-1}(t)-j)}2\right)
=
\frac L2\frac{1-u(2t,L-1)}2.
\end{eqnarray}
Therefore, it is sufficient to prove the heat-equation estimate
\begin{eqnarray}
  \label{eq:143}
  1-u(t,L-1)\le c' L \,e^{-L^2/(32 t)},
\end{eqnarray}
 uniformly for $t<2L^2/(c\log L)$ and $L$ large. 
While \eqref{eq:143} can be obtained directly via Fourier analysis, we give a simple and more probabilistic argument.
First it is well known (cf. for instance \cite{FSS}) that, for $x=1,\ldots,L$, the quantity $[1+u(t,x)-u(t,x-1)]/2$
 coincides with the probability that there is a particle at site $x$
 at time $t$,
 for a symmetric simple exclusion process on $\{1,\ldots,L\}$ with
 initial condition at time zero such that sites $x\le L/2$ are
 occupied by a particle, while sites $\{L/2+1,\ldots,L\}$ are
 empty (each particle attempts with rate one to jump to one of its two
 neighboring sites with equal probability $1/2$ and the jump is
 rejected if either the site is already
 occupied or if it lies outside $\{1,\ldots,L\}$). In particular (recall that $u(t,L)=0$) one has that 
$[ 1-u(t,L-1)]/2$ is the probability that there is a particle at $L$
at time $t$. By duality (cf. \cite[Section II.3]{cf:Liggett}), this equals
the probability that a continuous-time simple random walk of rate $1$
on $\{1,\ldots,L\}$, started from site $L$, is in $\{1,\ldots,L/2\}$
at time $t$. The bound \eqref{eq:143} then follows from standard
random walk estimates: if $P^{a,b}_x$ is the law of the
continuous-time simple random walk $X_t$ on $\{n\in \NN: a-1< n< b+1\}$ started from
$x$, one has
\begin{eqnarray}
  \label{eq:144}
  P^{1,L}_L\bigl(X_t\le \frac L2\big)\le   P^{-\infty,L}_L\big(X_t\le \frac L2\big)\le 
P^{-\infty,L}_{3L/4}\big(X_t\le\frac L2\big)\le
P^{-\infty,+\infty}_0\big(\exists s<t: |X_s|\ge\frac L4\big) 
\end{eqnarray}
and the latter expression is easily seen (e.g. using the local central
limit theorem) to be upper bounded by the
r.h.s. of \eqref{eq:143}.

\hfill$\stackrel{\tiny\mbox{Prop. \ref{prop:coldyn}}}{\qed}$

\section{Proof of Theorem \ref{th:d2}}
\label{sec:TM2D}
As discussed in Remark 
\ref{rem:d2tmix}, we only have to prove that, for the $\beta=+\infty$ dynamics,
\begin{eqnarray}
  \label{eq:provanda}
\bP\left(\tau_+<c_0 L^2\right)\le \exp(-\gamma L) 
\end{eqnarray}
for suitable positive constants
$c_0,\gamma$.
Also, thanks to  monotonicity, it is enough to prove this fact for the
dynamics $\sigma^-(t)$  in the domain 
\begin{eqnarray}
  \label{eq:88}
\tilde \Lambda:=\Lambda_L\cap 
\{x\in \ZZ^2:|x^{(1)}|+|x^{(2)}|\le L+1\},  
\end{eqnarray}
 with ``$+$'' boundary conditions on $\partial \tilde \Lambda$.
The advantage of looking at the dynamics in $\tilde\Lambda$ instead of $\Lambda_L$ will
be apparent in the proof of Theorem \ref{th:restogood}.

\smallskip

We need a certain number of geometric definitions:
\begin{definition}
\label{def:Gset}
  Given $\sigma\in\Omega_{\tilde \Lambda}$, let
  \begin{enumerate}[(a)]

\item $$
M(\sigma):=\{x:\;\sigma_x=-1 \};$$

  \item $$\Gamma(\sigma):=\cup_{x\in M(\sigma)}B_x,$$
where $B_x$ is the unit square of side $1$ centered at $x$, with sides
parallel to the coordinate axes. The boundary of $\Gamma(\sigma)$
is denoted as $\partial\Gamma(\sigma)$  and its geometric length as $|\partial\Gamma(\sigma)|$;
\item  for $i=1,2$ $$u_{max}^{(i)}(\sigma):=\max\{x^{(i)}:x=(x^{(1)},x^{(2)})\in\partial\Gamma(\sigma)\} $$ and
similarly
$$u^{(i)}_{min}(\sigma):=\min\{x^{(i)}:x=(x^{(1)},x^{(2)})\in\partial\Gamma(\sigma)\}; $$
\item $p(\sigma)$ be the configuration in $\Omega_{\tilde \Lambda}$
obtained by flipping every ``$-$'' spin in $\tilde\Lambda$ such that
a strict majority of its nearest neighbors are ``$+$'', and repeating the
same operation as long as such a site exists. We call $p(\cdot)$ the
``majority transformation'';
\item $D:=\{x\in \ZZ^2:|x^{(1)}|+|x^{(2)}|\le (9/10)L\}\subseteq \tilde\Lambda$;
\item $\mathcal G $ be the subset of $\Omega_{\tilde\Lambda}$ defined by
  \begin{eqnarray}
    \mathcal G&:=& \bigl\{\sigma: \partial\Gamma(\sigma)\mbox{\;
is a simple curve (i.e.\ without loops),\;} \\&&|\partial\Gamma(\sigma)|=2\sum_{i=1}^2(u_{max}^{(i)}(\sigma)-u_{min}^{(i)}(\sigma)),
D\subseteq M(\sigma) \mbox{\;and\;} p(\sigma)=\sigma\bigr\}.
  \end{eqnarray}
  \end{enumerate}
\end{definition}

Note 
that the constraint $
|\partial\Gamma(\sigma)|=2\sum_{i=1}^2(u_{max}^{(i)}(\sigma)-u_{min}^{(i)}(\sigma))$
in the definition of $\mathcal G$ is simply the requirement that
$\sigma$ minimizes the
Hamiltonian $H^+_{\tilde\Lambda}(\cdot)$, given the values $\{u_{max}^{(i)}(\sigma),u_{min}^{(i)}(\sigma)\}_{i=1,2}$.

\smallskip

Next we introduce a modified dynamics $\{\tilde \sigma^\xi(t)\}_{t\ge0}$ on
$\tilde \Lambda$ , with initial
condition $\tilde\sigma^\xi(0)=\xi$, as follows. Let
 $ \tau^\xi_D:=\inf\{t>0: M(\tilde\sigma^\xi(t))\nsupseteq
(D\cup\partial D)\},$
and set $\tilde\sigma^\xi(t):=\tilde\sigma^\xi(\tau_D^\xi)$ for $t\ge \tau_D^\xi$.
Whenever the Poisson clock
labeled $x$ rings at a time $t<\tau^\xi_D$, first refresh the current value of
the spin at $x$ according to the 
distribution $\pi(\cdot|\sigma_y=\tilde \sigma^\xi_y(t),y\ne x)$, and then apply the 
majority transformation $p(\cdot)$ of Definition \ref{def:Gset}(d)  to the configuration thus obtained.

We call $\tilde{\mathcal L}$ the generator of such dynamics and $\tilde\mu_t^\xi$ its law at time $t$.
When the initial condition is $\xi\equiv -$, it is immediate to see that, by the monotonicity of the usual Glauber
dynamics, one has
\begin{eqnarray}
  \label{eq:77}
   \left\{M(\sigma^-(t))\nsupseteq (D\cup\partial D)\right\} \Longrightarrow \left\{ M(\tilde\sigma^-(t))\nsupseteq (D\cup\partial D)\right\}
\end{eqnarray}
 for every $t>0$, so that \eqref{eq:provanda} is proven if we show that
\begin{eqnarray}
  \label{eq:78}
  \tilde\bP(\tau^-_D\le c_0 L^2)\le e^{-\gamma L}
\end{eqnarray}
for some suitable $c_0$, where $\tilde\bP$ is the law of the process $\{\tilde\si^-(t)\}_{t\ge0}$.

The advantage of the modified dynamics $\tilde\sigma^-(t)$ with
respect to the usual Glauber dynamics is that it belongs to the 
``good set'' $\mathcal G$ for all times:
\begin{theo}\label{th:restogood}
  For every $t>0$ one has that $\tilde\sigma^-(t)\in{\mathcal G}$.
Moreover, there exists a deterministic positive constant $\psi$ such that
\begin{eqnarray}
\label{eq:MM}
  |M(\tilde\sigma^-(0))|- |M(\tilde\sigma^-({\tau_D^-}))|\ge \psi L^2.
\end{eqnarray}
\end{theo}
{\sl Proof of Theorem \ref{th:restogood}.}
For this, 
we need some additional notions:
\begin{definition}
\label{def:mvw}
  Given $\sigma\in{\mathcal G}$ and $x\in\tilde\Lambda$, we say that
  $x$ is  a ``flippable site'' if two neighbors of $x$, at mutual distance
  $\sqrt 2$, are ``$+$'' and the other two
  neighbors are ``$-$''.
If $x$ is flippable, recall that $\sigma^{(x)}$ is the configuration obtained by flipping $\sigma_x$ to 
$-\sigma_x$. Then, for $x$ flippable we say that
  \begin{itemize}
  \item $x$ is a ``mountain'' if $x\in M(\sigma)$ and $\sigma^{(x)}\in {\mathcal G}$;
\item $x$ is a ``vertex'' if $x\in M(\sigma)$ and 
$p(\sigma^{(x)})\ne\sigma^{(x)}$;
  \item $x$ is a ``valley'' if $x \in \partial M(\sigma)$.

 \end{itemize}
\end{definition}

\begin{figure}[htp]
\begin{center}
\includegraphics[width=0.6\textwidth]{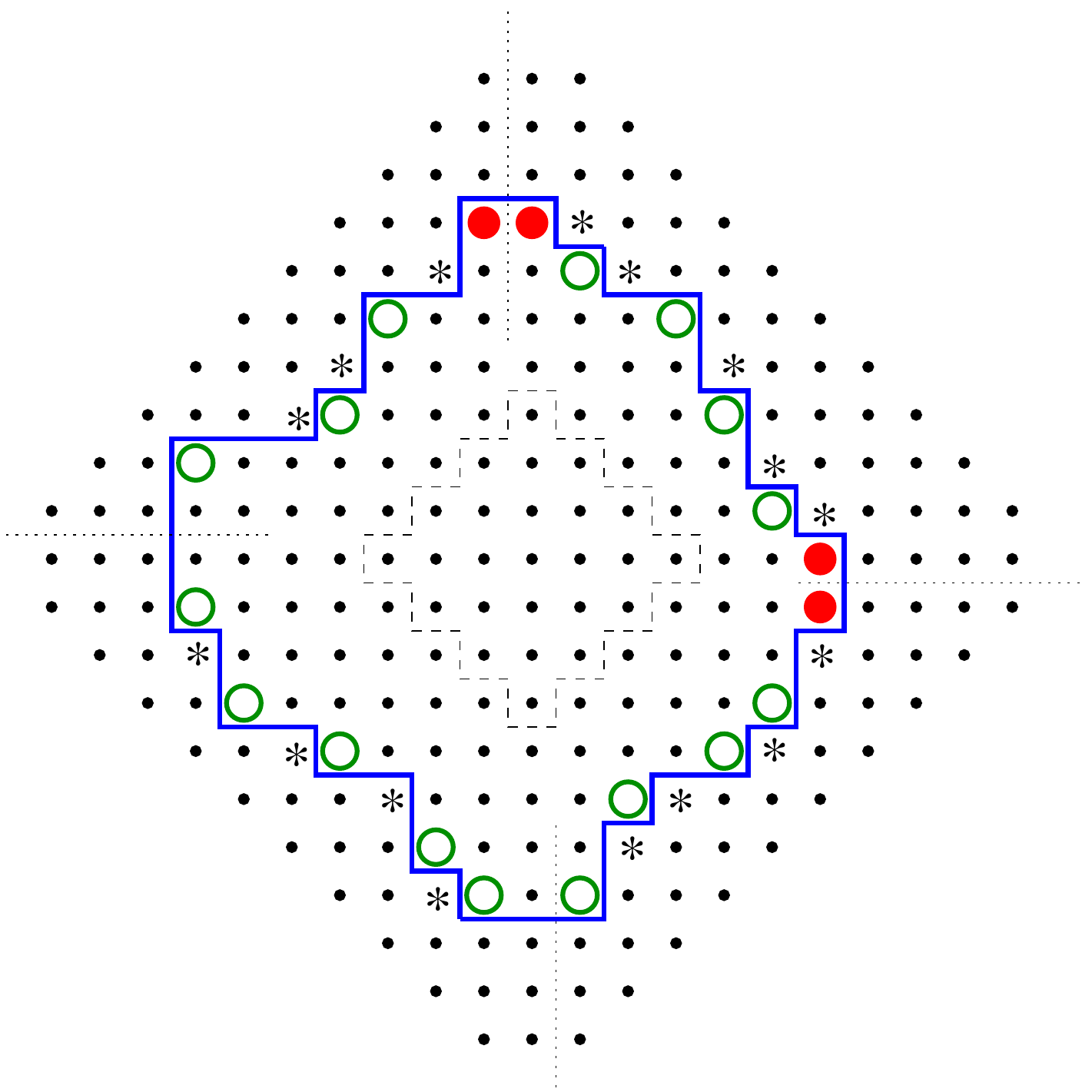}
\end{center}
\caption{The domain $\tilde\Lambda$, the sub-domain $D$ (enclosed by the dashed line; proportions are not respected) and the contour $\Gamma(\sigma)$ (thick line) of 
a configuration $\sigma\in\mathcal G$. Big full dots denote ``vertex'' sites (remark that they occur in nearest-neighboring pairs), empty dots denote
``mountains'' and $*$ denote ``valleys''. Vertices can occur only in the right-most or left-most column which intersects $M(\sigma)$, or in the highest or lowest 
row which intersects $M(\sigma)$, so there are plainly at most $8$ vertices. The curve $\Gamma(\sigma)$ can be seen as the union of four
monotone curves, which in the figure are delimited by the dotted lines. For each monotone curve, one has $m(\sigma)+w(\sigma)-v(\sigma)=1$.}
\label{fig:DL} 
\end{figure}

\begin{rem}\rm
\label{rem:vertici}
Observe that, if $x$ is a ``vertex'', then exactly one of its
neighbors, call it $y$, is also a vertex.
It is immediate to see that
$M(p(\sigma^{(x)}))=M(\sigma)\setminus\{x,y\}$, cf. Figure \ref{fig:DL}.
\end{rem}

\medskip

We prove
that $\tilde\sigma^-(t)\in\mathcal
G$ for $t\le \tau_D^-$  by induction: clearly the statement is true at
time zero, and we show that if it is true until some time $s$
then it remains true after the next update.

\begin{figure}[htp]
\begin{center}
\includegraphics[width=0.6\textwidth]{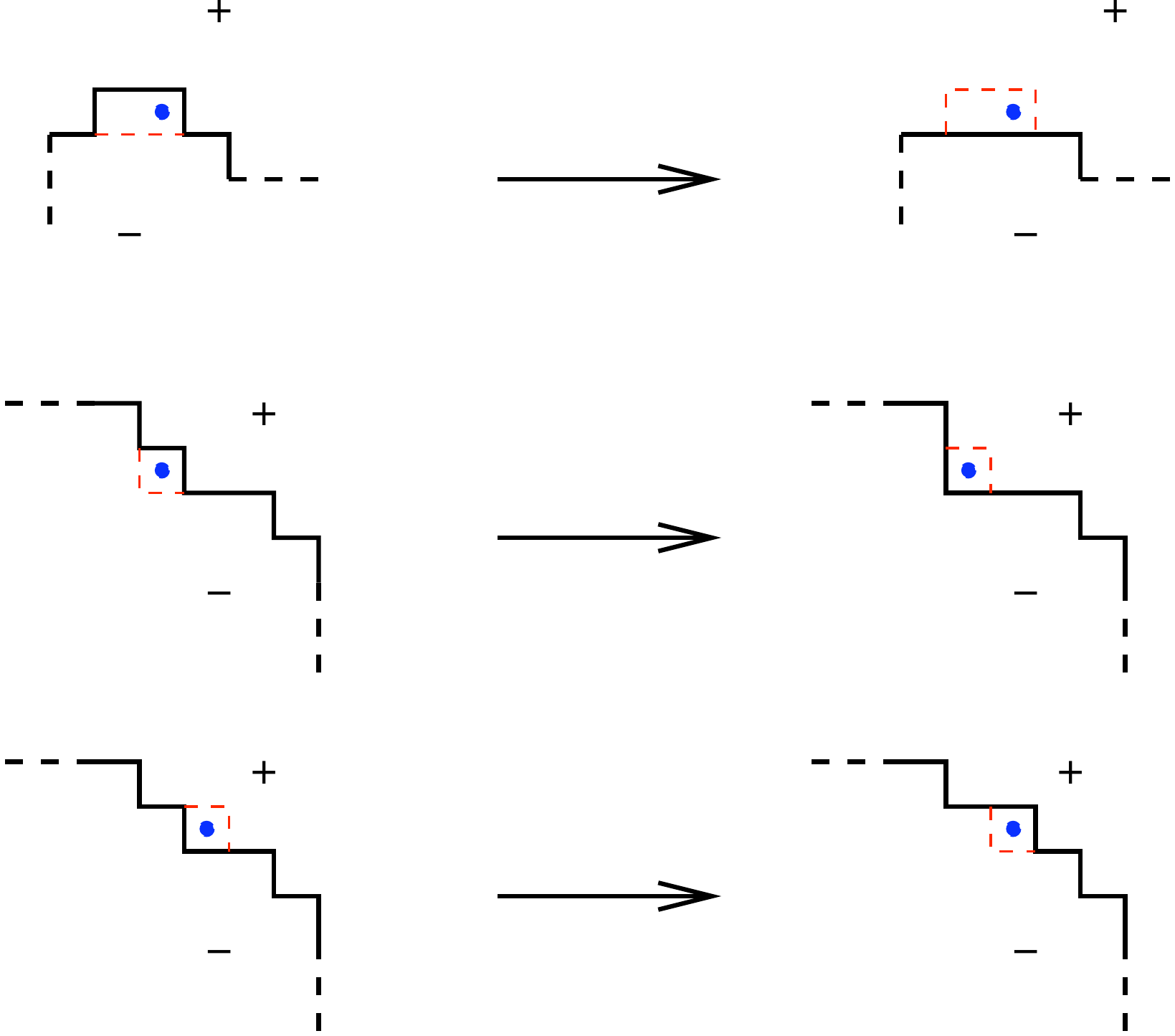}
\end{center}
\caption{In the top drawing, an update consisting in the flip of the ``vertex'' spin $\sigma_x$ ($x$ is marked by a dot) and of its neighboring vertex site from the value ``$-$'' to ``$+$''. 
In this case, $u_{max}^{(2)}$ decreases by $1$ (while $u_{max}^{(1)}$ and $u_{min}^{(i)},i=1,2$ are unchanged) and $\Gamma(\sigma)$ remains a simple curve, but its length decreases by $2$.
In the middle drawing, the effect of flipping from ``$-$'' to ``$+$'' a ``mountain'' $\sigma_x$.  $\Gamma(\sigma)$ remains a simple curve and neither its length nor
the values $\{u_{max}^{(i)},u_{min}^{(i)}\}_{ i=1,2}$ vary.
Similarly, the third drawing shows the effect of flipping a ``valley'' spin. In the three cases, only a portion of $\partial \Gamma(\sigma)$ is drawn.}
\label{fig:mvw} 
\end{figure}

The first observation 
is that, if $\chi:=\tilde\sigma^-(s)\in\mathcal G$, then
nothing happens in the evolution until the Poisson clock of a {\sl
  flippable site $x$} rings (the occurrence of sites with three neighbors of opposite
sign, or sites with exactly two neighbors of opposite sign at mutual distance
$2$ is forbidden by the condition $\chi\in\mathcal G$). When such a ring happens, we have three
possibilities:
\begin{enumerate}
\item $x$ is a valley. Then, with probability $1/2$ the configuration remains unchanged,
and with probability $1/2$ we change $\chi$ to $\chi^{(x)}$,
cf. Definition \ref{def:mvw}. We do not need to apply the
transformation $p(\cdot)$, since one sees easily that $p(\chi^{(x)})=\chi^{(x)}$.
Also, it is easy to see that $|\Gamma(\chi)|=|\Gamma(\chi^{(x)})|$, and that $\Gamma(\chi^{(x)})$ remains a simple curve, see Figure \ref{fig:mvw}.

\item $x$ is a vertex. With probability $1/2$ the configuration remains unchanged,
and with probability $1/2$ we change $\chi$ to $\chi^{(x)}$; then, the application of the transformation $p(\cdot)$ has the effect of flipping also the vertex neighbor of $x$, cf. Remark \ref{rem:vertici}. Altogether, $\Gamma$ remains a simple
curve; its
length decreases by $2$, but also does the sum
$2\sum_{i=1}^2(u_{max}^{(i)}(\chi)-u_{min}^{(i)}(\chi))$,
see Figure \ref{fig:mvw}.

\item $x$ is a mountain. This case is more subtle, since
  it is not obvious apriori that $\Gamma(\chi^{(x)})$  is a simple
  curve. Just to fix ideas, assume that the ``$+$'' neighbors of $x$ in
  $\chi$ are $x+(0,1),x+(1,0)$. If $\Gamma(\chi^{(x)})$ were not a
  simple curve, it would mean that $\chi_{x-(1,1)}=+$. Since
  $\chi\in\mathcal G$, one has that the set
  \begin{eqnarray}
    \label{eq:87}
    Y(x):=\tilde\Lambda\cap \left(\{x-(1,1)-(i,j),i,j\ge 0\}
\cup\{x+(i,j),i,j\ge 0,i+j>0\}\right)
  \end{eqnarray}
 belongs to $\tilde\Lambda\setminus M(\chi)$ (so in particular it has
 no intersection with $D$), otherwise the condition
$|\partial\Gamma(\chi)|=2\sum_{i=1}^2(u_{max}^{(i)}(\chi)-u_{min}^{(i)}(\chi))$
would be violated. However, by the 
 definition of the domains $\tilde\Lambda$ and $D$, there exists no
 site $x\in\tilde \Lambda$ such that $Y(x)\cap D=\emptyset$. Indeed,
 for that to happen one would need that $x=(x^{(1)},x^{(2)})$ with 
either $x^{(1)}\ge (9/10)L$ and $ x^{(2)}\le -(9/10)L$ or $x^{(1)}\le
-(9/10)L$ and $ x^{(2)}\ge (9/10)L$,
which is clearly incompatible with $x\in\tilde\Lambda$, cf. \eqref{eq:88}.
This shows that $\Gamma(\chi^{(x)})$ is a simple curve. Observe that
it is for this issue that it was important to change the shape of the domain from
$\Lambda $ to $\tilde \Lambda$. Of course, the value $(9/10)$ in 
the definition of $D$ could be changed to any number larger than $1/2$.

\end{enumerate}
In all cases, the configuration belongs to $\mathcal G$ after the
move, and the proof of $\tilde\sigma^-(t)\in\mathcal G$ for all $t\ge
0$ is complete.

\medskip

To prove \eqref{eq:MM}, assume that the last update before $\tau_D^-$
consisted in flipping from ``$-$'' to ``$+$'' a ``mountain'' site $x\in \partial D$ (if
instead the move consisted in flipping from ``$-$'' to ``$+$'' a ``vertex''
site, the proof which follows would be
very similar). Call $\chi$ the spin configuration just before
the last update and, just to fix ideas, assume that the two ``$-$'' neighbors of $x$
in $\chi$ are $x-(0,1)$ and $x-(1,0)$. Since $x$ is a mountain and
$\chi\in\mathcal G$, it is immediate to realize
that the set
\begin{eqnarray}
  \label{eq:86}
J(x):=  \tilde\Lambda\cap \{x+(i,j), i,j\ge 0,i+j>0\}
\end{eqnarray}
is a subset of $\tilde \Lambda\setminus M(\chi)$, otherwise the condition
$|\partial\Gamma(\chi)|=2\sum_{i=1}^2(u_{max}^{(i)}(\chi)-u_{min}^{(i)}(\chi))$
would be violated. 
Next, from the definition of the set $D$ one sees that 
$J(x)$ has cardinality at least $\psi L^2$, uniformly in
$x\in\partial D$ (explicitly, one can take $\psi$ to be slightly less
than $\frac12(1-9/10)^2$).
The estimate
\eqref{eq:MM} is then proven.

\hfill$\stackrel{\tiny\mbox{Theorem \ref{th:restogood}}}{\qed}$

The simplification of considering a dynamics which evolves in the set ${\mathcal G}$ is that for such configurations the numbers of valleys, 
mountains and vertices satisfy simple relations:
\begin{lemma}
\label{th:diamante}
Given $\sigma\in{\mathcal G}$, let $v(\sigma)$ (resp. $m(\sigma),
w(\sigma)$) be the number of valleys (resp. mountains, vertices) in $\tilde\Lambda$.
Then,
 $ m(\sigma)+w(\sigma)-v(\sigma)=4$ and
$w(\sigma)\le 8$.
\end{lemma}
The proof of Lemma \ref{th:diamante} is best explained through a picture, and therefore we refer to the caption of Figure \ref{fig:DL}.

\medskip

Now we can finish the proof of \eqref{eq:provanda}.
Thanks to Theorem \ref{th:restogood}, we see that
\eqref{eq:78}
follows if we have
\begin{eqnarray}
  \label{eq:79}
 \tilde \bP( |M(\tilde\sigma^-(0))|- |M(\tilde\sigma^-({c_0 L^2}))|\ge \psi L^2)\le e^{-\gamma L}.
\end{eqnarray}
By the exponential Tchebyshev inequality one has for $\lambda>0$
\begin{eqnarray}
  \label{eq:80}
  \tilde \bP( |M(\tilde\sigma^-(0))|- |M(\tilde\sigma^-({c_0 L^2}))|\ge \psi L^2)&\le& e^{-\lambda\psi L^2}
\tilde{\bf E}
\left[e^{\lambda(|M(\tilde\sigma^-(0))|- |M(\tilde\sigma^-({c_0 L^2}))|)}
\right]\\
&=&e^{-\lambda\psi L^2}\phi(c_0 L^2)
\end{eqnarray}
where $\phi(t):=\tilde{\bf E}
[f(\tilde \sigma^-(t))]$
and
$$
f(\sigma):=e^{\lambda(|M(\tilde\sigma^-(0))|- |M(\sigma)|)}.
$$ Note that
one has $\phi(0)=1$ and
\begin{eqnarray}
  \label{eq:81}
  \frac{d \phi(t)}{d t}=\tilde\mu_t^-\left(\tilde {\mathcal L}f
\right),
\end{eqnarray}
where we recall that $\tilde\mu_t^-$ is the law of $\tilde\sigma^-(t)$
and $\tilde {\mathcal L}$ is its generator.
By the very definition of the modified dynamics, if $M(\sigma)\nsupseteq (D\cup\partial D)$ then $\tilde{\mathcal L}(\sigma,\sigma')$ vanishes for every $\sigma'$. If instead $M(\sigma)\supseteq (D\cup\partial D)$, we know from Theorem \ref{th:restogood} that we need only to consider the case $\sigma\in\mathcal G$. Therefore, for
$M(\sigma)\supseteq (D\cup\partial D)$  one has (recall the discussion
after Remark \ref{rem:vertici})
\begin{eqnarray}
  \label{eq:82}
  \tilde {\mathcal L}f(\sigma)=f(\sigma)\left[
v(\sigma)\left(\frac{e^{-\lambda}-1}2\right)+m(\sigma) \left(\frac{e^{\lambda}-1}2\right)+w(\sigma) \left(\frac{e^{2\lambda}-1}2\right)
\right].
\end{eqnarray}
Now we choose $\lambda=1/L$. Since clearly there exists a constant $c$ such that $v(\sigma),m(\sigma),w(\sigma)\le c L$ for every $\sigma\in{\mathcal G}$, 
we have
\begin{eqnarray}
  \label{eq:83}
   \tilde {\mathcal L}f(\si)\le f(\sigma)\left[\frac\lambda2(m(\sigma)+2w(\sigma)-v(\sigma))+\frac cL
\right]\le \frac {c'}L f(\sigma),
\end{eqnarray}
where we used Lemma \ref{th:diamante}.
Plugging this inequality into \eqref{eq:81}, we find
\begin{eqnarray}
  \label{eq:84}
  \frac{d \phi(t)}{dt}\le \frac {c'}L\phi(t)
\end{eqnarray}
which implies $\phi(c_0 L^2)\le e^{c'\,c_0 L}$ and,
together with \eqref{eq:80},
\begin{eqnarray}
  \label{eq:85}
  \tilde\bP\left( |M(\tilde\sigma^-(0))|- |M(\tilde\sigma^-({c_0 L^2}))|\ge \psi L^2\right)&\le& e^{-L(\psi-c'\,c_0)}.
\end{eqnarray}
Choosing $c_0<\psi/c'$ we obtain \eqref{eq:79} and therefore \eqref{eq:provanda}.

\hfill$\stackrel{\tiny\mbox{Eq. \eqref{eq:provanda}}}{\qed}$

\section{Proof of Theorem \ref{th:d2gap}}
\label{sec:GAP2D}

\subsection{$\Omega(1/L)$ lower bound on the gap: a perturbative argument}
\label{sec:LBgap3D}

Here we prove the lower bound  $\gap\ge c/L$ for $\beta\ge C\log L$ with 
$C$ large enough. The result is particularly interesting in $d=2$, in
view of the matching upper bound in Theorem \ref{th:d2gap}, but as we
mentioned the proof works also in $d=3$. Actually, we give the proof
in the three-dimensional case, which is slightly more complicated.

We introduce the matrix
$U:=\{U(\sigma,\sigma')\}_{\{\sigma,\sigma'\in\Omega_\Lambda\}}$, unitarily
equivalent to the matrix $\mathcal L$ (the generator \eqref{eq:106}), as
\begin{eqnarray}
  \label{eq:114}
  U(\sigma,\sigma'):=\sqrt{\pi(\sigma)}\mathcal L(\sigma,\sigma')\frac1{\sqrt{\pi(\sigma')}}.
\end{eqnarray}
The spectrum of $U$ coincides with the spectrum of $\cL$, so that we
have to prove that the smallest non-zero eigenvalue of $-U$ is lower
bounded by $c/L$. Note that $U$ is symmetric thanks to the reversibility
condition
$\pi(\sigma)\cL(\sigma,\sigma')=\pi(\sigma')\cL(\sigma',\sigma)$.

\begin{rem}
  \rm
The transformation \eqref{eq:114} is the analogue of {\sl the inverse} of
the 
``ground-state transformation'' which maps a Schr\"odinger
operator of the form $H=-\Delta +V$ (which acts on $\mathbb L^2(\RR^d)$) into the
operator $$-\cL=
A^{-1} H A=-\Delta-\frac2{\psi_0}\nabla \psi_0\cdot \nabla,$$
where $\psi_0(\cdot)$ is the ground state eigenfunction
(we assume for definiteness that $H \psi_0=0$, i.e.\ the ground state
energy is zero) and  the unitary operator $A$ acts as
$(A f)(x)=\psi_0(x)f(x)$
for $f\in  L^2(\RR^d)$.
Note that $\cL$ is the generator of a diffusion with drift.

\end{rem}

Looking at the definition of $U$ and $\cL$, one immediately realizes that
\begin{itemize}

\item 
\begin{eqnarray}
  \label{eq:115}
  U(\si,\si)=\cL(\si,\si)=-\sum_{\si'\ne\si}\cL(\si,\si').
\end{eqnarray}

\item if $\si'=\si^{(x)}$ for some
$x\in\Lambda$ and $H^+_\Lambda(\sigma)=H^+_\Lambda(\sigma')$, then
$$U(\si,\si')=\cL(\si,\si')=\frac12;$$

\item if $\si'=\si^{(x)}$ for some
$x\in\Lambda$ and $|H^+_\Lambda(\sigma')-H^+_\Lambda(\sigma)|>0$
(hence $\ge 4$)
then
\begin{gather}
  \label{eq:104bis}
  \cL(\si,\si')=\frac1{1+\exp(-\beta(H^+_\Lambda(\sigma)-H^+_\Lambda(\sigma')))}\\
\label{gaga}
U(\si,\si')=\frac{\exp(-(\beta/2) (H^+_\Lambda(\sigma)-H^+_\Lambda(\sigma')))}{1+\exp(-\beta(H^+_\Lambda(\sigma)-H^+_\Lambda(\sigma')))}\le e^{-2\beta};
\end{gather}

\item if $\si\ne\si'$ and there exists no $x$ such that $\si'=\si^{(x)}$, then $\cL(\si,\si')=U(\si,\si')=0$.

\end{itemize}
 If we write $U= U_\infty+R$ with $ U_\infty=\lim_{\beta\to\infty}U$ we see that
all the matrix elements of $R$ are smaller than $1/L^5$ if $\beta$
satisfies \eqref{eq:7} with $C$ large enough. Since each row of $R$
has at most $|\Lambda|=O(L^3)$ non-zero elements, one sees easily that the spectral
radius of $R$ is $O(L^{-2})$.

We are therefore left with the task of proving that the smallest non-zero
eigenvalue of 
 $- U_\infty$ is larger than $c/L$. 
Thanks to formulas \eqref{eq:115}-\eqref{gaga}, we
have that 
\begin{eqnarray}
 \label{eq:117}
 U_\infty(\si,\si)=
-\frac12|\{x\in\Lambda:H^+_\Lambda(\si^{(x)})=
H^+_\Lambda(\si)\}|-|\{x\in\Lambda:H^+_\Lambda(\si^{(x)})< H^+_\Lambda(\si)\}|
\end{eqnarray}
and, for $\si\ne\si'$,
\begin{eqnarray}
 U_\infty(\si,\si')&=&
\left\{
  \begin{array}{lllll}
\frac12 &\mbox{if} &\si'=\si^{(x)} &\mbox{and}&
 H^+_\Lambda(\si')=H^+_\Lambda(\si);\\
0  &\mbox{otherwise.}    
  \end{array}
\right.
\end{eqnarray}
If we decompose $\Omega_\Lambda$ into a finite number $M$ of
equivalence classes
$\mathcal C_i,1\le i\le M$, where two configurations belong to the same class
iff
they can be connected via a finite number of single spin-flips which
do not change the energy $H^+_\Lambda(\cdot)$, then $U_\infty(\si,\si')=0$
whenever $\si\in\cC_i,\si'\in\cC_j$ with $i\ne j$. In other words, one
can write $U_\infty$ in a block matrix form $U_\infty=
\oplus_{i=1}^M U_\infty^{(i)}$ 
where $U_\infty^{(i)}=\{U_\infty(\si,\si')\}_{\{\si,\si'\in
  \cC_i\}}$ and in particular, if $\cS(U_\infty)$ denotes the spectrum
of $U_\infty$, one has $\cS(U_\infty)=\cup_{i=1}^M \cS(U^{(i)}_\infty)$.
It is clear that, if $\cC_1=\{+\}$ denotes the equivalence class whose
unique element is the all ``$+$'' configuration, one has
$\cS(U^{(1)}_\infty)=\{0\}$ (cf. \eqref{eq:117} and observe that any
spin flip increases the energy if $\si\equiv +$).
We need to prove that, for every $i> 1$, 
$\cS(U^{(i)}_\infty)\subset (-\infty,-c/L)$ for some positive $c$.

Let us fix $i>1$, let $\lambda=\lambda_i$ be the smallest eigenvalue of
$-U_\infty^{(i)}$ and $g:\cC_i\mapsto \RR$ an associated
eigenfunction.
We want to show that
\begin{eqnarray}
  \label{eq:123}
  \lambda\ge \frac cL
\end{eqnarray}
with $c$ independent of $i$.
The key point is the following:
\begin{lemma}
\label{th:killed}
  For every $\eta,\eta'\in\cC_i$ and $t>0$, one has
  \begin{eqnarray}
    \label{eq:119}
    e^{t U^{(i)}_\infty}(\eta,\eta')=\bP^{\eta}\left(\sigma^\eta(t)=\eta'
;\tau>t
\right)
  \end{eqnarray}
where $\bP^\eta$ denotes the law of the Ising evolution
$\{\sigma^\eta(t)\}_{t\ge0}$ in $\Lambda$ at
$\beta=+\infty$ with $+$ boundary condition, started from the
configuration $\eta$, and $\tau$ is the random time 
\begin{eqnarray}
  \label{eq:120}
  \tau=\inf\{t>0:H^+_\Lambda(\sigma(t))<H^+_\Lambda(\eta)\}.
\end{eqnarray}
\end{lemma}
\proof
Just check that the time derivatives of left- and right-hand side are
the same.
\hfill$\stackrel{\tiny\mbox{Lemma \ref{th:killed}}}{\qed}$
\begin{rem}
  \rm Note that $U_\infty^{(i)}$  is just the
  generator of the ``killed'' Markov process which coincides with the Ising
  evolution except that it is killed when it exits the
  set $\mathcal C_i$, cf. \eqref{eq:119}.
\end{rem}

\smallskip

Since the matrix $\{\exp(t
U^{(i)}_\infty)(\eta,\eta')\}_{\eta,\eta'\in\cC_i}$ has strictly
positive entries for $t>0$
(this follows from \eqref{eq:119} and the definition of the
equivalence classes $\cC_i$),
the Perron-Frobenius theorem implies that the eigenvalue $\lambda$ is
non-degenerate and the eigenfunction $g$ can be chosen 
strictly positive on $\cC_i$. Normalizing $g$ so that
$\sum_{\si\in\cC_i}g(\si)^2=1$, we have
\begin{eqnarray}
  \label{eq:121}
  e^{-\lambda t}=\sum_{\si,\si'\in\cC_i}g(\si)  e^{t U^{(i)}_\infty}(\si,\si')g(\si').
\end{eqnarray}
The desired estimate \eqref{eq:123} then follows from Lemma
\ref{th:killed} by letting $t\to\infty$, once we prove the following result:
\begin{theo}
\label{th:blocchi}
  There exists $a(L)<\infty$ and $c>0$ independent of $L$ such that, for every $i>1$,
  $\eta,\eta'\in\cC_i$ and $t>0$ one has
  \begin{eqnarray}
    \label{eq:122}
    \bP^{\eta}\left(\sigma^\eta(t)=\eta'
;\tau>t
\right)\le a(L)\exp\left(-c\,\frac{t}L\right).
  \end{eqnarray}
\end{theo}
\proof
Let $V$ be the smallest parallelepiped which contains the set
$\{x\in\Lambda:\eta_x=-\}$, and let $V^+$ be the rectangular layer of points
of $V$ with maximal vertical coordinate. Plainly, 
$\tau\le \tau_V$ where
\begin{eqnarray}
  \label{eq:124}
  \tau_V=\inf\{t>0:\sigma^\eta_x(t)=+\mbox{\;for every\;} x\in V^+\}:
\end{eqnarray}
this is because when the last ``$-$'' spin in $V^+$ flips to ``$+$'',
the energy decreases by at least $8$.
Then, 
\begin{eqnarray}
  \label{eq:125}
   \bP^{\eta}\left(\sigma^\eta(t)=\eta'
;\tau>t
\right)\le \bP^{\eta}\left(
\tau_V>t
\right)\le \bP^{\eta_V}\left(
\tau_V>t
\right)
\end{eqnarray}
where $\eta_V$ is the configuration where spins take value ``$-$'' in
$V$ and ``$+$'' in $\Lambda\setminus V$, and we used monotonicity of
the dynamics (the event $\tau_V>t$ is decreasing, and $\eta_V\le \eta$). Again by monotonicity, we can assume that $V$ is
the entire domain $\Lambda$, and that spins in $V\setminus V^+$
are frozen at the value
``$-$'' during the entire evolution:
both these operations make $\tau_V$ stochastically larger. But, in
this case, $V^+$ is just a $(2L+1)\times (2L+1)$
square, and the evolution in $V^+$ coincides with the 
evolution of the two-dimensional Ising model with $+$ boundary
conditions and $\beta=+\infty$ (the ``$+$'' boundary conditions on the
upper face of $V^+$ and the ``$-$'' boundary conditions on the
lower face compensate exactly).
We have then
\begin{eqnarray}
  \label{eq:126}
   \bP^{\eta_V}\left(
\tau_V>t
\right)\le \bP^{(d=2)}\left(
\tau_+>t\right)
\end{eqnarray}
where $\bP^{(d=2)}$ is the law of the evolution of the 
two-dimensional Ising model in the $(2L+1)\times (2L+1)$ square  with ``$+$'' boundary
conditions started from the ``$-$'' configuration, and  $\tau_+$ was
defined in \eqref{eq:4} as the first time when all spins in the square
are ``$+$''.
Finally, one has
\begin{eqnarray}
  \label{eq:127}
  \bP^{(d=2)}\left(
\tau_+>t\right)\le \left[ \bP^{(d=2)}\left(\tau_+\ge\frac {L^2}c\right)
\right]^{\lfloor tc/L^2\rfloor }\le a(L)e^{-c\gamma \frac tL}
\end{eqnarray}
where in the first inequality we used monotonicity (if $\tau_+$ has
not been reached at time $i\,L^2/c,i=1,\ldots,\lfloor tc/L^2\rfloor -1$, we restart the dynamics from the all
``$-$'' configuration)
and in the second one the result \eqref{eq:2} of Theorem \ref{th:d2}.
This ends the proof of Theorem \ref{th:blocchi}.
\hfill$\stackrel{\tiny\mbox{Theorem \ref{th:blocchi}}}{\qed}$

\begin{rem}\rm
The asymptotic behavior $\exp(-c \,t/L)$ can be understood by comparison
with the symmetric simple exclusion process (SSEP), as already noticed
in \cite{FSS}. In fact, by monotonicity, the tail of the law of $\tau_+$ is bounded from above by
the tail of the hitting time $\tau_0$ defined as follows. Consider the SSEP on $[-L,L]$
starting with all the negative sites occupied and all the non-negative
ones empty and define $\tau_0$ to be the first time at which the site
``$L$'' is occupied. 
In turn, using beautiful results by Liggett and by Arratia
\cite{Arratia}, the tail of $\tau_0$ is controlled from above by the tail of the same hitting time
but for the process in which the $L$ particles evolve as independent
random walks without the exclusion constraint. Finally the latter has
a tail $\exp(-t/L)$ simply because the tail for a single random walk
is $\exp(-t/L^2)$ and there are $L$ of them.
\end{rem}

\subsection{Upper bound on the gap} 
We will prove here the upper bound $\gap\le 1/(cL)$ for the dynamics in the square box $\Lambda=\{-L,\ldots,L\}^2$.
We use the variational characterization \eqref{eq:72} of the spectral
gap and we note that, if $\mathcal Y$ is some subset of the configuration space
$\Omega_\Lambda$, one has
\begin{eqnarray}
  \label{eq:73}
  \gap\le\frac1{\pi(\mathcal Y^{\tt c})}\inf_{f:f|_{\mathcal Y^{\tt c}}=0}\frac{\pi(f(-\mathcal L)f)}{\pi(f^2)},
\end{eqnarray}
where the infimum is taken over functions which vanish on $\mathcal Y^{\tt c}$: just observe that
\begin{eqnarray}
  \label{eq:74}
\Var_\pi(f)&=&\frac12\int(f(\sigma)-f(\tau))^2\pi(d\sigma)\pi(d\tau)
\\&\ge&
\frac12\int(f(\sigma)-f(\tau))^2\pi(d\sigma) \pi(d\tau)({\bf
  1}_{\tau\in \mathcal Y^{\tt c}} +{\bf
  1}_{\sigma\in \mathcal Y^{\tt c}}
) 
= \pi(\mathcal Y^{\tt c})\pi(f^2).
\end{eqnarray}
The bound $\gap\le 1/(cL)$  is therefore proven if one exhibits a set
$\mathcal Y$ such that $\pi(\mathcal Y^{\tt c})>1/2$ and a function $f$
which vanishes on $\mathcal Y^{\tt c}$ and such that
\begin{eqnarray}
  \label{eq:75}
\frac{\pi(f(-\mathcal L)f)}{\pi(f^2)}\le \frac cL.  
\end{eqnarray}

For $\sigma\in\Omega_\Lambda$, we let $M(\sigma)$ and $\Gamma(\sigma)$ be as in Definition \ref{def:Gset} and we let $Q\subset \RR^2$ be 
the square $[-L-1/2,L+1/2]^2$. We consider the eight points of
$\partial Q$ which are at distance $L/2$ from a corner of $Q$
and we call them $p_i, i=1,\ldots,8$, with the convention that we order them clockwise, starting from the right-most one on the north side of $Q$.
Let us assume for definiteness that
$L$ is even.

\begin{figure}[htp]
\begin{center}
\includegraphics[width=0.6\textwidth]{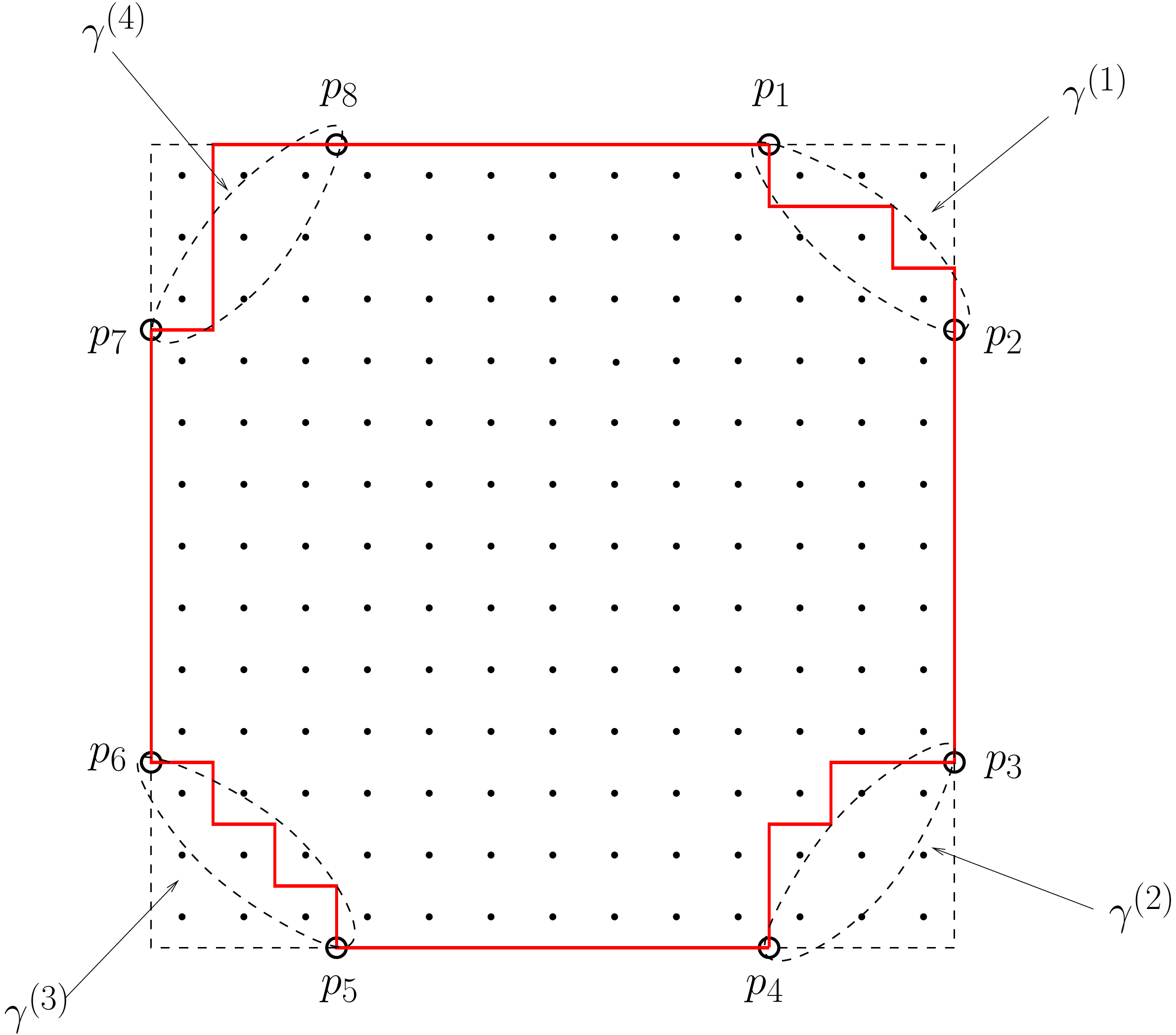}
\end{center}
\caption{The domain $\Lambda$ with $L=6$ and a configuration $\sigma\in\mathcal
  Y$: only the boundary $\partial \Gamma(\sigma)$ is drawn (thick
  line). The four portions $\gamma^{(i)},i=1,\ldots,4$ of $\partial
  \Gamma(\sigma)$ cannot intersect and, if suitably
  rotated, are just lattice paths. According to formula
  \eqref{eq:dlta}, one has in this example
  $\nabla^{(1)}_{-3}=\nabla^{(1)}_{0}=\nabla^{(1)}_{2}=-1$
and  $\nabla^{(1)}_{-2}=\nabla^{(1)}_{-1}=\nabla^{(1)}_{1}=+1$.}
\label{fig:paths} 
\end{figure}

\begin{definition}
\label{def:Y}
  We let $\mathcal Y$ be the subset of configurations in $\Omega_\Lambda$ such that $\Gamma(\sigma)$ is a simple curve of length $4(2L+1)$ and such that
$p_i\in\partial \Gamma(\sigma),i=1,\ldots,8$.
\end{definition}
Note that $4(2L+1)$ is just the length of $\partial Q$. Also, if
$\sigma\in\mathcal Y$ then the following properties are immediately
verified, cf. Figure
 \ref{fig:paths} :
\begin{enumerate}
\item the portion of $\partial \Gamma(\sigma)$ which connects $p_{2i}$
  to $p_{2i+1},i=1,2,3$ or $p_{8}$ to $p_1$ is a straight segment of
  length $L+1$
included in $\partial Q$;

\item the  portion of $\partial \Gamma(\sigma)$ which connects $p_{2i-1}$ to $p_{2i},i=1,\ldots,4$ (call it $\gamma^{(i)}$) is a ``lattice
path''  of length $L$ (in the sense of Section \ref{app:wilson}). More precisely, if $\gamma^{(i)}$ is rotated counter-clockwise by an angle $\pi/4+\pi/2(i-1)$, then
expanded by a factor $\sqrt 2$  and finally 
is  suitably translated, then it becomes the graph of a function
$x\in[-L/2,L/2]\mapsto 
\phi^{(i)}_x:=\phi^{(i)}_x(\sigma)\in[-L/2,L/2]$ such that
\begin{gather}
\label{eq:28}
\phi^{(i)}_{\pm L/2}=0\\
\label{eq:dlta}
\nabla^{(i)}_x:=\phi^{(i)}_{x+1}-\phi^{(i)}_x\in\{-1,+1\} \mbox{\;for\;}x\in \ZZ\cap [ -L/2, L/2-1]\\
\label{eq:uu}
\phi^{(i)}_\cdot \mbox{\;is linear in each interval\;} (x,x+1), x\in \ZZ\cap [ -L/2, L/2-1].
\end{gather}
\end{enumerate}

The test function $f$ in \eqref{eq:75} is then
\begin{eqnarray}
  \label{eq:27}
  f(\sigma):={\bf 1}_{\{\sigma\in\mathcal Y\}}\,\prod_{i=1}^4 g(\phi^{(i)}(\sigma)),
\end{eqnarray}
where 
\begin{eqnarray}
  \label{eq:103}
  g(\phi^{(i)}):=\prod_{x=-L/2}^{-1}\cos\left(\frac{\pi x}{L}\right)^{(1-\nabla^{(i)}_x)/2}\prod_{x=0}^{(L/2)-1}\cos\left(\frac{\pi (x+1)}{L}\right)^{(1+\nabla^{(i)}_x)/2}
\end{eqnarray}
with the convention that $0^0=1$. Note that $f$ vanishes on $\mathcal
Y^{\tt c}$ and that $$\pi(\mathcal Y^{\tt c})\ge \pi(\sigma_x=+ \mbox{\;for every\;} x\in \Lambda)=1+o(1)>1/2$$ as required above.

Let $\rho$ be the uniform measure over the set of lattice paths $\phi$ satisfying \eqref{eq:28}-\eqref{eq:uu}, and let $\hat {\mathcal L}$ be
the generator 
\begin{eqnarray}
  \label{eq:30}
\hat{\mathcal L}g(\phi)=\sum_{x=-L/2+1}^{L/2-1}\bigl[\rho(g|\phi_{y,y\in\ZZ,y\ne x})
-g(\phi)\bigr ].
\end{eqnarray}

\begin{rem}\rm
\label{rem:otimes}
 Conditionally on
$\sigma\in\mathcal Y$, the four lattice paths $\phi^{(i)}$ are independent and,
since
all their allowed configurations have the same length,
$\pi(\cdot|\mathcal Y)$ is just 
the product uniform measure $\rho^{\otimes 4}$, where 
each copy of $\rho$ acts on a path $\phi^{(i)},i=1,\ldots,4$. Indeed, the Ising Hamiltonian \eqref{eq:33} equals
\begin{eqnarray}
  \label{eq:110}
  H^+_\Lambda(\sigma)=const+2|\Gamma(\sigma)|
\end{eqnarray}
which does not depend on $\sigma$ if $\sigma\in\mathcal Y$, cf. Definition \ref{def:Y}.
\end{rem}

\begin{lemma}
\label{th:lemmapiro}
  If $f$ is defined as in \eqref{eq:27}, one has 
  \begin{eqnarray}
    \label{eq:29}
    \frac{\pi(f(-\mathcal L)f)}{\pi(f^2)}\le 4\frac{\rho(g(-\hat
      {\mathcal L})g)}{\rho(g^2)}+\frac c{L^2}.
  \end{eqnarray}
\end{lemma}

\proof
Let us consider first the denominator of the left-hand side: 
\begin{eqnarray}
  \label{eq:104}
  \pi(f^2)=\pi(\mathcal Y)\pi\left(\left.\prod_{i=1}^4
      g^2(\phi_i)\right| \mathcal  Y\right)=
\pi(\mathcal Y)\left[\rho(g^2)
\right]^4
\end{eqnarray}
where we used the fact that $\pi(\cdot|\mathcal Y)=\rho^{\otimes 4}$,
see
Remark \ref{rem:otimes}.

The numerator requires more work. From the definition \eqref{eq:106}
of the generator
one has
\begin{eqnarray}
  \label{eq:105}
 \pi(f(-\mathcal L) f)=\sum_{\sigma\in\mathcal
    Y}\pi(\sigma)\prod_{i=1}^4
g(\phi^{(i)}(\sigma))\sum_{z\in\Lambda}\left(f(\sigma)-\pi_{z,\sigma}(f)\right)
\end{eqnarray}
where $\pi_{z,\sigma}(\cdot)$ is defined in \eqref{eq:113}.
The first observation is that, if $\beta\ge C\log L$ with $C$ large
enough, then
\begin{eqnarray}
  \label{eq:107}
  |\pi_{z,\sigma}(f)-f(\sigma)|\le \frac1{L^4}f(\sigma)
\end{eqnarray}
whenever $\sigma\in\mathcal Y$ and $z$ is such that the configuration
$\sigma^{(z)}$  satisfies
$|\Gamma(\sigma^{(z)})|\ge |\Gamma(\sigma)|+1$ (in particular,
$\sigma^{(z)}$ 
does not
belong to $\mathcal Y$). The reason is that, since
$f(\sigma^{(z)})=0$, one has
\begin{eqnarray}
  \label{eq:111}
  \pi_{z,\sigma}(f)-f(\sigma)=-\pi_{z,\sigma}(-\sigma_z)f(\sigma),
\end{eqnarray}
where $\pi_{z,\sigma}(-\sigma_z)$ is the probability,
under $\pi_{z,\sigma}$, that the spin at $z$ equals $-\sigma_z$, and
this probability is smaller than $\exp(-2 \beta)\le L^{-4}$ by the
assumption that $|\Gamma(\sigma^{(z)})|\ge |\Gamma(\sigma)|+1$
(cf. \eqref{eq:110}).

Call $B(\sigma)$ the set of $z\in\Lambda$ such
that $|\Gamma(\sigma^{(z)})|= |\Gamma(\sigma)|$. One has then
\begin{eqnarray}
  \label{eq:108}
   \pi(f(-\mathcal L) f)\le \sum_{\sigma\in\mathcal
    Y}\pi(\sigma)\prod_{i=1}^4
g(\phi^{(i)}(\sigma))\sum_{z\in
  B(\sigma)}\left(f(\sigma)-\pi_{z,\sigma}(f)\right)+\frac
c{L^2}\pi(f^2)
\end{eqnarray}
since, if $\sigma\in\mathcal Y$, there are no spin flips 
which decrease $|\Gamma(\sigma)|$ (i.e.\ the energy).
Finally, recalling Remark \ref{rem:otimes} it is an easy task to check
the identity
\begin{eqnarray}
  \label{eq:112}
  \sum_{\sigma\in\mathcal
    Y}\pi(\sigma)\prod_{i=1}^4
g(\phi^{(i)}(\sigma))\sum_{z\in
  B(\sigma)}\left(f(\sigma)-\pi_{z,\sigma}(f)\right)=4\,\pi(\mathcal Y)
\left[\rho(g^2)\right]^3 \rho(g(-\hat
      {\mathcal L})g),
\end{eqnarray}
the reason being that the flips which leave
$|\Gamma(\sigma)|$
unchanged just correspond to the local updates in the generator
$\hat {\mathcal L}$ of \eqref{eq:30}.
The factor $4$ is due to the
symmetry among the lattice paths $\phi^{(i)},i=1,\ldots,4$.
\hfill$\stackrel{\tiny\mbox{Lemma \ref{th:lemmapiro}}}{\qed}$

The proof of \eqref{eq:75} is concluded once we prove
\begin{theo}
\label{th:ensembles}
  There exists $c>0$ such that
\begin{eqnarray}
  \label{eq:109}
  \frac{\rho(g(-\hat
      {\mathcal L})g)}{\rho(g^2)}\le \frac cL.
\end{eqnarray}
\end{theo}
 \proof
Using the form \eqref{eq:30} of the generator and the fact that $\rho$
is the uniform measure over lattice paths, one sees that 
\begin{eqnarray}
  \label{eq:116}
  \rho(g(-\hat
      {\mathcal L})g)&=&\frac12\sum_{x=-L/2+1}^{-1}\frac{\left[
\cos(\pi x/L)-\cos(\pi (x-1)/L)
\right]^2}{\cos^2(\pi x/L)}\rho\left(g^2{\bf
1}_{\{\nabla_{x-1}=+1,\nabla_x=-1\}}\right)\\
\nonumber
&&+ \frac12\sum_{x=-L/2+1}^{-1}\frac{\left[
\cos(\pi (x-1)/L)-\cos(\pi x/L)
\right]^2}{\cos^2(\pi (x-1)/L)}\rho\left(g^2{\bf
1}_{\{\nabla_{x-1}=-1,\nabla_x=+1\}}\right)
\end{eqnarray}
where we recall that $\nabla_x=\phi_{x+1}-\phi_x$ and we used the
symmetry $x\leftrightarrow -x$ to restrict the sum to $x<0$.
Clearly, one has 
\begin{eqnarray}
  \label{eq:129}
\sum_x\left[
\cos(\pi x/L)-\cos(\pi (x-1)/L)
\right]^2\le c/L.  
\end{eqnarray}
Moreover, by an ``equivalence of ensembles''
argument, one has
\begin{eqnarray}
  \label{eq:128}
  \frac{\rho\left(g^2{\bf
1}_{\{\nabla_{x-1}=+1,\nabla_x=-1\}}\right)
}{\rho(g^2)}\le c' \cos^2(\pi x/L)
\end{eqnarray}
and a similar estimate for $ \rho\left(g^2{\bf
1}_{\{\nabla_{x-1}=-1,\nabla_x=+1\}}\right)/\rho(g^2)$
where $c'$ is finite uniformly in $x<-1$.
The statement of the Theorem then immediately follows from
Eqs. \eqref{eq:116}, \eqref{eq:129} and \eqref{eq:128}.

The proof of \eqref{eq:128} is a direct adaptation of the proof of
\cite[Prop. 3.8]{CM}. Loosely speaking, the idea is to observe that
the law of $\{\nabla_x\}_{x=-L/2,\ldots,L/2-1}$ under $\r(\cdot\,
g^2)/\rho(g^2)$ is a product law of independent (but not identically distributed) random variables
conditioned to the event $\sum_{x=-L/2}^{L/2-1}\nabla_x=0$. Next, one
shows that, when computing the average of a local function like ${\bf
  1}_{\{\nabla_{x-1}=+1,\nabla_x=-1\}}$, the conditioning can be
eliminated, at the price of losing a multiplicative constant.  Under
the unconditioned product law, the expectation of ${\bf
  1}_{\{\nabla_{x-1}=+1,\nabla_x=-1\}}$ is easily computed and turns
out to be proportional to the r.h.s. of \eqref{eq:128}.

\hfill$\stackrel{\tiny\mbox{Theorem \ref{th:ensembles}}}{\qed}$

\section*{Acknowledgements}
We wish to thank Rick Kenyon for very  helpful e-mail exchanges about
dimer coverings and height fluctuations and David Wilson for a precious
suggestion on the proof of the mixing time lower bound in three
dimensions. 
This work has been carried out while F. Simenhaus and F. L. Toninelli
were visiting the Department of
Mathematics of the University of Roma Tre under the ERC Advanced Research
Grant ``PTRELSS''.
They both gratefully acknowledge the kind hospitality and
support.

\end{document}